\documentclass[10pt,a4paper]{scrartcl}

\pdfoutput=1

\usepackage[utf8]{inputenc}

\usepackage{textcomp,marvosym}

\usepackage{fixltx2e}

\usepackage{amsmath,amssymb}

\usepackage{cite}

\usepackage{nameref,hyperref}

\usepackage{rotating}

\usepackage[section]{placeins}

\usepackage{subcaption}

\date{}

\newcommand{\argmax}{\operatornamewithlimits{argmax}}

\begin{document}
\vspace*{0.35in}

\begin{flushleft}
{\Large
\textbf\newline{Inter-protein sequence co-evolution predicts known physical interactions in bacterial ribosomes and the trp operon}
}
\newline
\\
Christoph Feinauer\textsuperscript{1},
Hendrik Szurmant\textsuperscript{2},
Martin Weigt\textsuperscript{3,4,*},
Andrea Pagnani\textsuperscript{1,5,*}
\\
\bigskip
\bf{1} Department of Applied Science and Technology, and Center for Computational Sciences, Politecnico di Torino, Torino, Italy. 
\\
\bf{2}  Department of Molecular and Experimental Medicine, The Scripps Research Institute, 
La Jolla, CA, USA
\\
\bf{3} Sorbonne Universit\'es, UPMC, UMR 7238, Computational and Quantitative Biology, Paris, France
\\
\bf{4} CNRS, UMR 7238, Computational and Quantitative Biology, Paris, France
\\
\bf{5} Human Genetics Foundation, Molecular Biotechnology Center (MBC), Torino, Italy
\bigskip

%
%





* andrea.pagnani@polito.it, martin.weigt@upmc.fr

\end{flushleft}
\section*{Abstract}
Interaction between proteins is a fundamental mechanism that underlies
virtually all biological processes. Many important interactions are
conserved across a large variety of species. The need to maintain
interaction leads to a high degree of co-evolution between residues in
the interface between partner proteins. The inference of
protein-protein interaction networks from the rapidly growing sequence
databases is one of the most formidable tasks in systems biology
today. We propose here a novel approach based on the {\em
  Direct-Coupling Analysis} of the co-evolution between inter-protein
residue pairs. We use ribosomal and trp operon proteins as test cases:
For the small resp. large ribosomal subunit our approach predicts
protein-interaction partners at a true-positive rate of 70\%
resp. 90\% within the first 10 predictions, with areas of 0.69
resp. 0.81 under the ROC curves for all predictions. In the trp
operon, it assigns the two largest interaction scores to the only two
interactions experimentally known. On the level of residue
interactions we show that for both the small and the large ribosomal
subunit our approach predicts interacting residues in the system with
a true positive rate of 60\% and 85\% in the first 20 predictions.  We
use artificial data to show that the performance of our approach
depends crucially on the size of the joint multiple sequence
alignments and analyze how many sequences would be necessary for a
perfect prediction if the sequences were sampled from the same model
that we use for prediction.  Given the performance of our approach on
the test data we speculate that it can be used to detect new
interactions, especially in the light of the rapid growth of available
sequence data.


\section*{Introduction}

Proteins are the major work horses of the cell. Being part of all
essential biological processes, they have catalytic, structural,
transport, regulatory and many other functions. Few proteins exert
their function in isolation. Rather, most proteins take part in
concerted physical interactions with other proteins, forming networks
of protein-protein interactions (PPI). Unveiling the PPI organization
is one of the most formidable tasks in systems biology today.
High-throughput experimental technologies, applied for example in
large-scale yeast two-hybrid \cite{ito2001PNAS} analysis and in
protein affinity mass-spectrometry studies \cite{ho2002NATURE},
allowed a first partial glance at the complexity of organism-wide PPI
networks. However, the reliability of these methods remains
problematic due to their high false-positive and false-negative rates
\cite{braun2008NATUREMETHODS}.

Given the fast growth of biological sequence databases, it is tempting
to design computational techniques for identifying protein-protein
interactions \cite{harrington2008predicting}. Prominent techniques to
date include: the genomic co-localization of genes
\cite{dandekar1998conservation,galperin2000s} (with bacterial operons
as a prominent example), the Rosetta-stone method
\cite{marcotte2002predicting} (which assumes that proteins fused in
one species may interact also in others), phylogenetic profiling
\cite{pellegrini1999assigning} (which searches for the correlated
presence and absence of homologs across species), and similarities
between phylogenetic trees of orthologous proteins
\cite{Valencia2002A,Valencia2002B, Haussler2007,juan2008high}. Despite
the success of all these methods, their sensitivity is limited due to
the analysis of coarse global proxies for protein-protein
interaction. An approach that exploits more efficiently the large
amount of information stored in multiple sequence alignments (MSA)
seems therefore promising.

Recently, a breakthrough has been achieved using genomic sequences for
the related problem of inferring residue contacts from sequence data
alone.  \cite{valencia2013NATREVGEN}. The so-called Direct-Coupling
Analysis (DCA) \cite{weigt2009identification,morcos2011direct} allows
to construct statistical models that are able to describe the sequence
variability of large MSA of homologous proteins \cite{PFAM}. More
precisely, these models reproduce the empirically measured
covariations of amino acids at residue pairs. The parameters of the
models unveil networks of direct residue co-evolution, which in turn
accurately predict residue-residue contacts.


These models are computationally hard to infer and several
approximations have therefore been developed
\cite{weigt2009identification,morcos2011direct,aurell2013pl,Baldassi2014PLOSONE}. While
models based on the mean-field approximation borrowed from statistical
physics \cite{morcos2011direct,Baldassi2014PLOSONE} are fast,
approximations based on pseudo-likelihood maximization
\cite{balakrishnan2011Protein,aurell2013pl} are more accurate and used
throughout this paper.

When applied to two interacting protein families, DCA and related
methods are able to detect inter-protein contacts
\cite{weigt2009identification,ovchinnikov2014robust,hopf2014elife} and
thereby to guide protein complex assembly \cite{Schug2009PNAS,
  dago2012PNAS}. This is notable since contact networks in protein
complexes are strongly modular: There are many more
intra-protein contacts than inter-protein contacts. Moreover, DCA
helps to shed light on the sequence-based mechanisms of PPI
specificity
\cite{nimwegen2008MSB,procaccini2011dissecting,cheng2014toward}.

\begin{figure}[h]
\centering
\includegraphics[width=0.99\textwidth]{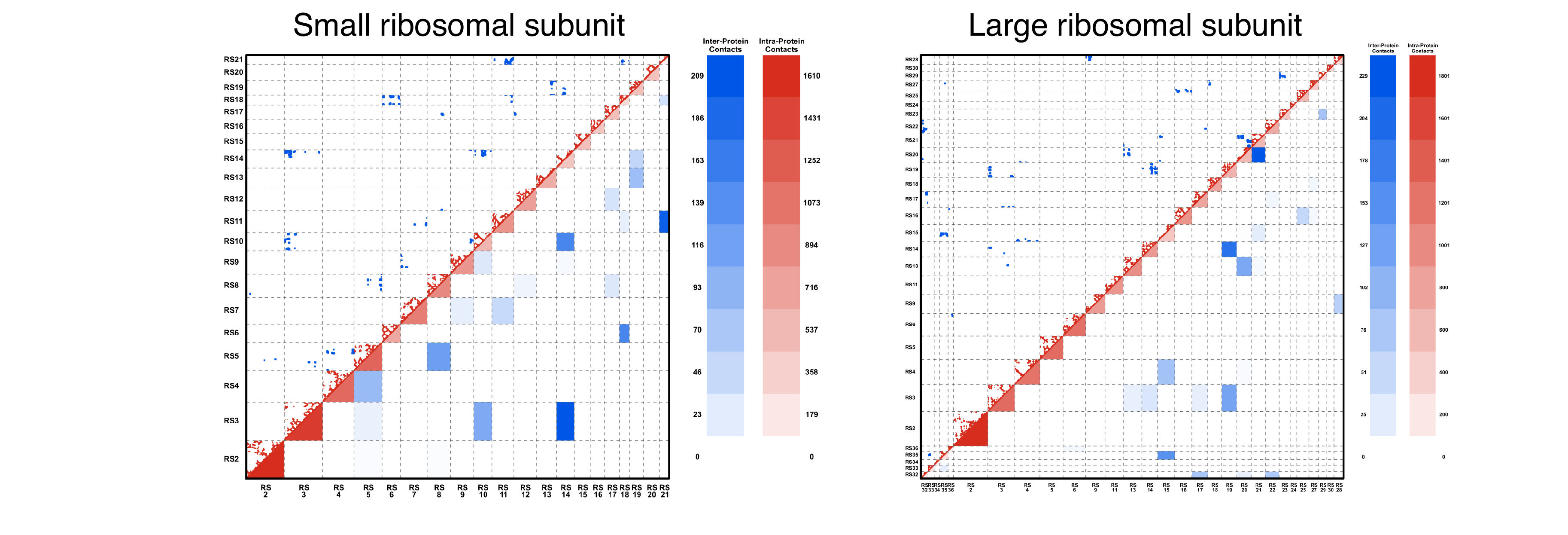}
\caption{{\bf Contact map and protein-protein interaction network of
    small and large ribosomal subunits.} The contact
  map and the protein-protein interaction network for $\mathbf{A}$ the
  small ribosomal subunit and $\mathbf{B}$ the large ribosomal subunit
  (proteins only), using a distance cutoff of 8\AA~between heavy
  atoms.  The upper diagonal part shows the contact map, with red dots
  indicating intra-protein contacts, and blue dots inter-protein
  contacts. The lower triangular part shows the coarse graining into
  the corresponding protein-protein interaction networks, with the
  color levels indicating the number of intra- resp. inter-protein
  contacts, cf. the scales.  The sparse character of both the contact
  network and the interaction network is clearly visible.}
\label{fig1}
\end{figure}

Here we address an important question: Is the strength of
inter-protein residue-residue co-evolution sufficient to {\em
  discriminate interacting from non-interacting pairs of protein
  families}, i.e. to infer PPI networks from sequence information?  A
positive answer would lever the applicability of these statistical
methods from structural biology (residue contact map inference) to
systems biology (PPI network inference). An obvious problem in this
context is the sparsity of PPI networks, illustrated by the bacterial
ribosomal subunits used in the following, cf.~Fig.~\ref{fig1}: The small
subunit contains 20 proteins and 21 protein-protein interfaces (11\%
of all 190 possible pairs). In the large subunit, 29 proteins form 29
interfaces (7\% of all 406 pairs). We see that while the number of
potential PPI between $N$ proteins is ${N \choose 2}$, the number of
real PPI grows only linearly as ${\cal O}(N)$.  Furthermore, the
number of potentially co-evolving residue-residue contacts across
interfaces is much smaller than the number of intra-protein
contacts. In the case of ribosomes, only 5.8\% of all contacts in the
small subunit are inter-protein contacts. In the large subunit this
fraction drops down to 4.5\%.  So the larger the number of proteins,
the more our problem resembles the famous search of a needle in a
haystack. The noise present in the large number of non-interacting
protein family pairs might exceed the co-evolutionary signal of
interacting pairs.

It should also be mentioned that the ribosomal structure relies on the
existence of ribosomal RNA, which is not included in our analysis. We
therefore expect many of the small PPI interfaces to be of little
importance for the ribosomal stability and that only large interfaces
constrain sequence evolution and thus become detectable by
co-evolutionary studies.

Ribosomal proteins and their interactions are essential and thus
conserved across all bacteria, and it appears reasonable to wonder
whether this makes them a specialized example of a protein complex
more amenable to co-evolutionary bias. As a second and smaller interaction
network, we therefore considered the enzymes of the tryptophan
biosynthesis pathway comprising a set of seven proteins in which only
two pairs are known to interact (PDB-ID 1qdl for the TrpE-TrpG complex
\cite{trpGE} and 1k7f for the TrpA-TrpB complex \cite{trpAB}). Also 
here the PPI network is very sparse; most pairs are not known to
interact, but might show some degree of coordinated evolution due to
the fact that in many organisms these genes show a common spatial
co-localization in a single operon and also due to a number of gene fusion
events, cf. the discussion below. While widespread, the tryptophan
biosynthesis pathway is not essential for viability when environmental
tryptophan is present. 


  
In this paper we report the performance of DCA in the prediction of
protein-protein interaction partners in the systems tested.  In a first step,
we analyze the performance on data from an artificial model. This allows for a
systematic analysis of the performance of different approaches and of the
influence of the number of sequences in the alignment. With this artificial
data set we are able to establish a lower-bound on the number of sequences that
would make our predictions on the PPI scale completely accurate if the
generating model was the same model we use for inference.  Given the
growth-rate of current protein sequence databases (notably UniProt
\cite{UniProt}), we expect that such a lower bound could be met in few years.
In a second step, we apply the method to the proteins of the bacterial ribosome
and to the proteins of the trp operon, and show that the results obtained for
simulated data translate well to the biological sequences of this test-set.

\section*{Materials and Methods}

The goal of the present work is to analyze each of the $\binom{N}{2}$
possible pairs of multiple sequence alignments from a given set of $N$
single-protein family alignments, and to extract a pairwise score that
measures the co-evolution between the proteins in the alignments. A
high co-evolutionary score is then taken as a proxy for
interaction. In the spirit of \cite{gap_paper} we describe in this
section consecutively the \textit{data generation and matching}, the
\textit{model} used for analyzing data and the \textit{inference and
  scoring} mechanism.

\subsection*{Data extraction and matching for the ribosomal and trp 
operon proteins}

\begin{figure}[h]
\centering
\includegraphics[width=0.8\textwidth]{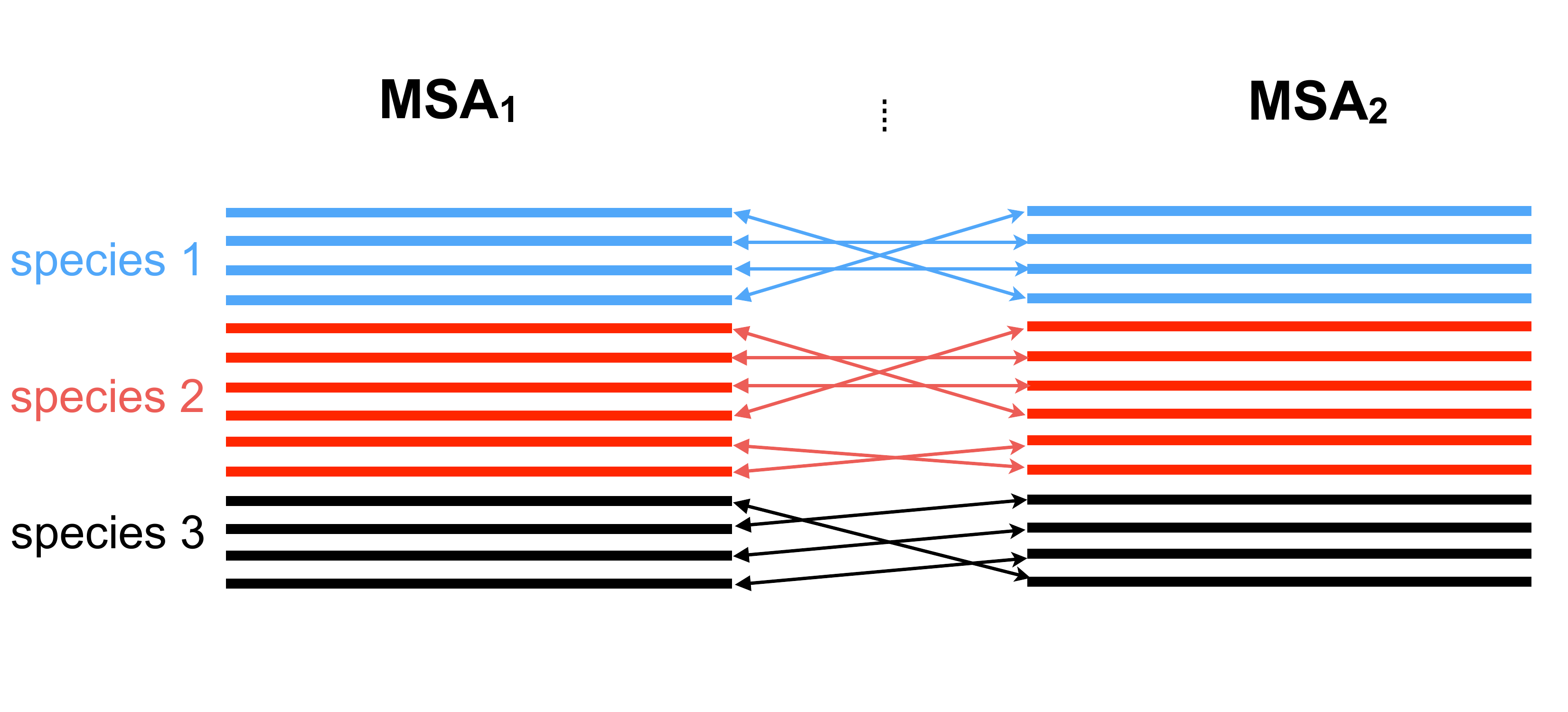}
\caption{{\bf Concatenating two multiple sequence alignments} Figure
  Caption Sketch of the matching procedure that allows us to
  concatenate two different MSAs, here MSA$_1$,MSA$_2$. $\pi$
  represents the optimal permutation of the sequences on the second
  MSA computed using a standard linear programming routine.}
\label{fig2}
\end{figure}

The input data is given by $N$ multiple sequence alignments $D_p$
consisting of $M_p$ sequences of length $L_p$ for every protein family
$p$. These alignments are extracted from UniProt \cite{UniProt} using
standard bioinformatics tools, in particular Mafft
\cite{katoh2002mafft} and HMMer \cite{finn2011hmmer} ({\em cf.}
Section~1 in \nameref{S1:Text} for details on the extraction pipeline and
Tables~S1,S2 in \nameref{S1:Text} listing the values of $N$, $M_p$, $L_p$ for
ribosomal and trp-operon proteins). For the analysis, it is necessary
to concatenate the MSAs of two putative co-evolving protein
families.  This means to create, for each pair of protein families
$(p,p')$, a new alignment $D_{p,p'}$ of sequence length $L_p +
L_{p'}$. Each line contains the concatenation of two potentially
interacting proteins. More precisely, in the case where families $p$
and $p'$ actually interact, each line should contain a pair of
interacting proteins. The general problem of producing a concatenated
alignment out of single MSAs of two protein families is
straightforward in two cases only: (i) we have prior knowledge which
pairs of sequences represent interaction partners; (ii) no paralogs
are present in the considered species ({\em i.e.} all species have at
most a single homolog of each of the sequences to be matched).  Often,
as displayed schematically in Fig.~\ref{fig2}, MSAs contain multiple protein
sequences within a given species and no prior knowledge can be used to
know who is (potentially) interacting with whom. In prokaryotes,
interacting proteins are frequently found to be coded in joint
operons. This suggests to use genomic co-localization as a matching
criterion. To do so, as explained in Section 2 in \nameref{S1:Text},
we approximated the {\em genomic distance} between sequences using
UniProt accession numbers. A better distance between sequences
could be defined in terms of their genomic location. Unfortunately, genomic
locations are available only in the context of whole genome sequencing
projects. The majority of sequences in Uniprot originate from fragments or
from incomplete genome sequencing projects. These difficulties lead us to
content ourselves with the proxy of accession numbers.

Having defined distances between each protein pair in the MSA, we calculate the
matching which minimizes the average distance between matched sequences by
linear programming. Additionally, we introduce a distance threshold used to
discard matched distal protein sequence pairs. The numeric value for this
threshold was determined using the small ribosomal subunit as a test case.

The average number of paralogs per species varies from system to system: For
both ribosomal subunits the proteins have between 1.5 and 3 paralogous
sequences per genome. The trp proteins on the other hand have considerably more
paralogous sequences and the number of such sequences per genome varies between
4 and 24. This means that especially in the trp operon the matching procedure
has the potential to generate much larger alignments than the competing
approach of excluding species with paralogous sequences.  In fact, using this
last approach (which corresponds to setting our threshold parameter to 0)
reduces the number of sequences in the alignments on the average by about 10\%
for the ribosomal proteins and by about 85\% for the proteins of the trp operon
(see Tables S3-S7 in \nameref{S1:Text} ).

However, our matching strategy based on genomic vicinity excludes proteins
coming from isolated genes; it identifies mostly protein pairs coded in gene
pairs colocalized inside operons. In agreement with \cite{Overbeek1999,
lathe2000, harrington2008predicting, Rogozin2004} we assume that in such a case
the maintainance of genomic colocalization is an indication for the
maintainance of interaction, if the original protein pair was also interacting.
While being somewhat speculative, we observe that this procedure removes most
paralogs in the systems under study: Even if many genomes contain a large
number of paralogs before matching (see above), in 99.8\% of all genomes in the
matched alignments for ribosomal protein pairs only a single sequence pair is
found, while for trp protein pairs the same holds for 82\% of all genomes.
In other words, if there are paralogs in a species the matching algorithm tends
to select one single pair, at least in the systems we studied.

We will show evidence that, in the interacting protein systems
investigated here, this strategy leads to a reinforced coevolutionary
signal as compared to including only genomes without
paralogs. However, an independent and direct test whether protein
pairs included in the alignment actually interact would constitute a
big step forward, in particular since the arguments for using genomic
colocalization hold chiefly for bacterial genomes.



Let us recall that the problem of finding a good matching
between sequences has already been studied in the past using different
strategies \cite{nimwegen2008MSB,
  procaccini2011dissecting}. Unfortunately, both methods are
computationally too demanding to be used in a case, where hundreds or
thousands of protein family pairs have to be matched.

\subsection*{Statistical sequence model}

Within DCA, the probability distribution over amino acid sequences $x = (x_1,...,x_L)$ of (aligned)
length $L$ is modeled by a so-called Potts model, or pairwise Markov  
Random Field,
\begin{equation}
{\cal P}(x) = \frac 1 Z \exp\left\{
\sum_{1\leq i<j\leq L} J_{ij}(x_i,x_j) + \sum_{1\leq i \leq L} h_{i}(x_i)
\right\}\  ,
\label{eq:potts1}
\end{equation}
which includes statistical couplings $J_{ij}(x_i,x_j)$ between residue pairs
and position-specific biases $h_{i}(x_i)$ of amino-acid usage
\cite{weigt2009identification}.  The number $Z$ is the normalization constant of
${\cal P}(x)$, which is a probability distribution over all amino-acid sequences
of length $L$. The variable $x_i$ represents the amino acid found at position
$i$ in the sequence and can take as values any of the $q=21$ different possible
letters in an MSA (gaps are treated as a 21st amino acid). The model
parameters are inferred using MSAs of homologous proteins.

In the case of two concatenated protein sequence $(x,x') =
(x_1,...,x_L,x'_1,...,x'_{L'})$, the joint probability takes the form
\begin{equation} {\cal P}(x,x') = \frac{1}{Z} e^{-H(x)-H'(x')-H^{int}(x,x')}.
\label{eq:Potts}
\end{equation}
The functions $H(x)$ and $H'(x')$ are the terms in the exponential in
Eq.~(\ref{eq:potts1}) referring to each single protein. The function
\begin{equation} 
H^{int}(x,x')=-\sum\limits_{i \in x, j \in x'} J_{ij}(x_i,x'_j) 
\label{eq:interaction}
\end{equation} 
describes the co-evolutionary coupling between the two protein
families. In the last expression, $x_i$ is the $i$th amino acid in
sequence $x$, and $x'_j$ the $j$th amino acid in sequence $x'$. The
sum runs over all inter-protein pairs of residue positions.  The $q
\times q$ matrices $J_{ij}$ in this term quantify how strongly sites
between the two proteins co-evolve in order to maintain their
physicochemical compatibility. The matrix contains a real number for
each possible amino acid combination at sites $i$ and $j$ and
contributes to the probability in Equation \ref{eq:Potts} depending on
whether an amino acid combination is favorable or not. The strongest
inter-protein couplings are enriched for inter-protein contacts
\cite{weigt2009identification,ovchinnikov2014robust}. The same kind of
model can be used to predict the interaction between more than two
proteins, with a corresponding number of interaction terms. However,
the number of parameters in the model is proportional to
$(L_1+L_2+..+L_N)^2$ for $N$ proteins while the number of samples in
the concatenated MSA $D_{p_1,....p_N}$ becomes smaller because one has
to find matching sequences for $N$ proteins
\textit{simultaneously}. This leads us to consider the case $N>2$ only
for artificial proteins where the total length and sample size are
controllable. 

\subsection*{Inference and Scoring} 

Following \cite{aurell2013pl}, the parameters of the model were inferred by
maximizing \textit{pseudo-likelihood functions}. This is an alternative to
directly maximizing the likelihood and considerably faster (see Section~3 in \nameref{S1:Text}
Text for details). Given that the model is mathematically equivalent to the one
used in \cite{aurell2013pl} we can use the output of the algorithm (plmDCA)
with default parameters as presented there directly for our purposes. This
output consists of scores $F_{ij}$ (the average-product corrected Frobenius
norm of the matrices $J_{ij}$) that quantify the amount of co-evolution between
sites $i$ and $j$ in the alignments. In order to quantify co-evolution between
\textit{proteins}, we took the $F_{ij}$ corresponding to inter-protein site
pairs (i.e. $i$ in $x$ and $j$ in $x'$) and calculated the mean of the 4
largest. These quantities, a real number for every protein pair, are used to
rank protein-protein interaction partners. The number 4 was chosen because it
performed well in the small ribosomal subunit, which we used as a test case
when designing the algorithm. Subsequent tests on larger systems showed that
any number between 1 and 6 performs almost equally well (see Section
{\em A Global View} in {\em Results}).

\subsection*{Simulated data} 

As the basis for the simulated data we used a fictitious protein
complex consisting of 5 proteins. Each protein has a length of 53
residues. The individual contact map of each one is given by the
bovine pancreatic trypsin inhibitor (PDB ID 5pti
\cite{wlodawer1984structure}), which is a small protein performing
well for the prediction of internal contacts by DCA. Each $P_i$ has
551 internal contacts. Moreover, each protein interacts with two
others in a circular way. The inter-protein contact matrices between
$P_{i}$ and $P_{i+1}$ (as well as between $P_1$ and $P_5$) are random
binary matrices with a density of 10\% of the internal contacts. This
models the sparsity of the inter-protein contacts as compared to the
intra-protein contacts. A contact map for the artificial complex can
be found in Figure S5 in \nameref{S1:Text}, There are no contacts between other
pairs of proteins.

In order to define as realistically as possible the coupling
parameters of the Potts model used for generating the artificial
sequences, we used the Pfam protein family PF00014 of the pancreatic
trypsin inhibitor \cite{PFAM}. Note that a member of this family was
also used to define the structure. The couplings describing the
co-evolution \textit{within} the single proteins were directly
extracted from the Pfam MSA using DCA. For the couplings
corresponding to the co-evolution \textit{between} the proteins, we
used a random subset of the internal parameters and used them to couple
sites that are in contact according the contact map as defined
above. Non-contacting pairs of sites remain uncoupled between
artificial proteins. Using this model, a joint MSA $D_{12345}$ of
sequences of length $265 = 5\times 53$ was generated using standard MC
simulations.

The process of defining the contact map, choosing the parameters and
generating the sequences is described in Section 5 \nameref{S1:Text}.



\section*{Results and Discussion}

\subsection*{Testing the approach using simulated data}
As a first test of our approach, we use {\em simulated data} generated
by Monte Carlo (MC) sampling of a Potts model of the form of
Eq.~(\ref{eq:Potts}), cf. {\em Materials and Methods}. 

The main simplifying assumptions in this context are: (i) We assume intra- and
inter-protein co-evolution strengths to be the same. (ii) We assume the
distribution of inter-protein residues contacts within the possible contacts to
be random. (iii) We assume the sequences to be identically and independently
distributed according to our model. This model includes the assumption that non-
contacting sites have zero couplings. The number of artificial sequences needed
for a good performance of our method should therefore be taken at most as a
lower bound for the number of biological sequences needed for a comparable
performance.

\begin{figure}[h]
\centering
\includegraphics[width=0.8\textwidth]{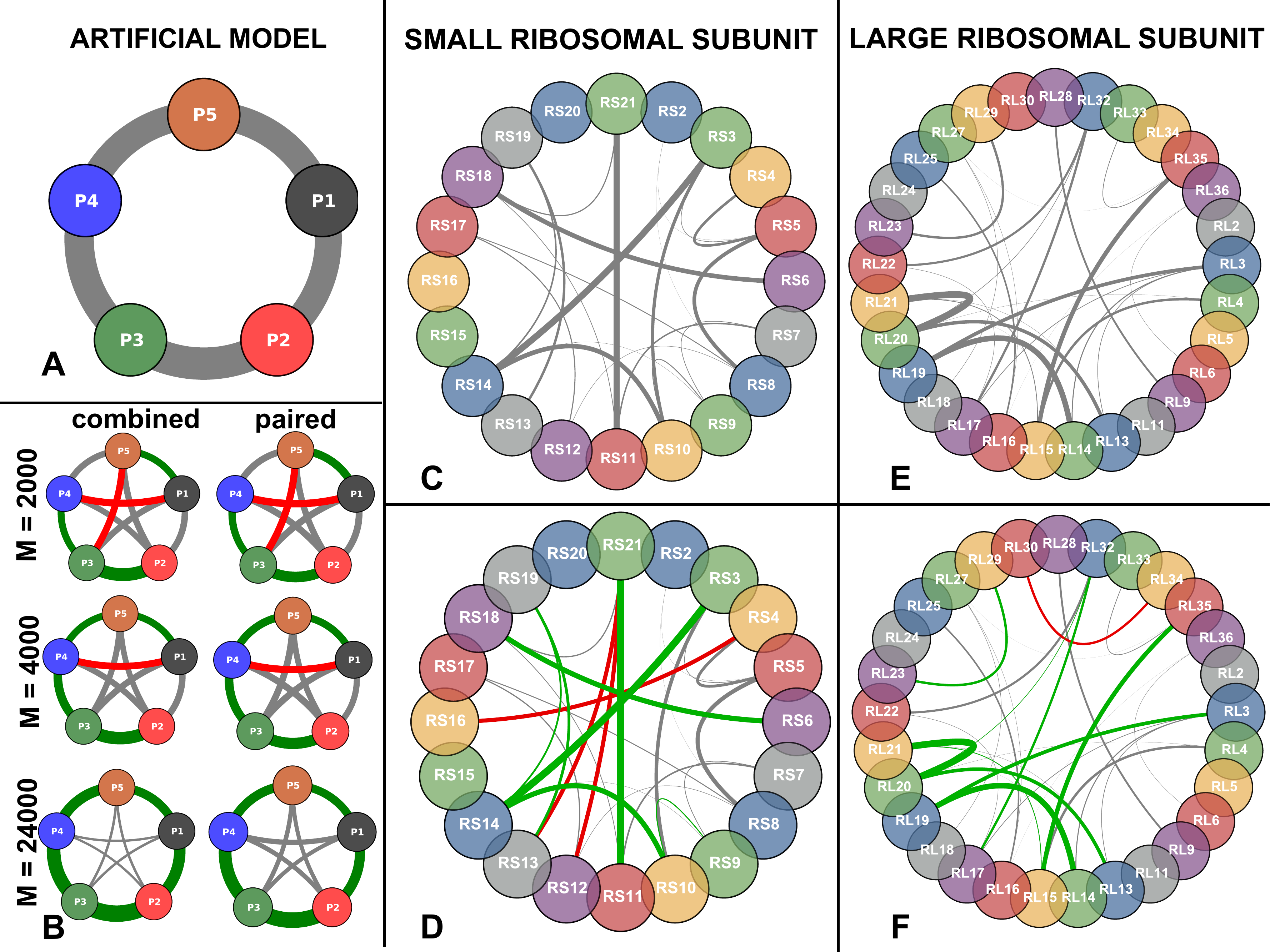}
\caption{{\bf Residue-residue structure of both artificial and
    ribosomal complex} $\mathbf{A}$ Architecture of the {\em
    artificial} protein complex. Arcs width are proportional to the
  number of inter-protein residue contacts. $\mathbf{B}$ Inferred PPI
  network for both {\em paired} and {\em combined} strategy for
  different number $M$ of sequences generated from the artificial
  model. Green arcs are true positives, red false positives, gray
  low-ranking predictions. Arc widths are proportional to the
  inter-protein interaction score. $\mathbf{C}$ SRU architecture (same
  color code as A). $\mathbf{D}$ Inferred PPI network (same color code
  as B).  $\mathbf{E}$ Same as C for LRU.  $\mathbf{F}$ Same as D for
  LRU.  Arc width in panels C-F is provided by the number of
  inter-protein contacts, as a measure of interface size. It becomes
  obvious that mainly large interfaces are recognized by our
  approach.}
\label{fig3}
\end{figure}

In panel ${\rm A}$ of Fig.~\ref{fig3} we show the architecture of our
artificial protein complex. It is composed of five fictitious,
structurally identical proteins $P_1,...,P_5$, each one consisting of
53 residues. In order to simulate co-evolution between the proteins,
we generate a \emph{joint} MSA $D_{12345}$ for all 5 proteins with a
model that contains couplings between inter-protein site pairs. These
couplings are modeled in a way to resemble couplings inferred from
real proteins (see \emph{Materials and Methods}).

To assess our capability to infer the PPI network of panel ${\rm A}$
from such data, we adopted two different strategies which we called
{\em combined} and {\em paired} in panel ${\rm B}$ of Fig.~\ref{fig3}. The {\em
  combined} strategy uses plmDCA on the full-length alignments of
length 265 and models the interaction between all proteins pairs
\textit{simultaneously}. Given that in this artificial setting we use
the same model to generate the data as to analyze it, the approach is
guaranteed to infer the model correctly for a large number of analyzed
sequences and therefore to assign a higher interaction score to any interacting
protein pair than to any non-interacting pair.

To assess the coupling strength between two proteins, we average the
four strongest residue coupling strengths between them. This leads to
a score oriented toward the strongest signal while also reducing noise
by averaging. In panel ${\rm B}$ of Fig.~\ref{fig3} we show the results for MSA
sizes $M=2000, 4000, 24,000$ while intermediate values are reported in
Figure S6 in \nameref{S1:Text}.  The two lower figures - $M=2000, 4000$ -
represent the lower and upper bound of what we can currently obtain
from databases for the proteins analyzed by us. The largest value
$M=24,000$ is what we expect to be available in a few years from now,
seen the explosive growth of sequence databases. The thickness of each
link in Fig.~\ref{fig3} is proportional to the inferred inter-protein
interaction score. The five strongest links are colored in green when they
correspond to actual PPI according to panel ${\rm A}$, and in red when
they correspond to non-interacting pairs. For increasing sample size the
predictions become more consistent and for $M=24,000$ any interacting
protein-pair has a higher interaction score than any non-interacting pair.

Due to the running time of plmDCA only alignments for sequences of
total length $L\lesssim 1000$ can be analyzed. This is exceeded
already by the sum of the lengths of the proteins of the small
ribosomal subunit. Additionally, creating a combined multiple sequence
alignment for more than two proteins would lead to very low sequence
numbers due to the necessary matching (see \emph{Materials and
  Methods}). Therefore, using the combined strategy is not generally
applicable. In the {\em paired} strategy we therefore analyze each
pair of proteins separately. This means that plmDCA is applied to all
$N \choose 2$ protein-pair alignments $D_{ab},\ 1\leq a<b\leq N$.  In
panel ${\rm B}$ of Fig.~\ref{fig3} we find that the paired strategy is also
able to detect the correct PPI network for large enough $M$.  We
observe, however, that the performance of the paired strategy is
slightly worse. Couplings between non-interacting proteins are
estimated significantly larger than using the combined strategy for
large $M$.  Even in the limit $M\to\infty$ we do not expect these
links to disappear: Correlations between, e.g., $P_1$ and $P_3$ are
generated via the paths $1-2-3$ and $1-5-4-3$, but in the paired
strategy these correlations have to be modeled by direct couplings
between $P_1$ and $P_3$ since the real direct coupling paths are not
contained in the data.

\begin{figure}[h]
\centering
\includegraphics[width=0.8\textwidth]{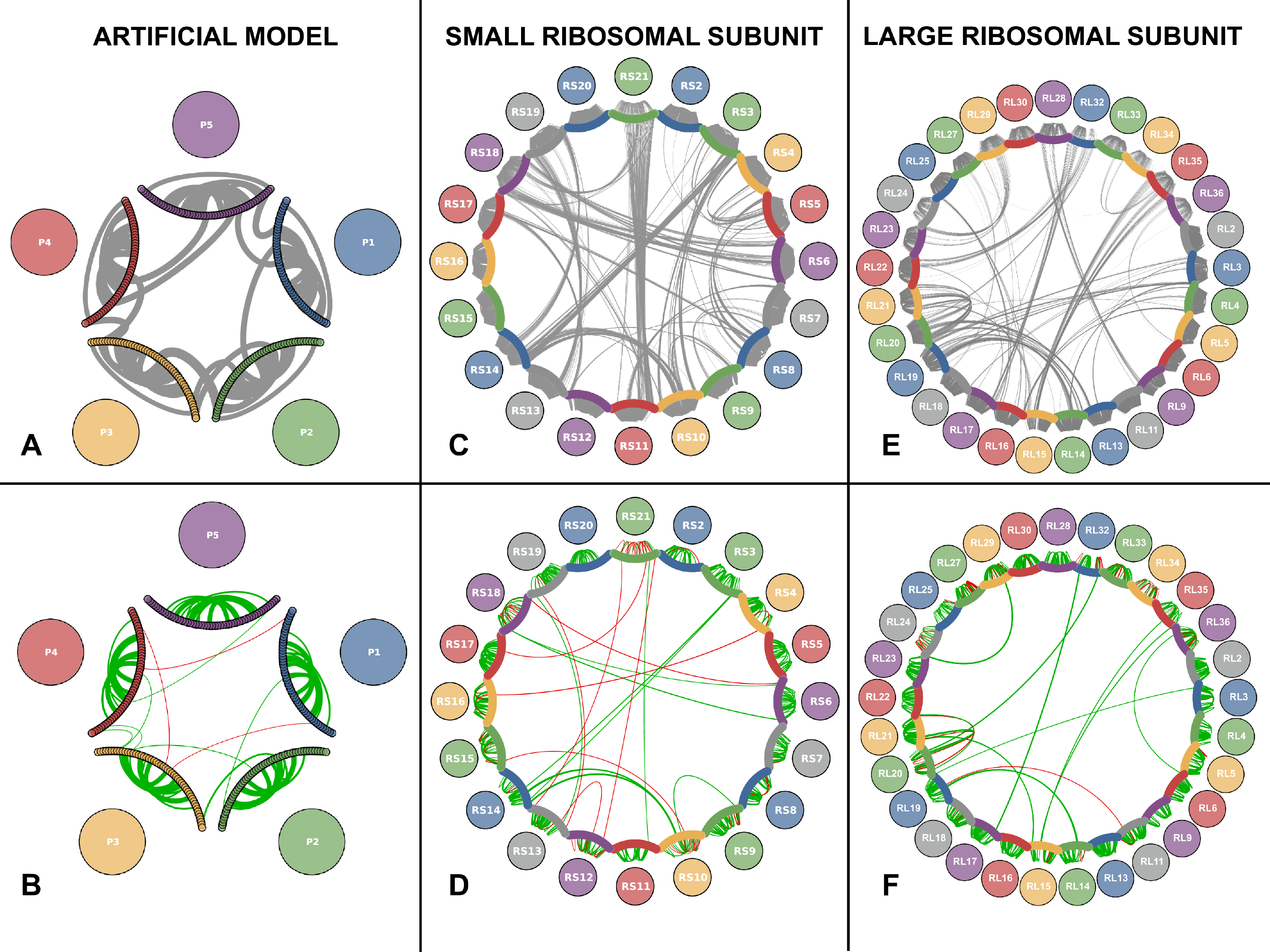}
\caption{{\bf Architecture and inferred protein-protein interaction
    network of the artificial protein complex} \textbf{A}
  Residue-residue interaction structure of the generating model for
  the artificial data. Colored arcs represent the protein
  chain. Non-zero couplings in the coupling matrix of the generating
  model are represented as curves between the nodes. The width of the
  curves is proportional to the interaction score. Only the 10
  strongest intra/inter-protein scores are shown. \textbf{B} Same as
  \textbf{A}, but based on the inferred couplings. Green arcs are true
  positives, red false positives. Note that not all green arcs have a
  corresponding arc in \textbf{A} due to our choice to display only
  the 10 strongest couplings, which not always correspond to the
  strongest score.  \textbf{C} Same as \textbf{A} for SRU. All links
  represent a contact in the PDB structure and have equal
  width. \textbf{D} Same as \textbf{B} for SRU. \textbf{E} Same as
  \textbf{C} for LRU. All links represent a contact in the PDB
  structure and have equal width. \textbf{F} Same as D for LRU.}
\label{fig4}
\end{figure}

After having answered the {\em 'who-with-whom'} question for the
artificial protein network, we address the {\em 'how'} question of
finding inter-protein contact pairs.  Fig.~\ref{fig4} panel ${\rm A}$ displays
individual residue contact pairs within and between proteins in the artificial
complex.  Panel ${\rm B}$ shows the 10 strongest intra-protein
couplings for each protein and the 10 strongest inter-protein
couplings inferred by plmDCA ($M=4000$, combined strategy). Green
links correspond to contact pairs and red links to non-contact
pairs. We see that the intra-protein prediction is perfect, whereas a
few errors appear for inter-protein predictions in agreement with the
results of Fig.~\ref{fig3}.

\subsection*{The PPI network of bacterial ribosomes}

As a more realistic test we apply the method to the bacterial large
and small ribosomal subunits (LRU, SRU). To define contacts and
protein interaction partners we used high-resolution crystal
structures with PDB-IDs 2z4k (SRU) and 2z4l (LRU)
\cite{borovinskaya2007structural}. The contact network is summarized
by the contact maps in Fig.~\ref{fig1}. The ribosomal RNA is ignored
in our analysis.

Panels ${\rm C,E}$ of Fig.~\ref{fig3} display the architectures of
both SRU and LRU. The SRU (LRU) complex consists of 20 (29) proteins
of lengths 51-218 (38-271); 21 (29) out of ${20 \choose 2} = 190$
(${29 \choose 2} = 406$) pairs are in contact. The interfaces contain
between 3-209 (1-229) residue pairs. The width of the inter-protein
links in the PPI network Fig.~\ref{fig3} in panels ${\rm C,E}$ are
proportional to these numbers.  The number of contacts within the
individual proteins ranges from 297 to 2337 (303-2687). Globally,
there are 22644 (30555) intra-protein and 1401 (1,439) inter-protein
contacts, so the contacts relevant for our study comprise only 5.8\%
(4.5\%) of all contacts.

Fig.~\ref{fig3} panel D shows the inferred SRU PPI architecture.  As
expected, the biological case is harder than the artificial case where
the data are independently and identically distributed according to
the generating model.  Even though the histograms of the inferred
interaction scores for both cases are very similar (see Figure S2 in
\nameref{S1:Text}, biological data are expected to show non-functional
correlations due to the effect of phylogeny or sequencing efforts
which are biased to model species and known pathogens.  Nonetheless,
among the top ten predicted interacting protein pairs the method makes
only three errors (true-positive rate 70\% as compared to
$21/190\simeq 11\%$ true PPI between all protein pairs, with an
overall area under ROC curve (AUC) of 0.69 (see Fig.~\ref{fig7}).  The
method spots correctly the pairs with larger interaction surfaces
whereas the small ones are lost.  Two of the false-positive (FP)
predictions include protein RS21, which has the smallest paired
alignments with other proteins ($M$ between 1468 and 1931).  Also the
third FP, corresponding to the pair RS4-RS18, is probably due to a
small MSA with $M=2064$. At the same time, the interaction of RS21
with RS11, which is one of the largest interfaces (199 contacts), is
still detected despite the low $M=1729$. The same procedure for the
LRU (406 protein pairs) performs even better: 9 out of the 10 first
PPI predictions are correct (see Fig.~\ref{fig3} panel {\rm F}), and
the AUC is 0.81.

The results on the residue scale for both SRU and LRU are depicted in
panels D and F of Fig.~\ref{fig4}. Shown are the first 20
intra-protein residue contact predictions for each protein (excluding
contacts with linear sequence separations below 5 to concentrate on
non-trivial predictions) and the first 20 inter-protein residue
contact predictions. In the SRU case of panel D for example, the
results are qualitatively similar to the artificial case, albeit with
a slightly reduced true-positive rate of 60\% among the first 20
inter-protein residue contact predictions (compared to the ratio of
1401 actual inter-protein residue contacts and 2,403,992 possible
inter-protein residue contacts, i.e., 0.058\%).  Again 3 out of the 8
false positives are related to RS21, which due to the smaller MSA size
is also the only one having a considerable false-positive rate in the
intra-protein residue contact prediction. About 95\% of the displayed
400 highest intra-protein residue contacts are actually contacts (see
Figure~1 in \nameref{S1:Text}). Analogous considerations with a
somewhat larger accuracy (85\%) hold for LRU as displayed in
Fig.~\ref{fig4} panel F.

\subsection*{The PPI network of the tryptophan
  biosynthetic pathway}

As a distinct test case for our methodology we analyzed the 7 enzymes
(TrpA, B,C,D,E,F,G) that comprise the well characterized tryptophan
biosynthesis pathway. In contrast to the ribosomal proteins, these
enzymes are only conditionally essential in the absence of
environmental tryptophan and their genes are only expressed under
deplete tryptophan conditions. In this particular system, only two
protein-protein interactions are known and resolved structurally:
TrpA-TrpB (PDB-ID 1k7f \cite{trpAB}) and TrpG-TrpE (PDB-ID 1qdl
\cite{trpGE}). 
Whereas the TrpG-TrpE pair catalyzes a
single step in the pathway and their interaction is thus essential for
correct functioning, the TrpA-TrpB pair catalyzes the last two steps in
tryptophan biosynthesis. Both enzymes function in isolation but their
interactions are known to increase substrate affinity and reaction
velocity by up to two orders of magnitude. All other proteins catalyze
individual reactions, but one might speculate that the efficiency of
the pathway could benefit from co-localization of enzymes involved in
subsequent reactions. Interestingly, the Pfam database \cite{PFAM}
reports that in many species pairs of genes in the operon appear to be
fused, suggesting that some of the fused pairs are actually PPI
candidates. An example is the TrpCF protein, which is fused in {\em
  Escherichia coli} and related species (but not in the majority of
species).

\begin{figure}[h]
\centering
\includegraphics[width=0.6\textwidth]{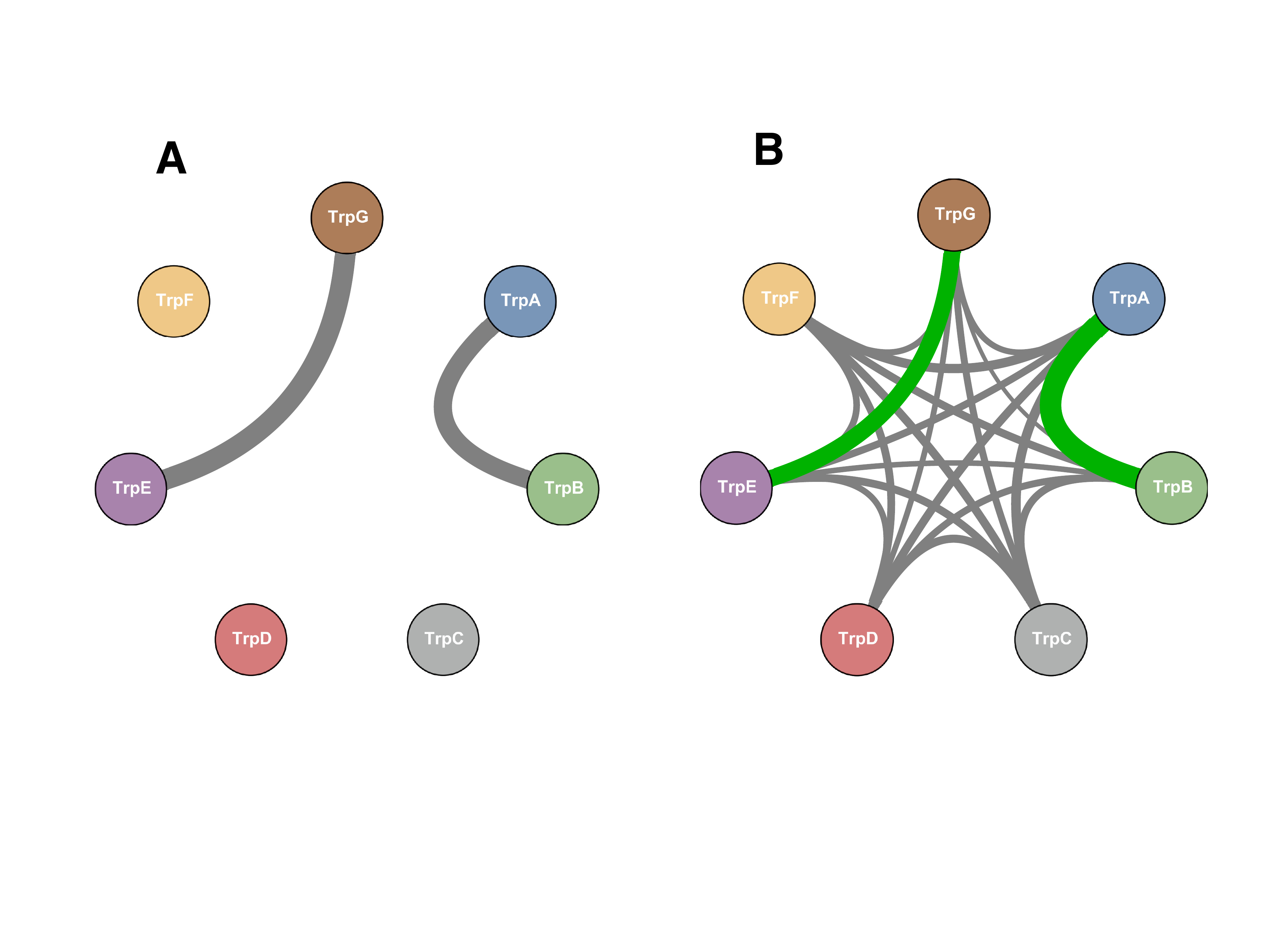}
\caption{{\bf Tryptophan biosynthesis pathway} \textbf{A} Architecture
  of the known protein-protein interaction among the 7 enzymes which
  are coded in the Trp operon. The widths of the arcs are proportional
  to the number of inter-protein residues (which in this case is
  almost equal for the two interacting pairs). \textbf{B} Inferred PPI
  network, here the width of the arcs is proportional to the
  interaction score. Green arc correspond to the protein pairs for
  which a known structure exist.}
\label{fig5}
\end{figure}

After applying our method to all 21 protein pairs we find elevated
interaction scores only for TrpA-TrpB and TrpE-TrpG, which are the
only known interacting pairs (see Fig.~\ref{fig5} and Table S10 in \nameref{S1:Text} for the interaction scores of all pairs). Those two pairs have
interaction scores of 0.375 and 0.295, while the other pairs are
distributed between 0.071 and 0.167. Even though we do not define a
significance threshold for prediction (see Section \emph{A global
  view}), these two pairs would be discernible as interesting
candidates even if we did not have the 3D structures.

We speculate therefore that the fusions in many species do not imply
strong inter-protein co-evolution.  To further investigate this
aspect, we took a closer look at the protein pair TrpC-TrpF. For this
protein pair, a high resolution structure of a fused version exists
(PDB-ID 1pii \cite{trpCF}). We ran our algorithm on the complete
multiple sequence alignment, the multiple sequence alignment with
fused sequence pairs removed and only on the fused sequences. In none
of these cases did we observe a statistically significant interaction
score or a statistically significant prediction of inter-protein
contacts present in the structure of the fused protein.

Our results are corroborated by the finding that all scores measuring
the co-evolution between a ribosomal protein and an enzyme from the
tryptophan synthesis pathway are small (see the following
subsection). No indication for an interaction between the two systems
is found, as to be expected from the disjoint functions of the two
systems.

\begin{figure}[h]
\centering
\includegraphics[width=0.8\textwidth]{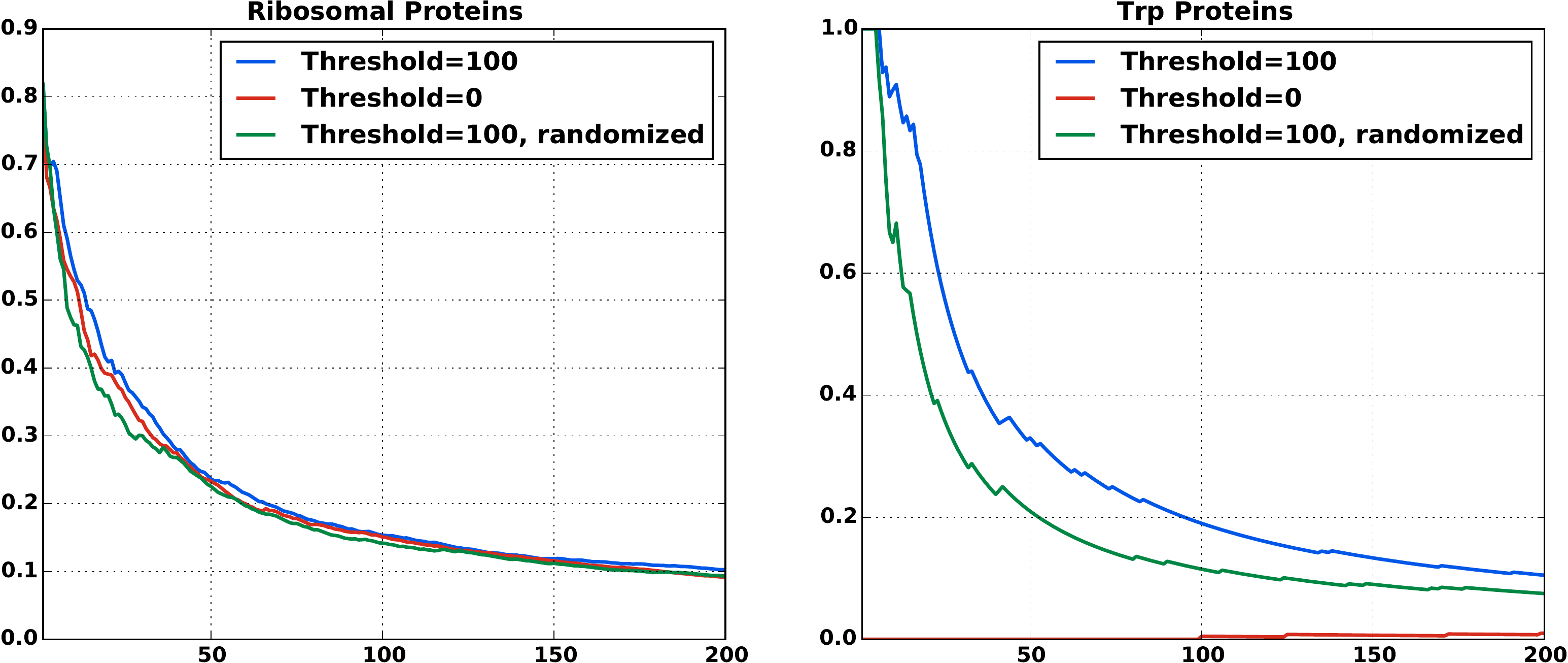}
\caption{{\bf Efficacy of the different matching procedures}
True-positive rates for inter-protein residue contact prediction for different
matching procedures. Shown are means for all protein pairs that have at least
100 residue pairs in contact. The ribosomal and the trp proteins were tested
independently. The red curves correspond to a matching including only protein
sequences without paralogs inside the same species ("matching by uniqueness in
genome"). The low performance of this approach on Trp proteins is due to a very
low number of species without homologs, which leads to very small matched
alignments. The blue curves show the results for our matching procedure as
described in the text. The green curves correspond to alignments that have
been obtained by first applying our matching procedure and then randomizing
the matching within individual species. The definition of "contact" was the
same as used above (a distance of less than 8.0\AA between two heavy atom
in the residues).}
\label{fig6}
\end{figure}

\subsection*{A global view}
It is interesting to assemble a larger-scale system out of the three systems
(SRU, LRU, Trp). To this end, we created all possible pairings between the
proteins used in the present study (SRU vs. RU, SRU vs. Trp, LRU vs Trp, SRU vs
SRU, LRU vs. LRU, and Trp vs. Trp).  This leads to a total of 1540 pairs, out
of which only 49 pairs are known to interact (which we defined as true
positives). We present the findings in Fig.~\ref{fig7} and in Figs. S7-S9 in \nameref{S1:Text}.
Fig. S7 in \nameref{S1:Text} shows the true-negative rate, which is the
fraction of true negatives in the indicated number of predictions with the
{\em lowest} interaction scores.  As it can be seen our scoring produces a false
negative just after 420 true negatives.  Figs. ~\ref{fig7} and S8 show true positive rates
for the complete system and the individual systems. We also show true positive
rates for alternative ways to calculate the interaction score between protein
pairs, i.e. a different number of inter-protein residue-residue interaction
scores to average. We notice that in the complete system, the performance is
similar to the performance in individual systems. All of the 10
highest-scoring protein pairs are known to interact, and 75\% of the
first 20 protein-pairs. After these first 20 pairs,
the true positive rate drops to around 45\% in the first 40 predictions. This
is analogous to the case of protein contact prediction, where methods based on
the same model are able to extract a number of high confidence contacts but see
a large drop in performance afterwards \cite{morcos2011direct}. The area under
the AUC for the whole system is 0.83 (see Fig.~\ref{fig7}). This is
stable when averaging different numbers of residue-contact scores to arrive at
a protein-protein interaction score, but the performance seems to worsen when
using more than 6. This is probably because only a few inter-protein residue
contacts have a large score and averaging over too many only adds noise. It can
also be seen that averaging over 4 performs very well in the small ribosomal
subunit, which is why we have chosen this value for the large part of the
analysis. On the larger-scale system, though, any number between 1 and 6
performs almost identically.

A further question is whether it is possible to define a threshold
allowing to reliably discriminate between interacting and non
interacting pairs in terms of the interaction score. Figure S9 in \nameref{S1:Text} shows two normalized histograms of the interaction scores. 
The rightmost tail of the interacting pairs
distribution is well separated from the rightmost tail of
non-interacting one, but the highest scores of non-interacting pairs
are strongly overlapping with the lowest scores of the interacting
ones. The situation is therefore analogous to what is observed in the
case of the inference of contacts within single protein families
\cite{weigt2009identification,morcos2011direct,aurell2013pl,Baldassi2014PLOSONE},
where the same technique is known to produce relatively few high
confidence contacts in the topmost scoring residue pairs. To
conclude, while high scores seem to reliably predict interacting
pairs, and low scores non-interacting pairs, there is a 
zone where a clear discrimination between interacting and
non-interacting pairs is not possible.

\subsection*{Limitations of the method}

The main limitation of the method we present in this paper is arguably the
concatenation of the two MSAs. In the test cases we analyzed this step was
rendered difficult due to the presence of paralogs in the majority of species.
The solution based on the minimization of the genomic distances within the
concatenated MSA as outlined in subsection ``Data extraction and matching for
the ribosomal and trp operon proteins'' can be problematic when used naively.


An example is the homo-dimerization of OmpR-class Response Regulators
\cite{weigt2009identification}. In this class of response regulators, DCA
discovers a very clear homo-dimerization signal. However, the corresponding
Pfam family PF00072 also contains a large number of response regulators not
belonging to the OmpR class, functioning either in the monomeric form or in
different dimeric structures. Using the full Pfam MSA, the OmpR-class
homodimerization signal fades out due to the mixing of these different classes.
The analysis as presented in this paper would therefore not capture this
protein-protein interaction.

This suggests that more robust methods than genomic proximity should be
developed. This is of course especially true if one is interested in
eukaryiotic systems, were we have no evidence that genomic colocalization is an
indication for interaction.

A second limitation of our method is the apparent inability to actually discern
between interacting and non interacting pairs. Everything that the method can
do at this point is to state that some interactions are more likely than others
and show that this ordering is considerably better than a random one. This is a
problem related to all DCA methods as reported for instance in
\cite{weigt2009identification, morcos2011direct, aurell2013pl, Baldassi2014PLOSONE} in
the context of inference of contact maps in single proteins. It is probable,
though, that a cutoff can be chosen if more data becomes available and the
method is applied to a data-set much larger in scale than the one presented
here.

\section*{Conclusions}

\begin{figure}[h]
\centering
\includegraphics[width=0.8\textwidth]{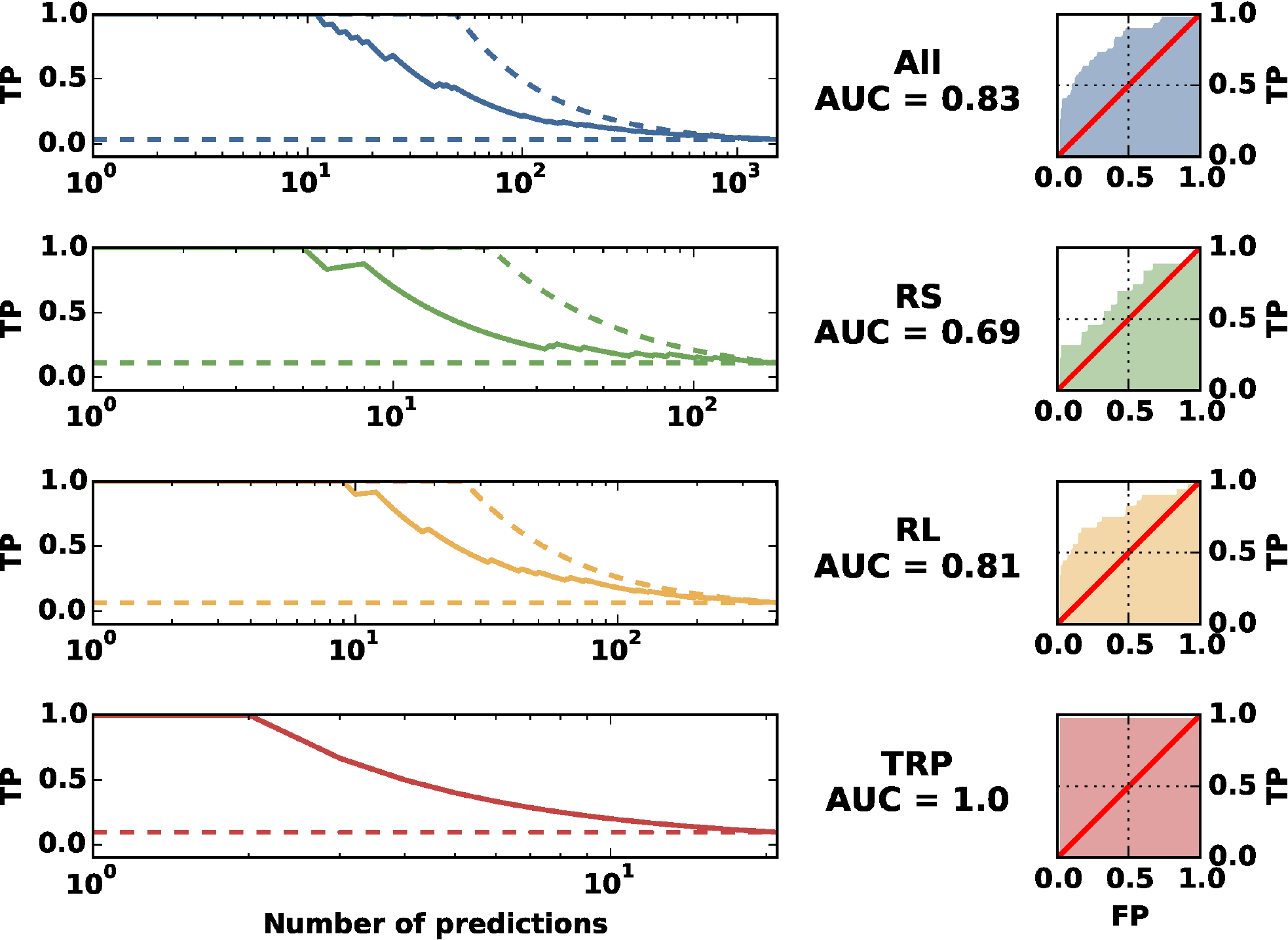}
\caption{{\bf Performance Summary}
The plots illustrate the performance in predicting protein interaction
partners. The left panels show the fraction of true positives among
the first $n$ PPI predictions, with $n$ being the number indicated on the
horizontal axis (solid lines). The dashed lines show the best possible (upper dashed line) and the mean of a random prediction (lower dashed line). The right panels show ROC-curves, which indicate the
dependence of the true-positive predictions (TP/P) from the false
positive predictions (FP/N). The area under the curve (AUC) is a
global global measure for the prediction quality; it is 1/2 for a
random, and 1 for a perfect prediction. A protein
pair is identified as an interacting (true positive) pair, if at least one PDB
structure with at least one inter-protein contacts exists. }
\label{fig7}
\end{figure}

To conclude, we have shown that DCA performs well in the
systems tested when used to predict protein-protein interaction
partners. In the small and large ribosomal subunit our tests resulted
in a true positive rate of 70\% and 90\% in the first 10 predictions
(AUC of 0.69 and 0.81) while in the trp operon the two largest
interaction scores corresponded to the only two interactions
experimentally known (AUC 1). The performance is summarized in
Fig.~\ref{fig7}. The Figure shows both the high quality of the first
predictions, but also a drop in performance after a fraction of all
interacting pairs (about 40\% in our test case). This is analogous to
the case of protein contact prediction by DCA and related methods,
where the performance drops after a limited number of high-confidence
predictions \cite{morcos2011direct}. In the same context and with the
same caveat, a good performance in predicting inter-protein
contacts on the residue level has been shown. The artificial data have
shown that the performance of our approach depends crucially on the
size of the alignments. Only for very large MSA ($M=24,000$ sequences
in our data) a perfect inference of the artificial PPI network was
achieved. MSA for real proteins pairs are typically much smaller.
Even for pairs of ribosomal proteins, which exist in all bacterial
genomes, only about 1500-3200 sequence pairs could be recovered. This
places these data towards the lower detection threshold of PPI. We
therefore expect the performance of the presented approach to improve
in the near future thanks to the ongoing sequencing efforts (the
number of sequence entries in Uniprot \cite{UniProt} has been growing
from about 10 millions in 2010 to 90 millions in early 2015) and
improved inference schemes.  The performance of the same
algorithm on different and dissimilar systems suggests
that the approach could be used to detect interactions
experimentally unknown so far. In fact, if we trust our results on the
trp operon we can already draw some speculative biological inferences.
While there are many high-resolution structures of the ribosome
available, one might have expected that in the trp operon there could
be more transient previously unreported interactions in the tryptophan
biosynthesis pathway beyond the two interactions that have been
structurally characterized.  As mentioned, various enzyme fusions can
be observed in the databases, suggesting that there is an evolutionary
benefit to co-localizing the enzymes of the pathway in the cell.  An
obvious benefit of such co-localization would be that the pathway
intermediates do not have to diffuse throughout the cell from one
enzyme component to the next. In the tryptophan biosynthesis pathway
in particular, there are numerous phosphorylated intermediates that
need to be protected from unspecific cellular phosphatase
activity. Organizing the enzymes in the pathway in a multi-protein
complex would seem like an efficient way to protect the intermediates
from decay. However, our data indicate that the only statistically
relevant co-evolutionary signals that can be observed are restricted
to the known strong interactions between TrpA with TrpB and TrpE with
TrpG. This could be interpreted in a number of ways: {\em (i)} The
most obvious explanation is that there are no additional
protein-protein interactions beyond those that are known and that no
multi-enzyme complex exists for the tryptophan biosynthesis
pathway. Alternatively {\em (ii)} it seems plausible that there are
numerous structural solutions to form a tryptophan biosynthesis
complex and that there is no dominant structure from which a
co-evolutionary pattern can be observed in the sequence databases.
Lastly {\em (iii)} it is not out of question that the enzymes of the
pathway do not directly form a complex but that they are jointly
interacting with an unidentified scaffold component. Of course we
cannot exclude that our method is not able to capture other
potentially present interactions.

From a methodological point of view, one possible algorithmic
improvement is creating better MSAs for protein pairs.  The vast
majority of protein families show genomic amplification within
species.  This raises the issue of which sequence in one MSA should be
matched with which sequence in the other MSA when concatenating the
two MSAs, as shown in Fig.~\ref{fig2}. In the absence of prior
knowledge and as long as only prokaryotes are concerned, we showed
that it is possible to use the simple criterion of {\em matching by
  genomic proximity}.  This criterion is based on the observation that
two sequences are more likely to interact if they are genomically
co-localized. Our results have shown that in the case of the ribosomal
network better inference results can be obtained by using this
matching criterion than by using a random matching or using a
conservative matching taking only species with a single sequence in
both MSAs into account, cf.~Fig.~\ref{fig6}. However, we found it
beneficial for the predictive performance to introduce a threshold
distance above which we simply discarded candidate sequences. This is
not based on biological principles.

We believe that our {\em naive} matching strategy can be improved
substantially. Even if closeness of sequence pairs on the genome is a
good proxy for interaction in some cases, for example if they belong
to the same operon, excluding all distal pairs is a very crude
criterion.  This criterion is known to be erroneous in many cases, for
example in the bacterial two component signal transduction system
\cite{nimwegen2008MSB,procaccini2011dissecting,cheng2014toward}. It
would therefore be interesting to include the matching into the
inference procedure itself, {\em e.g.} to find a matching that
maximizes the inter-protein sequence covariation,
cf. \cite{nimwegen2008MSB} for a related idea. However, for highly
amplified protein families this leads to a computationally hard
optimization task. Simple implementations get stuck in local minima
and do not lead to improvements over the simple and straight-forward
scheme proposed here.


\section*{Acknowledgments}
CF and AP are supported from the EU Marie Curie Training Network
NETADIS, (FP7 Grant 290038). HS was supported by grants GM106085 and
GM019416 from the US National Institute of General Medical Sciences,
National Institutes of Health. MW was supported by the Agence
Nationale de la Recherche project COEVSTAT (ANR-13-BS04-0012-01). CF
and AP acknowledge Riccardo Zecchina and Carlo Baldassi for many
interesting discussions.



\clearpage

\section*{Supporting Informations}

\subsection*{S1 Text}
\label{S1:Text}

\section{Multiple Sequence Alignments}

\label{sec:riboAlign}

\subsection{Multiple Sequence Alignments}
The data we use are multiple sequence alignments (MSA). Each such MSA
is a rectangular matrix, with entries coming from a 21-letter alphabet
containing the 20 standard amino acids and a gap symbol ``-''. In the
following we denote this alignment by a matrix
\begin{equation}
X = \left( x_i^a \right),\ \ i=1,...,L,\ \ a=1,...,M
\end{equation}
with $L$ being the number of residues of each MSA row, i.e., the
number of residues in each considered protein, and $M$ the number of
MSA rows, i.e., the number of proteins collected in the alignment. For
simplicity of notation we assume that the 21 amino acids are translated
into consecutive numbers 1,...,21.

\subsection{Alignment Generation}

For all proteins of the small ribosomal subunit (SRU) and the large
ribosomal subunit (LRU) the sequence names were extracted from the
corresponding PFAM alignments \cite{PFAM}. Using these names, the following
procedure was used to create the alignments for the single proteins:
\begin{enumerate}
	\item Extract sequences corresponding to names from Uniprot \cite{UniProt}
	\item Run MAFFT \cite{MAFFT} on them using
          \verb+mafft --anysymbol --auto+
	\item Remove columns from the alignment that contain more than 80\% gaps
	\item Create an Hidden Markov Model (HMM) 
          using {\verb+hmmbuild+} from the hmmer suite \cite{HMMER}
	\item Search Uniprot using \verb+hmmsearch+ \cite{HMMER}
	\item Remove inserts
	\item If there exist in one species two or more sequences that
          are more than 95\% identical, remove all but one.
\end{enumerate}

The number of sequences for the single files can be found in Table
\ref{table:singleSizes}

The alignments for the proteins of the Trp Operon where constructed analogously
with some modifications to ensure that only full-length sequences where
extracted. Also, we chose the \verb+linsi+ program of the MAFFT package to
create the initial MSAs. The number of sequences for the Trp alignments can be
found in Table \ref{tab:trp_singleSizes}.

\begin{table}
\begin{tabular}{|c|c|c|c|c|}
\hline
 & L & M & P & S \\
\hline
RS2 & 219 & 6053 & 1.743 & 5.978 \\ 
RS3 & 216 & 6235 & 1.716 & 7.761 \\ 
RS4 & 171 & 8522 & 2.175 & 11.305 \\ 
RS5 & 164 & 5075 & 1.678 & 5.845 \\ 
RS6 & 105 & 4132 & 1.563 & 6.630 \\ 
RS7 & 147 & 5733 & 1.595 & 4.962 \\ 
RS8 & 127 & 5761 & 1.700 & 5.992 \\ 
RS9 & 127 & 4983 & 1.663 & 5.917 \\ 
RS10 & 100 & 4560 & 1.511 & 4.232 \\ 
RS11 & 120 & 5136 & 1.520 & 4.019 \\ 
RS12 & 124 & 5607 & 1.581 & 4.036 \\ 
RS13 & 116 & 5729 & 1.856 & 5.763 \\ 
RS14 & 96 & 5555 & 1.689 & 4.780 \\ 
RS15 & 89 & 5361 & 1.646 & 6.036 \\ 
RS16 & 83 & 4463 & 1.507 & 5.851 \\ 
RS17 & 82 & 4774 & 1.616 & 5.481 \\ 
RS18 & 73 & 4512 & 1.483 & 4.879 \\ 
RS19 & 89 & 5364 & 1.537 & 4.700 \\ 
RS20 & 88 & 3848 & 1.676 & 7.460 \\ 
RS21 & 65 & 3209 & 1.456 & 4.188 \\ 
\hline
\end{tabular}
\begin{tabular}{|c|c|c|c|c|}
\hline
 & L & M & P & S \\
\hline
RL3 & 205 & 6077 & 2.025 & 6.522 \\ 
RL4 & 198 & 5671 & 1.906 & 6.810 \\ 
RL5 & 177 & 5032 & 1.636 & 6.245 \\ 
RL6 & 178 & 5308 & 1.765 & 6.894 \\ 
RL9 & 149 & 4199 & 1.698 & 7.621 \\ 
RL11 & 141 & 5027 & 1.683 & 5.517 \\ 
RL13 & 147 & 5091 & 1.717 & 6.458 \\ 
RL14 & 120 & 5145 & 1.528 & 4.358 \\ 
RL15 & 140 & 5926 & 1.964 & 6.754 \\ 
RL16 & 133 & 5673 & 1.604 & 4.904 \\ 
RL17 & 121 & 4345 & 1.612 & 7.637 \\ 
RL18 & 111 & 4961 & 1.674 & 6.570 \\ 
RL19 & 116 & 4079 & 1.511 & 6.454 \\ 
RL20 & 119 & 4476 & 1.554 & 5.864 \\ 
RL21 & 102 & 4123 & 1.551 & 6.486 \\ 
RL22 & 108 & 6378 & 1.918 & 5.790 \\ 
RL23 & 87 & 5632 & 1.711 & 6.292 \\ 
RL24 & 99 & 9062 & 3.073 & 12.820 \\ 
RL25 & 186 & 3272 & 1.680 & 6.109 \\ 
RL27 & 89 & 3989 & 1.486 & 5.419 \\ 
RL28 & 74 & 4051 & 1.584 & 5.694 \\ 
RL29 & 66 & 4456 & 1.540 & 6.024 \\ 
RL30 & 60 & 4356 & 1.671 & 5.313 \\ 
RL32 & 60 & 4206 & 1.463 & 4.997 \\ 
RL33 & 49 & 4604 & 1.678 & 4.943 \\ 
RL34 & 45 & 3195 & 1.346 & 4.280 \\ 
RL35 & 65 & 3691 & 1.502 & 5.889 \\ 
RL36 & 38 & 3779 & 1.408 & 3.103 \\ 
\hline

\end{tabular}

\caption{Alignment sizes (M) and lengths (L) for proteins of the small (RSXX) and large (RLXX) ribosomal subunit. (P) indicates the average number of paralogs per species and (S) the standard deviation of this number.}
\label{table:singleSizes}
\end{table}

\begin{table}
\begin{center}
\begin{tabular}{|c|c|c|c|c|}
\hline
 & L & M & P & S \\
\hline
TrpA & 259 & 10220 & 4.457 & 32.604 \\ 
TrpB & 399 & 46557 & 16.992 & 145.826 \\ 
TrpC & 254 & 10323 & 4.536 & 39.868 \\ 
TrpD & 337 & 17582 & 7.130 & 59.693 \\ 
TrpE & 460 & 28173 & 11.749 & 124.933 \\ 
TrpF & 197 & 8713 & 4.122 & 32.400 \\ 
TrpG & 192 & 78265 & 24.713 & 187.331 \\ 
\hline
\end{tabular}
\end{center}
\caption{Alignment sizes (M) and lengths (L) for proteins of the Trp Operon. (P) indicates the average number of paralogs per species and (S) the standard deviation of this number.}
\label{tab:trp_singleSizes}
\end{table}

\subsection{Internal Sensitivity Plots}

\begin{figure}
\includegraphics[width=\textwidth]{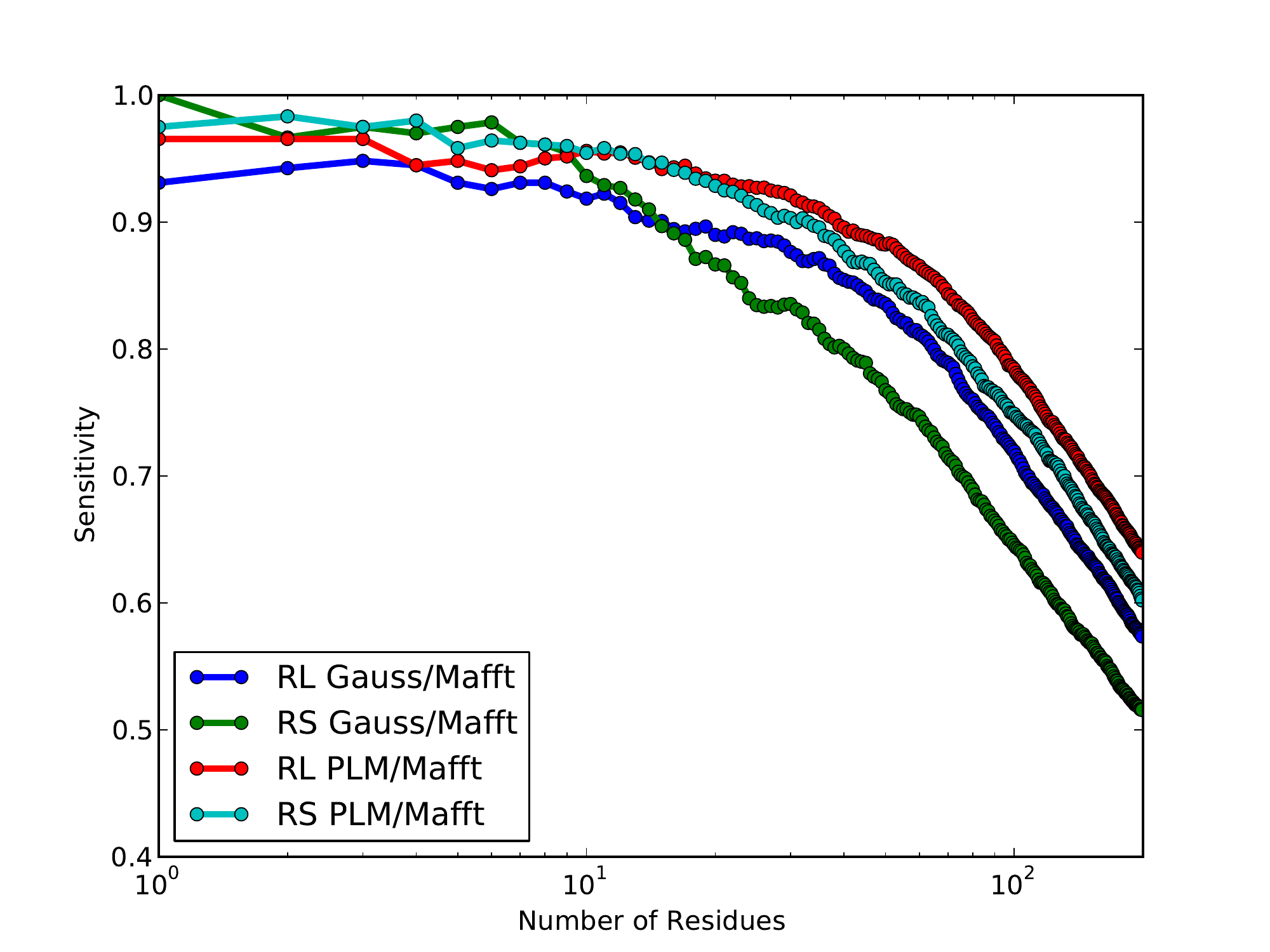}
\caption{Intra-Protein Sensitivity Plots. On the alignments for the
  single ribosomal proteins the plmDCA algorithm was run and an
  ordered list of residue pairs obtained. For every number $n$ on the
  abscissae the fraction of the number of true positives (the
  sensitivity) in the first $n$ pairs on this list was calculated for
  every protein. The plot shows the mean of these values for the
  Gaussian algorithm of \cite{Baldassi2014PLOSONE} and the plmDCA
  algorithm run on the proteins of the large and small ribosomal
  subunit.}
\label{fig:rsSens}
\end{figure}

As an assessment of quality for the alignments, sensitivity plots
using the pdb files \verb+2Z4K+ and \verb+2Z4L+ were made. Figure
\ref{fig:rsSens} shows results for contact predictions based on the
GaussDCA \cite{Baldassi2014PLOSONE} and plmDCA alghorithm
\cite{ekeberg2014fast}.

\section{Matching Procedure}
\label{sec:matching}

\subsection{Pipeline for Matching}

The problem of generating a concatenated alignment from two MSAs of two
different protein families (say $\mathrm{MSA}_1$ and $\mathrm{MSA}_2$) is to
decide which sequence from the first alignment should be concatenated to which
sequence from the other alignment. This means to find for any protein $p_i^1$
in $\mathrm{MSA}_1$ 
a matching partner $p_j^2$ in $\mathrm{MSA}_2$ belonging to the same species.
The problem is trivially solved in the case when no paralogs are present and
each species has one and only one sequence in each individual MSA. In this case
we can simply concatenate these two sequences (we term this case \emph{matching
by uniqueness}). The problem is that species often have several paralogs. In
this case, given that we would like to observe a co-evolutionary signal between
protein interaction partners, one would like to match sequences of proteins
that are (possibly) interacting.

As long as Prokaryotes are concerned, it turns out empirically that proteins
are more likely to interact if their genes are {\em co-localized} on the DNA
\cite{weigt2009identification, nimwegen2008MSB}. This suggests to try to match
proteins that are close on the genome when creating a concatenated MSA. 

As a proxy to the genomic distance we use a \emph{distance} between Uniprot
accession numbers (UAN). This UAN consists of a 6 digit alphanumeric sequence
for every sequence and can be extracted from the sequence annotation, e.g. the
"\verb+D8UHT6+" part of the sequence annotation "\verb+D8UHT6_PANSA+". 


We define the distance between UANs as follows: Different positions in the UAN
can take on different values, some only numeric (0-9) and some alphanumeric
values (0-9,A-Z). We define for every position $i\in 1\dots 6$ the number $B_i$
as the number of different values position $i$ can take, i.e. $B_i=10$ for the
numeric positions and $B_i=36$ for the alphanumeric positions.

We further map the possible single position values in the UAN to the
natural numbers in ascending order, i.e. we assign to the numeric
symbols $0-9$ the natural numbers $0-9$ and to the letters the natural
numbers following $9$ (so to A we assign $10$, to B we assign $11$
etc.). This leads for example for the the UAN \verb+L9XG27+ to the
numeric sequence $A=(21,9,33,16,2,7)$.

Now we can define a unique number $N$ for any UAN that has been mapped
to the sequence of natural numbers $A_i$ as
\begin{equation}
N = A_6 + \sum\limits_{i=1}^{5} A_i \left(\prod\limits_{j=i+1}^{6} B_j\right)
\end{equation}
The distance between two UANs that have been mapped to the numbers
$N_1$ and $N_2$ can now be defined as
\begin{equation}
	D_{12} = \left|N_1-N_2\right|
\end{equation}
This procedure induces a distance $D_{ij}$ for any sequence $p_i \in
\mathrm{MSA}_1$ and $p_j \in\mathrm{MSA}_2$, where both $p_i, p_j$ belong to
the same species. In this way we define a complete weighted bipartite graph,
and the problem of finding the proper pairing can thus be translated into a
minimum weighted bipartite matching problem.  This problem can be readily
solved using a standard linear programming techniques. Finally we discard from
the optimal solution sequence pairs whose distance is above a given threshold
of 100 (manually optimized on the small ribosomal subunit). In the cases we
analyzed, such a threshold moderately increases the quality of the prediction
of interaction partners.

\section{Inference technique}
As a simple but meaningful statistical model, we consider a pairwise
generalized 21 states (to mimic the 20 amino acids + 1 insert symbol
alphabet of MSAs) Potts model with the following Hamiltonian
\begin{equation}
{\cal H} = - \sum_{0 \leq i < j \leq L} J_{i,j}(x_i,x_j) - \sum_{i=1}^L h_i(x_i)
\label{eq:hamilton}
\end{equation}

We can now assume to have a dataset $D=\{x^1,\dots,x^M \}$, where $x$
represents one sequence, either artificially generated, or extracted using the
bioinformatic pipeline discussed above. Notice that if the sequences $x$ are
concatenations of two sequences $(x,x')$, the sums in Equation
\ref{eq:hamilton} can be split into three parts: One in which appear only sites
in $x$, one in which appear only sites in $x'$ and one interaction part with
$J_{ij}$ for which $i$ is in $x$ and $j$ in $x'$. By labeling the first part
$H(x)$, the second $H'(x')$ and the third $H^{int}(x,x')$ one arrives at the
representation referred to in the main text. Given that the representations are
mathematically equivalent, we will here in supplemental information treat the
sequence as one simple sequence $x$.

The inference proceeds by assuming as a working hypothesis that the dataset $D$
is composed by configuration sampled uniformly from the equilibrium
Boltzmann-Gibbs distribution $P(\vec x) = \exp(-{\cal H}) / Z$ (as an inference
process, we are free to consider $T=\beta=1$). We are now ready to use $D$ to
infer the topology of the network. To do so -- as discussed in the main text --
in the last years different maximum-likelihood techniques have been proposed
\cite{weigt2009PNAS, morcos2011direct, marks2011direct,
balakrishnan2011Protein, jones2012PSICOV, aurell2013pl}. So far the most
promising in terms of accuracy seems to be the pseudo-likelihood maximization
introduced in \cite{aurell2013pl} where from the previously defined
Boltzmann-Gibbs measure we consider the following conditional probability
distribution:
\begin{equation}
P_i(x_i | x_{\backslash i} ) = 
\frac{\exp\left(\sum_{j \neq i} J_{ij}(x_i,x_j) + h_i(x_i)\right)}
{\sum_{a=1}^{21} \exp\left(\sum_{j \neq i} J_{ij}(x_i, a) + h_i(a)\right)}
\end{equation}
Given a data set $D$ we can thus maximize the conditional likelihood
by maximizing
\begin{equation}
L_i(J_{i,\backslash i},h_i) = \frac 1M \sum_{\alpha=1}^M 
\log P_i(x^\alpha_i | x^\alpha_{\backslash i} )\,\,\,\,\,\,,
\label{eq:pl}
\end{equation}
as a function of $J_{i,\backslash i},h_i$.  As customary in many
maximum-likelihood inference techniques, we add to the maximization an
${\cal L}_2$ regularization term, so that eventually the extremization
procedure turns out to be:
\begin{equation}
\{J^*_{i,\backslash j} , h^*_i\} = \argmax_{J_{i,\backslash i},h_i} \{
L_i - \lambda_J \sum_{j\neq i}\| J_{ij}\|_2 - \lambda_h \|h_i\|_2 \}\,,
\label{eq:extreme}
\end{equation}
with $\| J_{ij}\|_2 = \sum_{a,b=1}^{21}J_{ij}^2(a,b)$, and $\| h_i\|_2
= \sum_{a=1}^{21}h^2(a)$.  We refer to the original paper
\cite{aurell2013pl} for the details of the implementation. We only
mention that beside the original MATLAB \cite{MATLAB2014a}
implementation available at {\verb+http://plmdca.csc.kth.se/+}, we
developed an efficient implementation of the pseudo-likelihood
implementation in a new open-source language called Julia
\cite{Julia}. The package can be downloaded at
     {\verb+https://github.com/pagnani/PlmDCA+}.

\section{Ribosomal Protein Interaction Partner Prediction}

Using the ribosomal alignments as described in Section
\ref{sec:riboAlign} and the matching as described in Section
\ref{sec:matching}, concatenated alignments for the ribosomal proteins
(small and large ribosomal subunit independently) were created. Tables
\ref{tab:matchedSmallRibo_tr0} and  \ref{tab:matchedSmallRibo_tr100} show the resulting alignment sizes for the
SRU and Tables \ref{tab:matchedLargeRibo_tr0} and \ref{tab:matchedLargeRibo_tr100} for the LRU.

The creation of the alignments for the Trp Proteins was analagous and the resulting alignment sizes can be found in Table
\ref{tab:matchedTrp}.

\begin{table}
\tiny
\hspace*{-2cm}\begin{tabular}{c| *{20}{p{0.3cm}}}
	&RL2	&RL3	&RL4	&RL5	&RL6	&RL7	&RL8	&RL9	&RL10	&RL11	&RL12	&RL13	&RL14	&RL15	&RL16	&RL17	&RL18	&RL19	&RL20	&RL21 \\ 
 \hline 
RL2 	&	&2914	&2537	&2458	&2224	&2825	&2833	&2491	&2457	&2839	&2664	&2342	&2511	&2748	&2462	&2373	&2515	&2842	&2109	&1740\\ 
RL3 	&	&	&2947	&2719	&2430	&3109	&3223	&2531	&2680	&3097	&2922	&2577	&2992	&2694	&2645	&2686	&2659	&3213	&2123	&1907\\ 
RL4 	&	&	&	&2411	&1837	&2719	&2812	&2214	&2314	&2802	&2528	&2463	&2522	&2319	&2064	&2354	&2182	&2765	&1774	&1468\\ 
RL5 	&	&	&	&	&2231	&2613	&2736	&2508	&2607	&2623	&2410	&2532	&2517	&2381	&2221	&2699	&2142	&2657	&2127	&1743\\ 
RL6 	&	&	&	&	&	&2206	&2251	&2216	&2200	&2204	&2041	&2117	&1938	&2169	&2430	&2226	&2590	&2263	&2116	&1931\\ 
RL7 	&	&	&	&	&	&	&3001	&2469	&2580	&2914	&3172	&2452	&2753	&2650	&2414	&2524	&2483	&2937	&2089	&1711\\ 
RL8 	&	&	&	&	&	&	&	&2539	&2782	&3098	&2831	&2654	&3004	&2707	&2494	&3037	&2497	&3402	&2114	&1786\\ 
RL9 	&	&	&	&	&	&	&	&	&2466	&2564	&2348	&2400	&2284	&2383	&2204	&2469	&2188	&2489	&2103	&1755\\ 
RL10 	&	&	&	&	&	&	&	&	&	&2579	&2423	&2460	&2443	&2378	&2212	&2711	&2144	&2784	&2100	&1734\\ 
RL11 	&	&	&	&	&	&	&	&	&	&	&2810	&2618	&2849	&2694	&2417	&2604	&2497	&3008	&2083	&1729\\ 
RL12 	&	&	&	&	&	&	&	&	&	&	&	&2295	&2646	&2507	&2224	&2369	&2303	&2828	&1925	&1542\\ 
RL13 	&	&	&	&	&	&	&	&	&	&	&	&	&2395	&2188	&2174	&2502	&2117	&2564	&2060	&1712\\ 
RL14 	&	&	&	&	&	&	&	&	&	&	&	&	&	&2420	&2169	&2510	&2398	&2920	&1804	&1529\\ 
RL15 	&	&	&	&	&	&	&	&	&	&	&	&	&	&	&2417	&2348	&2461	&2679	&2115	&1753\\ 
RL16 	&	&	&	&	&	&	&	&	&	&	&	&	&	&	&	&2212	&2532	&2474	&2116	&1925\\ 
RL17 	&	&	&	&	&	&	&	&	&	&	&	&	&	&	&	&	&2127	&2918	&2097	&1735\\ 
RL18 	&	&	&	&	&	&	&	&	&	&	&	&	&	&	&	&	&	&2484	&2043	&1867\\ 
RL19 	&	&	&	&	&	&	&	&	&	&	&	&	&	&	&	&	&	&	&2096	&1767\\ 
RL20 	&	&	&	&	&	&	&	&	&	&	&	&	&	&	&	&	&	&	&	&1683\\ 
RL21 	&	&	&	&	&	&	&	&	&	&	&	&	&	&	&	&	&	&	&	&\\ 
 \hline 
	&2520	&2740	&2370	&2439	&2191	&2612	&2726	&2348	&2424	&2633	&2463	&2349	&2453	&2422	&2306	&2447	&2328	&2689	&2036	&1738

\hspace*{-2cm}\end{tabular}
\caption{Matched Alignment Sizes for Small Ribosomal Subunit, at threshold 100}
\label{tab:matchedSmallRibo_tr100}
\end{table}

\begin{table}
\tiny
\hspace*{-2cm}\begin{tabular}{c| *{20}{p{0.3cm}}}
	&RL2	&RL3	&RL4	&RL5	&RL6	&RL7	&RL8	&RL9	&RL10	&RL11	&RL12	&RL13	&RL14	&RL15	&RL16	&RL17	&RL18	&RL19	&RL20	&RL21 \\ 
 \hline 
RL2 	&	&2594	&2143	&2343	&2149	&2608	&2611	&2342	&2333	&2592	&2379	&2095	&2256	&2533	&2318	&2303	&2311	&2599	&2051	&1692\\ 
RL3 	&	&	&2219	&2373	&2371	&2615	&2628	&2363	&2348	&2579	&2406	&2097	&2267	&2535	&2506	&2341	&2444	&2656	&2057	&1871\\ 
RL4 	&	&	&	&1895	&1722	&2178	&2140	&1893	&1888	&2117	&2010	&1707	&1886	&2072	&1877	&1858	&1877	&2146	&1653	&1394\\ 
RL5 	&	&	&	&	&2156	&2356	&2364	&2344	&2333	&2322	&2156	&2078	&1984	&2313	&2160	&2320	&2084	&2319	&2069	&1707\\ 
RL6 	&	&	&	&	&	&2135	&2189	&2153	&2146	&2134	&1960	&2063	&1840	&2138	&2376	&2150	&2251	&2180	&2071	&1879\\ 
RL7 	&	&	&	&	&	&	&2617	&2327	&2326	&2596	&2494	&2088	&2267	&2536	&2304	&2304	&2310	&2605	&2043	&1665\\ 
RL8 	&	&	&	&	&	&	&	&2338	&2341	&2623	&2379	&2113	&2302	&2570	&2385	&2336	&2333	&2669	&2057	&1743\\ 
RL9 	&	&	&	&	&	&	&	&	&2323	&2324	&2156	&2071	&1996	&2315	&2155	&2303	&2102	&2320	&2057	&1700\\ 
RL10 	&	&	&	&	&	&	&	&	&	&2327	&2153	&2090	&1996	&2301	&2159	&2302	&2096	&2330	&2055	&1693\\ 
RL11 	&	&	&	&	&	&	&	&	&	&	&2386	&2091	&2280	&2559	&2318	&2291	&2318	&2596	&2040	&1685\\ 
RL12 	&	&	&	&	&	&	&	&	&	&	&	&1920	&2145	&2324	&2094	&2120	&2069	&2395	&1866	&1508\\ 
RL13 	&	&	&	&	&	&	&	&	&	&	&	&	&1806	&2077	&2091	&2052	&2054	&2086	&2003	&1661\\ 
RL14 	&	&	&	&	&	&	&	&	&	&	&	&	&	&2213	&2037	&1980	&2109	&2290	&1735	&1485\\ 
RL15 	&	&	&	&	&	&	&	&	&	&	&	&	&	&	&2316	&2287	&2304	&2539	&2043	&1697\\ 
RL16 	&	&	&	&	&	&	&	&	&	&	&	&	&	&	&	&2149	&2451	&2373	&2066	&1877\\ 
RL17 	&	&	&	&	&	&	&	&	&	&	&	&	&	&	&	&	&2077	&2321	&2047	&1687\\ 
RL18 	&	&	&	&	&	&	&	&	&	&	&	&	&	&	&	&	&	&2308	&1998	&1827\\ 
RL19 	&	&	&	&	&	&	&	&	&	&	&	&	&	&	&	&	&	&	&2033	&1734\\ 
RL20 	&	&	&	&	&	&	&	&	&	&	&	&	&	&	&	&	&	&	&	&1617\\ 
RL21 	&	&	&	&	&	&	&	&	&	&	&	&	&	&	&	&	&	&	&	&\\ 
 \hline 
	&2329	&2383	&1930	&2193	&2109	&2335	&2355	&2189	&2186	&2325	&2154	&2013	&2046	&2299	&2211	&2170	&2175	&2342	&1977	&1691

\hspace*{-2cm}\end{tabular}
\caption{Matched Alignment Sizes for Small Ribosomal Subunit, at threshold 0 (matching by uniqueness)}
\label{tab:matchedSmallRibo_tr0}
\end{table}

\begin{table}
\tiny
\hspace*{-3cm}\begin{tabular}{c| *{29}{p{0.2cm}}}
	&RL2	&RL3	&RL4	&RL5	&RL6	&RL9	&RL11	&RL13	&RL14	&RL15	&RL16	&RL17	&RL18	&RL19	&RL20	&RL21	&RL22	&RL23	&RL24	&RL25	&RL27	&RL28	&RL29	&RL30	&RL32	&RL33	&RL34	&RL35	&RL36 \\ 
 \hline 
RL2 	&	&2699	&2720	&2875	&2824	&2142	&2505	&2461	&3077	&2658	&3101	&2438	&2672	&2190	&2509	&2174	&2957	&3075	&2435	&1739	&2164	&1932	&2904	&2471	&2296	&2163	&1970	&2094	&2328\\ 
RL3 	&	&	&2789	&2626	&2923	&2149	&2382	&2395	&2873	&2604	&2626	&2456	&2649	&2184	&2161	&2164	&3132	&2661	&2338	&1733	&2184	&1964	&2591	&2290	&2033	&1902	&1993	&2108	&1984\\ 
RL4 	&	&	&	&2639	&2709	&2167	&2407	&2418	&2637	&2676	&2647	&2438	&2871	&2209	&2167	&2168	&2788	&2695	&2805	&1747	&2195	&1962	&2652	&2333	&2040	&1894	&2001	&2134	&2011\\ 
RL5 	&	&	&	&	&2902	&2232	&2492	&2498	&2799	&2692	&3134	&2608	&2775	&2312	&2327	&2309	&2688	&3035	&2483	&1773	&2299	&2014	&2744	&2389	&2164	&1945	&2084	&2203	&2136\\ 
RL6 	&	&	&	&	&	&2216	&2551	&2506	&3043	&2768	&2839	&2651	&2828	&2283	&2277	&2275	&2990	&2773	&2495	&1785	&2286	&2005	&2828	&2455	&2114	&1937	&2039	&2207	&2101\\ 
RL9 	&	&	&	&	&	&	&2154	&2156	&2168	&2161	&2174	&2191	&2238	&2283	&2224	&2237	&2153	&2165	&502	&1792	&2259	&2025	&2230	&1877	&2099	&1810	&2106	&2190	&1768\\ 
RL11 	&	&	&	&	&	&	&	&2422	&2492	&2375	&2468	&2223	&2499	&2217	&2179	&2174	&2370	&2539	&2314	&1732	&2187	&1973	&2457	&2131	&2040	&2024	&1991	&2133	&1777\\ 
RL13 	&	&	&	&	&	&	&	&	&2491	&2482	&2498	&2246	&2493	&2208	&2197	&2198	&2340	&2482	&1127	&1755	&2217	&1980	&2445	&2110	&2053	&1852	&1999	&2155	&1800\\ 
RL14 	&	&	&	&	&	&	&	&	&	&2643	&3080	&2465	&2752	&2232	&2574	&2227	&3166	&3012	&2328	&1735	&2208	&1989	&2606	&2241	&2345	&2181	&2003	&2126	&2345\\ 
RL15 	&	&	&	&	&	&	&	&	&	&	&2616	&2509	&2740	&2189	&2169	&2160	&2714	&2700	&2354	&1760	&2196	&1964	&2706	&2388	&2024	&1848	&1970	&2109	&2040\\ 
RL16 	&	&	&	&	&	&	&	&	&	&	&	&2488	&2730	&2240	&2564	&2229	&2812	&3348	&2314	&1759	&2213	&1991	&2610	&2259	&2372	&2191	&2012	&2142	&2325\\ 
RL17 	&	&	&	&	&	&	&	&	&	&	&	&	&2755	&2385	&2176	&2341	&2465	&2530	&2207	&1726	&2380	&2146	&2689	&2180	&2190	&1917	&2131	&2181	&2302\\ 
RL18 	&	&	&	&	&	&	&	&	&	&	&	&	&	&2422	&2223	&2369	&2734	&2739	&2934	&1772	&2417	&2170	&2886	&2454	&2227	&1975	&2176	&2216	&2193\\ 
RL19 	&	&	&	&	&	&	&	&	&	&	&	&	&	&	&2331	&2437	&2188	&2262	&580	&1774	&2507	&2277	&2434	&1913	&2361	&1906	&2225	&2315	&1948\\ 
RL20 	&	&	&	&	&	&	&	&	&	&	&	&	&	&	&	&2311	&2483	&2518	&411	&1787	&2297	&2011	&2248	&1868	&2450	&2161	&2048	&2477	&2152\\ 
RL21 	&	&	&	&	&	&	&	&	&	&	&	&	&	&	&	&	&2202	&2242	&542	&1754	&2692	&2163	&2380	&1887	&2258	&1890	&2177	&2259	&1913\\ 
RL22 	&	&	&	&	&	&	&	&	&	&	&	&	&	&	&	&	&	&2942	&2380	&1739	&2208	&1970	&2595	&2251	&2297	&2160	&1989	&2120	&2294\\ 
RL23 	&	&	&	&	&	&	&	&	&	&	&	&	&	&	&	&	&	&	&2405	&1748	&2254	&2007	&2727	&2397	&2381	&2221	&2044	&2152	&2337\\ 
RL24 	&	&	&	&	&	&	&	&	&	&	&	&	&	&	&	&	&	&	&	&391	&503	&528	&2459	&2093	&449	&1111	&522	&437	&1468\\ 
RL25 	&	&	&	&	&	&	&	&	&	&	&	&	&	&	&	&	&	&	&	&	&1770	&1595	&1745	&1547	&1649	&1564	&1598	&1761	&1362\\ 
RL27 	&	&	&	&	&	&	&	&	&	&	&	&	&	&	&	&	&	&	&	&	&	&2234	&2427	&1915	&2300	&1928	&2232	&2295	&1931\\ 
RL28 	&	&	&	&	&	&	&	&	&	&	&	&	&	&	&	&	&	&	&	&	&	&	&2148	&1719	&2185	&1935	&2015	&2039	&1750\\ 
RL29 	&	&	&	&	&	&	&	&	&	&	&	&	&	&	&	&	&	&	&	&	&	&	&	&2584	&2223	&1957	&2163	&2251	&2160\\ 
RL30 	&	&	&	&	&	&	&	&	&	&	&	&	&	&	&	&	&	&	&	&	&	&	&	&	&1765	&1579	&1732	&1851	&1738\\ 
RL32 	&	&	&	&	&	&	&	&	&	&	&	&	&	&	&	&	&	&	&	&	&	&	&	&	&	&2183	&2074	&2130	&2132\\ 
RL33 	&	&	&	&	&	&	&	&	&	&	&	&	&	&	&	&	&	&	&	&	&	&	&	&	&	&	&1741	&1819	&1921\\ 
RL34 	&	&	&	&	&	&	&	&	&	&	&	&	&	&	&	&	&	&	&	&	&	&	&	&	&	&	&	&2089	&1779\\ 
RL35 	&	&	&	&	&	&	&	&	&	&	&	&	&	&	&	&	&	&	&	&	&	&	&	&	&	&	&	&	&1800\\ 
RL36 	&	&	&	&	&	&	&	&	&	&	&	&	&	&	&	&	&	&	&	&	&	&	&	&	&	&	&	&	&\\ 
 \hline 
	&2485	&2378	&2390	&2471	&2486	&2067	&2257	&2214	&2494	&2365	&2492	&2336	&2497	&2172	&2189	&2148	&2469	&2514	&1604	&1664	&2168	&1953	&2459	&2086	&2101	&1918	&1961	&2064	&1993

\hspace*{-3cm}\end{tabular}
\caption{Matched Alignment Sizes for Large Ribosomal Subunit, at threshold 100}
\label{tab:matchedLargeRibo_tr100}
\end{table}

\begin{table}
\tiny
\hspace*{-3cm}\begin{tabular}{c| *{29}{p{0.2cm}}}
	&RL2	&RL3	&RL4	&RL5	&RL6	&RL9	&RL11	&RL13	&RL14	&RL15	&RL16	&RL17	&RL18	&RL19	&RL20	&RL21	&RL22	&RL23	&RL24	&RL25	&RL27	&RL28	&RL29	&RL30	&RL32	&RL33	&RL34	&RL35	&RL36 \\ 
 \hline 
RL2 	&	&2144	&2173	&2333	&2307	&2079	&2277	&2286	&2568	&2115	&2547	&2041	&2313	&2120	&2325	&2095	&2325	&2552	&217	&1698	&2100	&1895	&2257	&1964	&2182	&1875	&1918	&2052	&1900\\ 
RL3 	&	&	&2139	&2179	&2162	&2087	&2152	&2169	&2165	&2102	&2168	&2033	&2178	&2126	&2109	&2095	&2112	&2177	&177	&1703	&2115	&1899	&2140	&1851	&1965	&1651	&1933	&2068	&1648\\ 
RL4 	&	&	&	&2205	&2188	&2099	&2175	&2191	&2190	&2118	&2197	&2045	&2207	&2140	&2119	&2102	&2117	&2193	&189	&1704	&2126	&1912	&2170	&1865	&1978	&1663	&1944	&2075	&1652\\ 
RL5 	&	&	&	&	&2425	&2176	&2316	&2319	&2388	&2151	&2370	&2162	&2379	&2255	&2257	&2221	&2150	&2369	&221	&1735	&2235	&1960	&2394	&2038	&2093	&1727	&2005	&2161	&1771\\ 
RL6 	&	&	&	&	&	&2164	&2307	&2310	&2344	&2149	&2337	&2134	&2368	&2221	&2214	&2187	&2130	&2324	&221	&1725	&2204	&1949	&2379	&2016	&2045	&1694	&1981	&2144	&1713\\ 
RL9 	&	&	&	&	&	&	&2088	&2106	&2110	&2095	&2122	&2142	&2178	&2232	&2176	&2187	&2102	&2119	&167	&1735	&2224	&1967	&2181	&1824	&2059	&1697	&2018	&2147	&1720\\ 
RL11 	&	&	&	&	&	&	&	&2305	&2317	&2117	&2300	&2061	&2319	&2143	&2129	&2115	&2109	&2299	&219	&1693	&2133	&1922	&2289	&1978	&1994	&1672	&1938	&2074	&1668\\ 
RL13 	&	&	&	&	&	&	&	&	&2312	&2130	&2306	&2064	&2323	&2152	&2141	&2129	&2121	&2305	&205	&1713	&2145	&1918	&2292	&1984	&1988	&1670	&1952	&2090	&1668\\ 
RL14 	&	&	&	&	&	&	&	&	&	&2146	&2600	&2089	&2349	&2165	&2392	&2155	&2388	&2606	&217	&1710	&2151	&1940	&2318	&2012	&2259	&1941	&1961	&2085	&1998\\ 
RL15 	&	&	&	&	&	&	&	&	&	&	&2166	&2062	&2162	&2137	&2120	&2107	&2120	&2132	&181	&1713	&2127	&1917	&2110	&1842	&1975	&1657	&1936	&2073	&1653\\ 
RL16 	&	&	&	&	&	&	&	&	&	&	&	&2089	&2335	&2171	&2370	&2144	&2347	&2539	&216	&1724	&2155	&1935	&2304	&2001	&2226	&1902	&1963	&2095	&1931\\ 
RL17 	&	&	&	&	&	&	&	&	&	&	&	&	&2302	&2346	&2146	&2280	&2049	&2121	&222	&1677	&2337	&2099	&2305	&1798	&2155	&1724	&2094	&2144	&1802\\ 
RL18 	&	&	&	&	&	&	&	&	&	&	&	&	&	&2366	&2177	&2310	&2144	&2381	&293	&1731	&2366	&2127	&2521	&2056	&2177	&1774	&2126	&2166	&1821\\ 
RL19 	&	&	&	&	&	&	&	&	&	&	&	&	&	&	&2260	&2370	&2138	&2208	&235	&1748	&2438	&2156	&2392	&1875	&2248	&1816	&2190	&2260	&1889\\ 
RL20 	&	&	&	&	&	&	&	&	&	&	&	&	&	&	&	&2231	&2345	&2379	&174	&1737	&2226	&1960	&2197	&1840	&2294	&1952	&2003	&2177	&2011\\ 
RL21 	&	&	&	&	&	&	&	&	&	&	&	&	&	&	&	&	&2125	&2179	&224	&1720	&2367	&2089	&2317	&1838	&2191	&1777	&2130	&2203	&1849\\ 
RL22 	&	&	&	&	&	&	&	&	&	&	&	&	&	&	&	&	&	&2373	&170	&1707	&2132	&1917	&2107	&1810	&2215	&1908	&1946	&2081	&1936\\ 
RL23 	&	&	&	&	&	&	&	&	&	&	&	&	&	&	&	&	&	&	&227	&1716	&2187	&1955	&2351	&2006	&2289	&1957	&1988	&2107	&1999\\ 
RL24 	&	&	&	&	&	&	&	&	&	&	&	&	&	&	&	&	&	&	&	&116	&238	&211	&288	&169	&224	&180	&207	&182	&195\\ 
RL25 	&	&	&	&	&	&	&	&	&	&	&	&	&	&	&	&	&	&	&	&	&1733	&1539	&1713	&1517	&1602	&1474	&1566	&1704	&1323\\ 
RL27 	&	&	&	&	&	&	&	&	&	&	&	&	&	&	&	&	&	&	&	&	&	&2164	&2376	&1863	&2243	&1814	&2194	&2236	&1879\\ 
RL28 	&	&	&	&	&	&	&	&	&	&	&	&	&	&	&	&	&	&	&	&	&	&	&2109	&1654	&2036	&1697	&1980	&1989	&1685\\ 
RL29 	&	&	&	&	&	&	&	&	&	&	&	&	&	&	&	&	&	&	&	&	&	&	&	&2052	&2188	&1771	&2132	&2210	&1835\\ 
RL30 	&	&	&	&	&	&	&	&	&	&	&	&	&	&	&	&	&	&	&	&	&	&	&	&	&1724	&1427	&1711	&1821	&1441\\ 
RL32 	&	&	&	&	&	&	&	&	&	&	&	&	&	&	&	&	&	&	&	&	&	&	&	&	&	&1988	&2048	&2084	&2033\\ 
RL33 	&	&	&	&	&	&	&	&	&	&	&	&	&	&	&	&	&	&	&	&	&	&	&	&	&	&	&1655	&1708	&1693\\ 
RL34 	&	&	&	&	&	&	&	&	&	&	&	&	&	&	&	&	&	&	&	&	&	&	&	&	&	&	&	&2051	&1730\\ 
RL35 	&	&	&	&	&	&	&	&	&	&	&	&	&	&	&	&	&	&	&	&	&	&	&	&	&	&	&	&	&1763\\ 
RL36 	&	&	&	&	&	&	&	&	&	&	&	&	&	&	&	&	&	&	&	&	&	&	&	&	&	&	&	&	&\\ 
 \hline 
	&2095	&1980	&1996	&2107	&2084	&2000	&2040	&2046	&2138	&1975	&2127	&2019	&2141	&2100	&2088	&2062	&2040	&2144	&207	&1613	&2090	&1878	&2132	&1785	&2018	&1695	&1904	&1998	&1722

\hspace*{-3cm}\end{tabular}
\caption{Matched Alignment Sizes for Large Ribosomal Subunit, at treshold 0 (matching by uniqueness)}
\label{tab:matchedLargeRibo_tr0}
\end{table}

\begin{table}
\begin{center}
\hspace*{-3cm}\begin{tabular}{|c | c |c | c |}
\hline
P1 & P2  & tr=100 & tr=0 \\
\hline
TrpC & TrpG & 4272 & 18 \\ 
TrpE & TrpF & 2519 & 830 \\ 
TrpA & TrpD & 2823 & 743 \\ 
TrpD & TrpG & 6249 & 28 \\ 
TrpB & TrpF & 3643 & 95 \\ 
TrpB & TrpD & 3737 & 95 \\ 
TrpB & TrpG & 8053 & 41 \\ 
TrpE & TrpG & 5324 & 8 \\ 
TrpD & TrpF & 2819 & 695 \\ 
TrpC & TrpF & 3825 & 1578 \\ 
TrpA & TrpC & 3198 & 1546 \\ 
TrpC & TrpD & 3392 & 748 \\ 
TrpA & TrpF & 3357 & 1433 \\ 
TrpA & TrpE & 3118 & 905 \\ 
TrpD & TrpE & 2681 & 482 \\ 
TrpB & TrpC & 3326 & 82 \\ 
TrpB & TrpE & 3911 & 53 \\ 
TrpC & TrpE & 2976 & 930 \\ 
TrpF & TrpG & 3635 & 32 \\ 
TrpA & TrpB & 4374 & 95 \\ 
TrpA & TrpG & 4646 & 22 \\ 

\hline
\end{tabular}
\caption{Matched Alignment Sizes for Trp for different matching thresholds (threshold 0 corresponds to matching by uniqueness)}
\label{tab:matchedTrp}
\end{center}
\end{table}

As discussed in the main text, in principle one would be interested in a MSA in
which a sequence is a concatenation of sequences from all proteins families in
the complex at once. A comparative glance at Tables \ref{tab:matchedLargeRibo_tr100}
and \ref{table:singleSizes} shows that in the matching procedure described
above a lot of sequences have to be discarded for not having a suitable
matching partner. This leads to a reduction of the predictive power of the
method. It is expected that extending the matching procedure to more than two
proteins would lead to very low sequence numbers in the matched alignments and
in turn reduce the predictive power of the method further. For this reason we
only performed the concatenation of pairs of proteins. 

In order to produce an interaction score for the two proteins, we run the PLM
algorithm \cite{aurell2013pl} on the concatenated alignments. This results in a
list of residue pairs of the alignment ordered by their interaction strength.
We filtered out the pairs that contain one residue of one protein and one of
the other. This results in a list of \emph{possibly} interacting inter-protein
residue pairs ordered by the interaction score. In order to arrive at an
interaction score for the two proteins we took the mean of the scores for the 4
highest scoring pairs (PPI-score). The number 4 was used because it performed
best on the small ribosomal subunit, but the predictive performance on a
larger-scale network is virtually identical for any value between 1 and 6 (see
Figure S8). The list of protein pairs ordered by this score was used for
prediction. The first few predictions are shown in Table \ref{tab:ppiRibo}. For
completeness, we show the same table but with the score calculated by the
Gaussian approximation of \cite{Baldassi2014PLOSONE} in Table
\ref{tab:ppiRiboGauss}. Finally in Table~\ref{tab:lscontSRU} we display for the
LSU the number of intra/inter-protein contacts, while in
Table~\ref{tab:lscontLRU} we do the same for the LRU.

Table \ref{tab:ppiTrp} shows the interaction scores for the protein pairs of the Trp Operon.

\begin{figure}
\centering
\includegraphics[width=\textwidth]{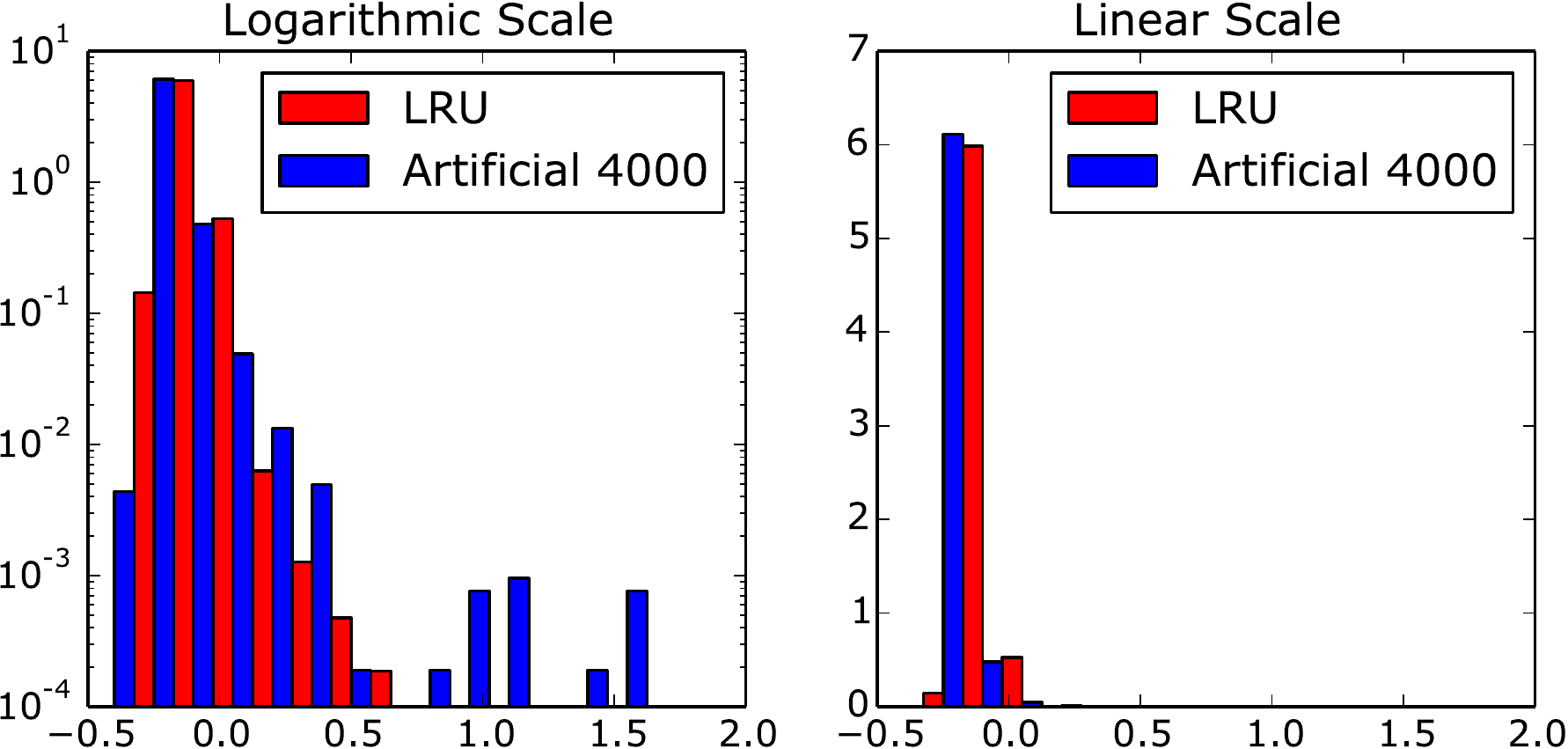}
\caption{Histograms of interaction scores resulting from the analysis
  of the LRU and the artificial complex (combined strategy). Both
  intra- and inter-protein scores are included. The plots are
  normalized such that the area of all bars of a given color sums to
  one. The data is shown both on a logarithmic (left) and on a linear
  scale (right). }
\label{fig:artSRUhist}
\end{figure}

\begin{table}
\begin{tabular}{|c|c|c|c|}
\hline
P1 & P2 & Score & Interacting\\
\hline
RS10 &RS14 &0.618890 &1\\
RS18 &RS6  &0.422457 &1\\
RS14 &RS3  &0.394753 &1\\
RS10 &RS9  &0.347508 &1\\
RS13 &RS19 &0.317640 &1\\
RS13 &RS21 &0.306248 &0\\
RS11 &RS21 &0.296700 &1\\
RS14 &RS19 &0.291335 &1\\
RS12 &RS21 &0.290965 &0\\
RS16 &RS4  &0.287438 &0\\
RS21 &RS7  &0.287102 &0\\
RS13 &RS15 &0.284783 &0\\
RS12 &RS16 &0.283105 &0\\
RS19 &RS21 &0.282142 &0\\
RS10 &RS18 &0.279595 &0\\
\hline
\end{tabular}
\begin{tabular}{|c|c|c|c|}
\hline
P1 & P2 & Score & Interacting\\
\hline
RL20 &RL21 &0.576795 &1\\
RL14 &RL19 &0.514107 &1\\
RL15 &RL35 &0.440323 &1\\
RL15 &RL21 &0.439233 &1\\
RL17 &RL32 &0.425920 &1\\
RL20 &RL32 &0.421733 &1\\
RL23 &RL29 &0.414060 &1\\
RL13 &RL20 &0.334348 &1\\
RL19 &RL3  &0.328640 &1\\
RL30 &RL34 &0.326368 &0\\
RL22 &RL32 &0.324540 &1\\
RL16 &RL36 &0.318915 &1\\
RL16 &RL33 &0.313083 &0\\
RL33 &RL36 &0.307188 &0\\
RL27 &RL34 &0.306283 &0\\
\hline
\end{tabular}
\caption{Ordered List of Interaction Candidates SRU (left) and LRU
  (right) based on plmDCA scores; the fourth column indicates whether
  the protein pair is indeed interacting}
\label{tab:ppiRibo}
\end{table}

\begin{table}
\begin{tabular}{|c|c|c|c|}
\hline
P1 & P2 & Score & Interacting\\
\hline
RS10 &RS9  &1.123465 &1\\
RS10 &RS14 &1.102428 &1\\
RS12 &RS21 &1.079407 &0\\
RS13 &RS18 &1.029537 &0\\
RS14 &RS17 &1.001716 &0\\
RS12 &RS15 &0.997813 &0\\
RS18 &RS6  &0.963688 &1\\
RS11 &RS13 &0.943144 &0\\
RS19 &RS21 &0.942921 &0\\
RS15 &RS18 &0.938286 &0\\
RS14 &RS15 &0.933949 &0\\
RS13 &RS15 &0.933337 &0\\
RS13 &RS19 &0.918528 &1\\
RS18 &RS21 &0.918101 &1\\
RS10 &RS13 &0.917482 &0\\
\hline
\end{tabular}
\begin{tabular}{|c|c|c|c|}
\hline
P1 & P2 & Score & Interacting\\
\hline
RL20 &RL21 &1.665182 &1\\
RL14 &RL19 &1.430611 &1\\
RL15 &RL21 &1.333611 &1\\
RL15 &RL35 &1.134808 &1\\
RL23 &RL29 &1.086992 &1\\
RL20 &RL32 &1.037364 &1\\
RL22 &RL32 &1.029724 &1\\
RL30 &RL34 &1.008776 &0\\
RL17 &RL32 &1.002790 &1\\
RL34 &RL36 &0.983223 &0\\
RL21 &RL2  &0.977507 &0\\
RL21 &RL34 &0.958441 &0\\
RL18 &RL34 &0.942494 &0\\
RL36 &RL6  &0.925895 &1\\
RL33 &RL36 &0.898444 &0\\
\hline
\end{tabular}
\caption{Ordered List of Interaction Candidates SRU (left) and LRU
  (right) based on Gaussian scores; the fourth column indicates
  whether the protein pair is indeed interacting}
\label{tab:ppiRiboGauss}
\end{table}

\begin{table}
\begin{center}
\begin{tabular}{|c|c|c|}
\hline
TrpA & TrpB & 0.375 \\ 
TrpE & TrpG & 0.295 \\ 
TrpA & TrpC & 0.167 \\ 
TrpA & TrpF & 0.162 \\ 
TrpC & TrpF & 0.146 \\ 
TrpA & TrpD & 0.144 \\ 
TrpC & TrpD & 0.141 \\ 
TrpB & TrpF & 0.136 \\ 
TrpC & TrpE & 0.135 \\ 
TrpD & TrpF & 0.135 \\ 
TrpB & TrpC & 0.132 \\ 
TrpA & TrpE & 0.126 \\ 
TrpC & TrpG & 0.121 \\ 
TrpB & TrpD & 0.120 \\ 
TrpE & TrpF & 0.115 \\ 
TrpD & TrpE & 0.107 \\ 
TrpF & TrpG & 0.107 \\ 
TrpA & TrpG & 0.104 \\ 
TrpD & TrpG & 0.100 \\ 
TrpB & TrpE & 0.096 \\ 
TrpB & TrpG & 0.071 \\ 
\hline
\end{tabular}
\end{center}
\caption{Ordered List of Interaction Scores for the Trp Operon
  based on plmDCA scores}
\label{tab:ppiTrp}
\end{table}

\begin{table}
\begin{minipage}[b]{0.40\textwidth}
\centering
\begin{tabular}{|c|c|c|}
\hline
\multicolumn{3}{|c|}{SRU Intra-Protein}\\
\hline
	&SEP=0& SEP=5\\
\hline
RS2	&2337&	1610\\
RS3	&2217&	1494\\
RS4	&1728&	1152\\
RS5	&1684&	1175\\
RS6	&1002&	666\\
RS7	&1494&	982\\
RS8	&1334&	903\\
RS9	&1240&	799\\
RS10	&878 &	557\\
RS11	&1220&	822\\
RS12	&1136&	731\\
RS13	&1024&	623\\
RS14	&790 &	440\\
RS15	&823 &	489\\
RS16	&685 &	436\\
RS17	&733 &	487\\
RS18	&482 &	293\\
RS19	&748 &	482\\
RS20	&792 &	464\\
RS21	&297 &	110\\
\hline
SUM:	&22644&	14715\\
\hline
\end{tabular}
\end{minipage}\qquad
\begin{minipage}[b]{.40\textwidth}
\begin{tabular}{|c|c|c|}
\hline
\multicolumn{3}{|c|}{SRU Inter-Protein}\\
\hline
RS2	&RS5  &	4     \\
RS2	&RS8  &	3     \\
RS3	&RS5  &	17    \\
RS3	&RS10 &	105   \\
RS3	&RS14 &	209   \\
RS4	&RS5  &	84    \\
RS5	&RS8  &	120   \\
RS6	&RS18 &	150   \\
RS7	&RS9  &	19    \\
RS7	&RS11 &	46    \\
RS8	&RS12 &	12    \\
RS8	&RS17 &	28    \\
RS9	&RS10 &	28    \\
RS9	&RS14 &	7     \\
RS10	&RS14 &	150   \\
RS11	&RS18 &	20    \\
RS11	&RS21 &	199   \\
RS12	&RS17 &	34    \\
RS13	&RS19 &	80    \\
RS14	&RS19 &	50    \\
RS18	&RS21 &	36    \\
\hline
SUM:	&&	1401  \\
\hline
FRACTION &SEP=0&0.058 \\
\hline
FRACTION &SEP=5&0.087 \\
\hline
\end{tabular}
\end{minipage}
\caption{Left table: number of intra-protein contacts below 8\AA{} of all
  residues (SEP=0 column), and considering only those with a distance
  on the sequence of at least 5 residues (SEP = 5 column) for the SRU.
  Right table: number of inter-protein contacts below 8\AA for the
  SRU. Fractions are defined as $\frac{\# Intra}{\# Intra
    + \# Inter}$ where $\#Inter$ is computed assuming SEP=0,5 respectively.}
\label{tab:lscontSRU}
\end{table}

\begin{table}
\begin{minipage}[b]{0.40\textwidth}
\centering
\begin{tabular}{|c|c|c|}
\hline
\multicolumn{3}{|c|}{LRU Intra-Protein}\\
\hline
	&SEP=0 &	SEP=5 \\
\hline
RL32	&324   &	157   \\
RL33	&399   &	256   \\
RL34	&303   &	145   \\
RL35	&495   &	268   \\
RL36	&332   &	208   \\
RL2	&2687  &	1801  \\
RL3	&1931  &	1263  \\
RL4	&1869  &	1199  \\
RL5	&1887  &	1257  \\
RL6	&1811  &	1217  \\
RL9	&1360  &	855   \\
RL11	&1390  &	903   \\
RL13	&1464  &	959   \\
RL14	&1266  &	869   \\
RL15	&920   &	481   \\
RL16	&1343  &	915   \\
RL17	&1194  &	767   \\
RL18	&1150  &	777   \\
RL19	&1043  &	669   \\
RL20	&1045  &	600   \\
RL21	&915   &	600   \\
RL22	&1085  &	720   \\
RL23	&735   &	461   \\
RL24	&386   &	233   \\
RL25	&893   &	597   \\
RL27	&692   &	442   \\
RL29	&538   &	303   \\
RL30	&511   &	321   \\
RL28	&587   &	351   \\
\hline
SUM:	&30555 &	19594 \\
\hline
\end{tabular}
\end{minipage}\qquad
\begin{minipage}[b]{.40\textwidth}
\begin{tabular}{|c|c|c|}
\hline
\multicolumn{3}{|c|}{LRU Inter-Protein}\\
\hline
RL32  &	RL17 &	78   \\	      
RL32  &	RL20 &	17   \\
RL32  &	RL22 &	73   \\
RL33  &	RL35 &	21   \\
RL35  &	RL15 &	149  \\
RL35  &	RL27 &	1    \\
RL36  &	RL6  &	10   \\
RL36  &	RL16 &	1    \\
RL3   &	RL13 &	20   \\
RL3   &	RL14 &	34   \\
RL3   &	RL17 &	21   \\
RL3   &	RL19 &	123  \\
RL4   &	RL15 &	83   \\
RL4   &	RL20 &	6    \\
RL9   &	RL28 &	63   \\
RL13  &	RL20 &	118  \\
RL13  &	RL21 &	8    \\
RL14  &	RL19 &	191  \\
RL15  &	RL20 &	2    \\
RL15  &	RL21 &	24   \\
RL16  &	RL25 &	53   \\
RL16  &	RL27 &	9    \\
RL17  &	RL22 &	12   \\
RL18  &	RL27 &	12   \\
RL20  &	RL21 &	229  \\
RL23  &	RL29 &	81   \\
\hline
SUM:  &&     	1439 \\
\hline
FRACTION &SEP=0&0.045\\
\hline
FRACTION &SEP=5&0.068\\
\hline
\end{tabular}
\end{minipage}
\caption{Left table: number of intra-protein contacts below 8\AA{} of all
  residues (SEP=0 column), and considering only those with a distance
  on the sequence of at least 5 residues (SEP = 5 column) for the LRU.
  Right table: number of inter-protein contacts below 8\AA for the
  LRU. Fractions are defined as $\frac{\# Intra}{\# Intra
    + \# Inter}$ where $\#Inter$ is computed assuming SEP=0,5 respectively.}
\label{tab:lscontLRU}
\end{table}

\subsection{Structural view of the Ribosomal Complex}

\begin{figure}[h]
\includegraphics[width=\textwidth]{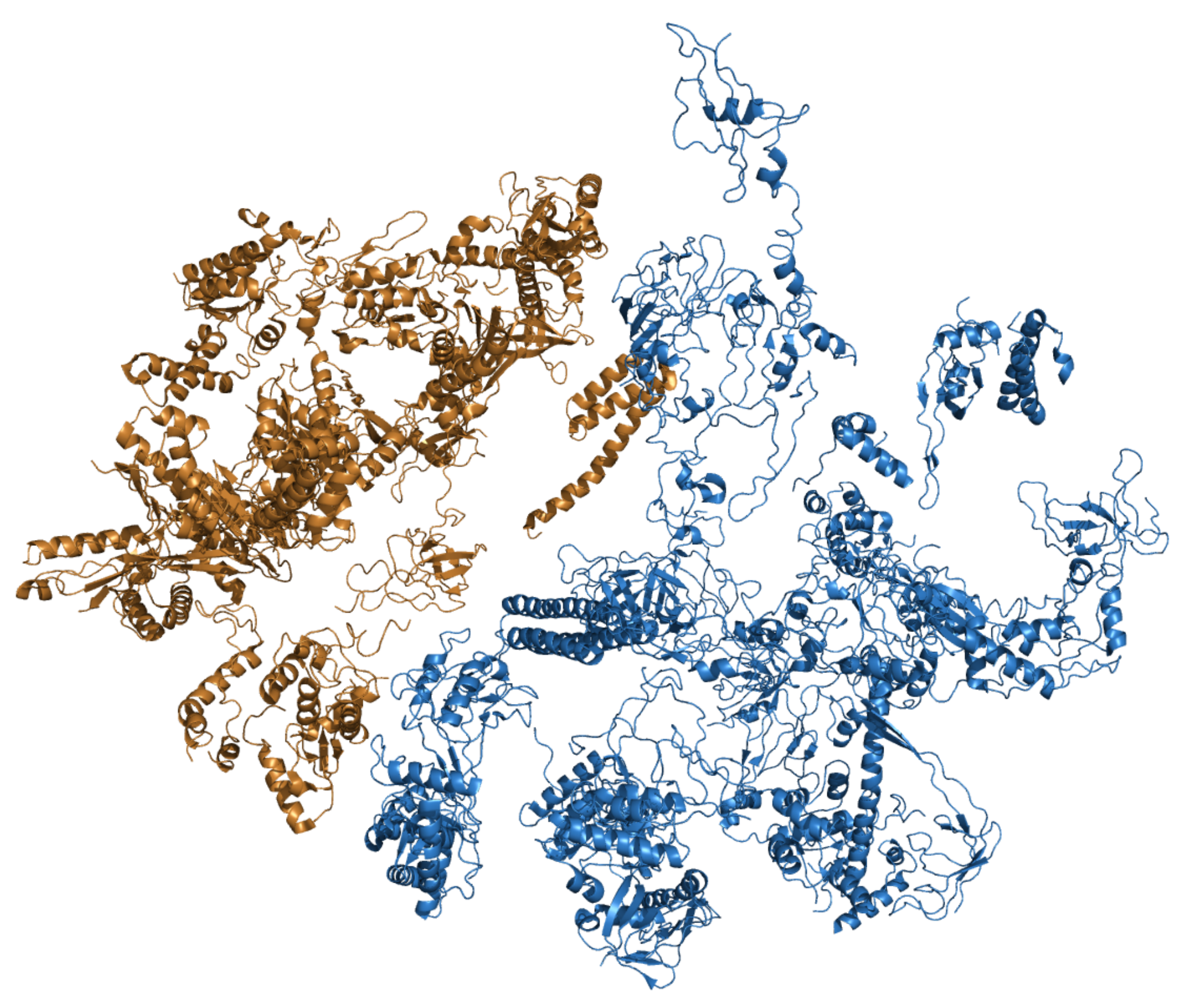}
\caption{Cartoon view of the small (brass color) and large (blue
  color) bacterial ribosomal complexes 2Z4K, 2Z4L. For the ease of
  visualization we have carved out the ribosomal RNAs strands.}
\label{fig:ribo3d}
\end{figure}

\begin{figure}[h]
\centering
\includegraphics[width=0.75\textwidth]{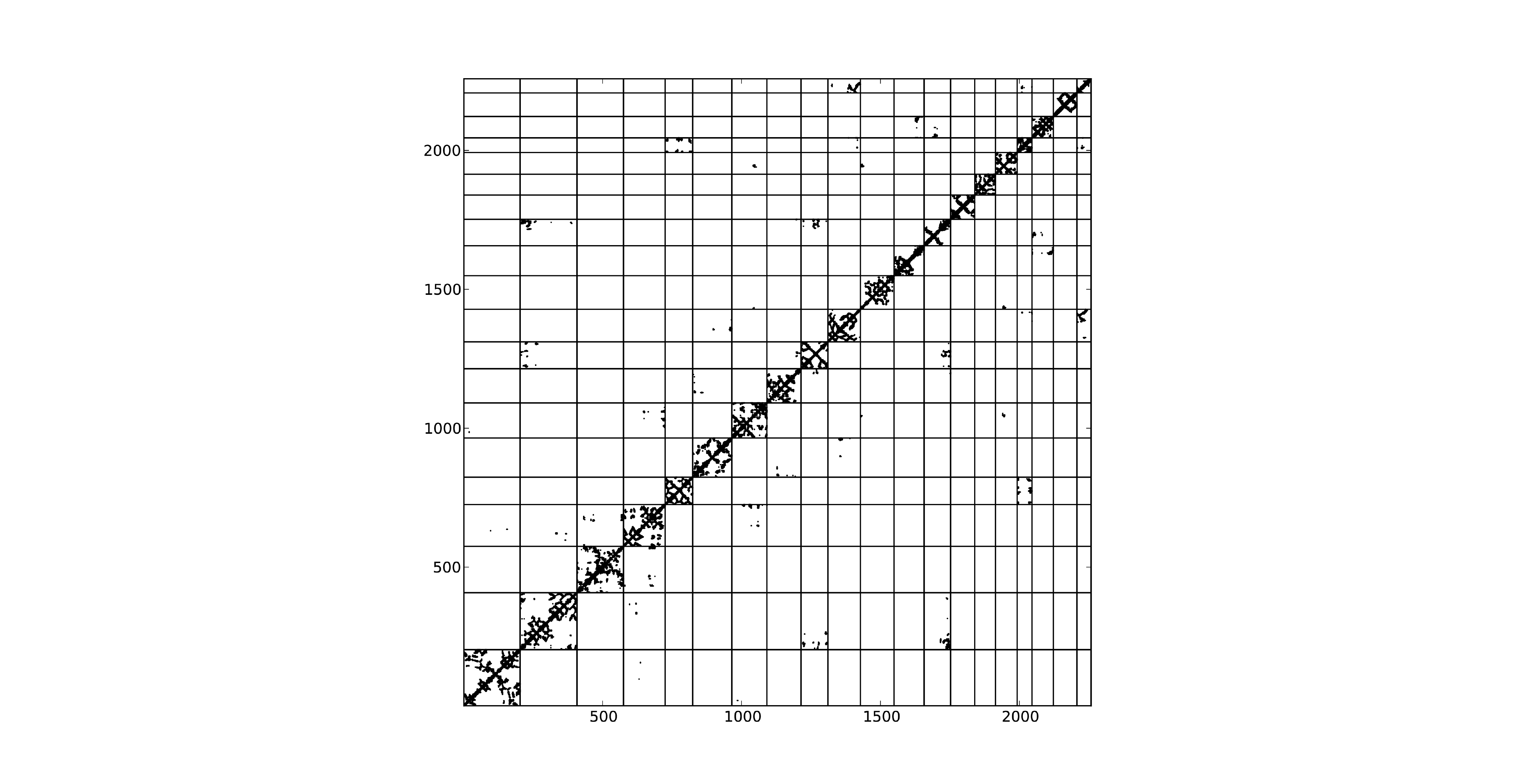}
\includegraphics[width=0.75\textwidth]{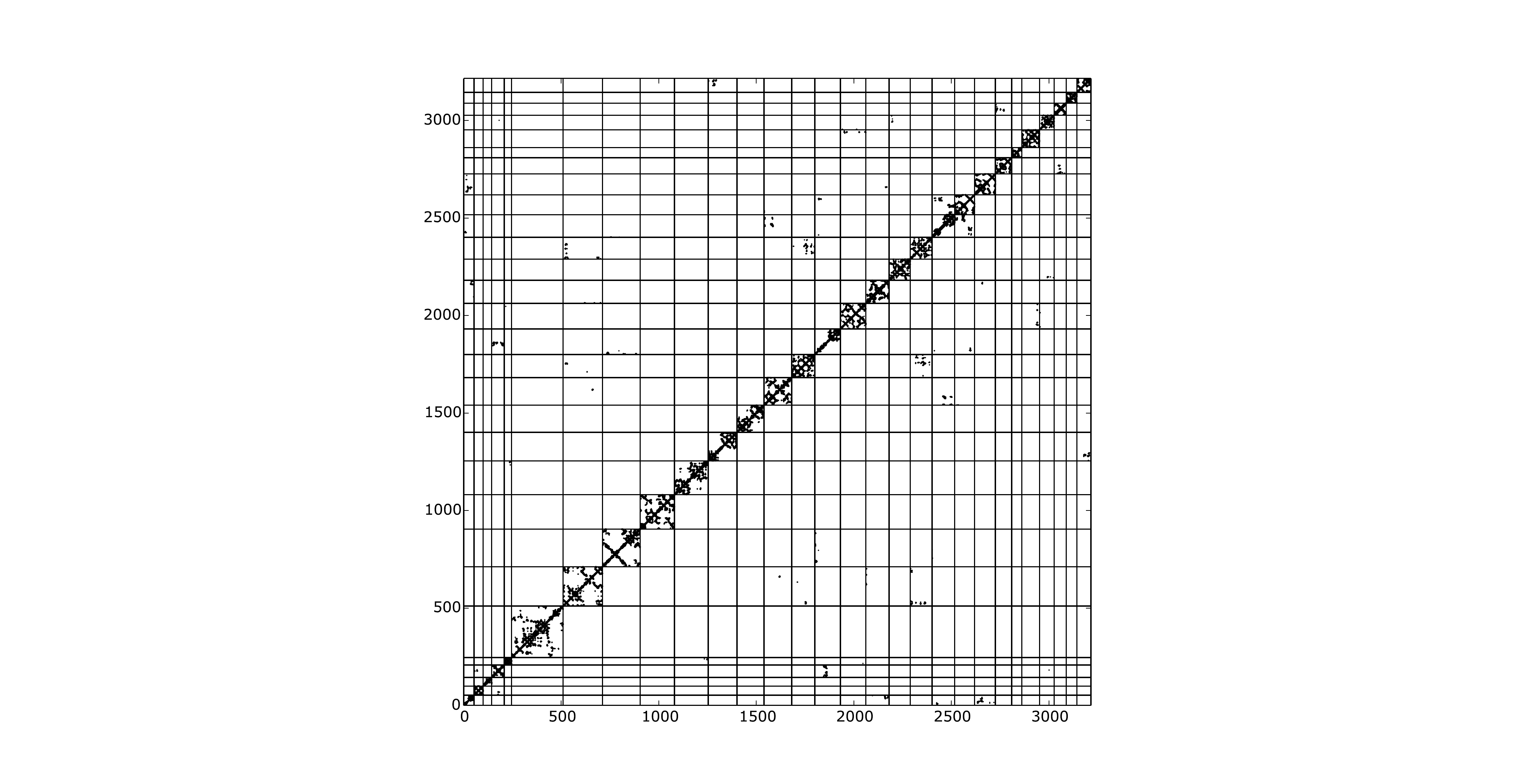}
\caption{Upper panel: contact map of the SRU (threshold distance
  8\AA). Lower panel: contact map of the LRU.}
\label{fig:riboCM}
\end{figure}

In Fig.~\ref{fig:ribo3d} we display a cartoon view of the ribosomal protein
network. The contact map for the the small and large ribosomal units
are displayed in Fig.~\ref{fig:riboCM}

\section{Artificial Data}

An artificial large network consisting of 5 proteins was created in two steps:

1) First, a contact map was defined. This contact map contains the information
which residues are in contact. This includes internal residue contacts (where
both residues belong to one of the 5 proteins) and inter-protein residue
contacts (where one residue belongs to one protein and the other to a different
protein). The contact map is therefore a binary, symmetric matrix of size
$N_{all} \times N_{all}$ with $N_{all} = N_1+N_2+N_3+N_4+N_5$ where $N_i$ is
the number of residues in the $i^{th}$ protein. We decided to use the Kunitz
domain (PF00014) as a model for the proteins and set all $N_i = 53$. The
$53\times53$ submatrices that define the contacts within each protein were
defined by extracting the contacts of the PDB structure 5pti of the Kunitz
domain. This implies that the internal structure of every protein is the same.

We defined as contacting proteins the protein pairs $1-2,2-3,3-4,4-5$ and
$1-5$.  For the $53\times53$ submatrices that define the contacts between contacting
protein pairs we used random binary matrices with 10\% of the number of
internal contacts. This was done individually for each contacting protein pair
such that no two contact matrices between two proteins were the same. For
non-contacting protein pairs all entries of the contact matrices were set to 0.

\begin{figure}
  \centering
  \includegraphics[width=0.95\columnwidth]{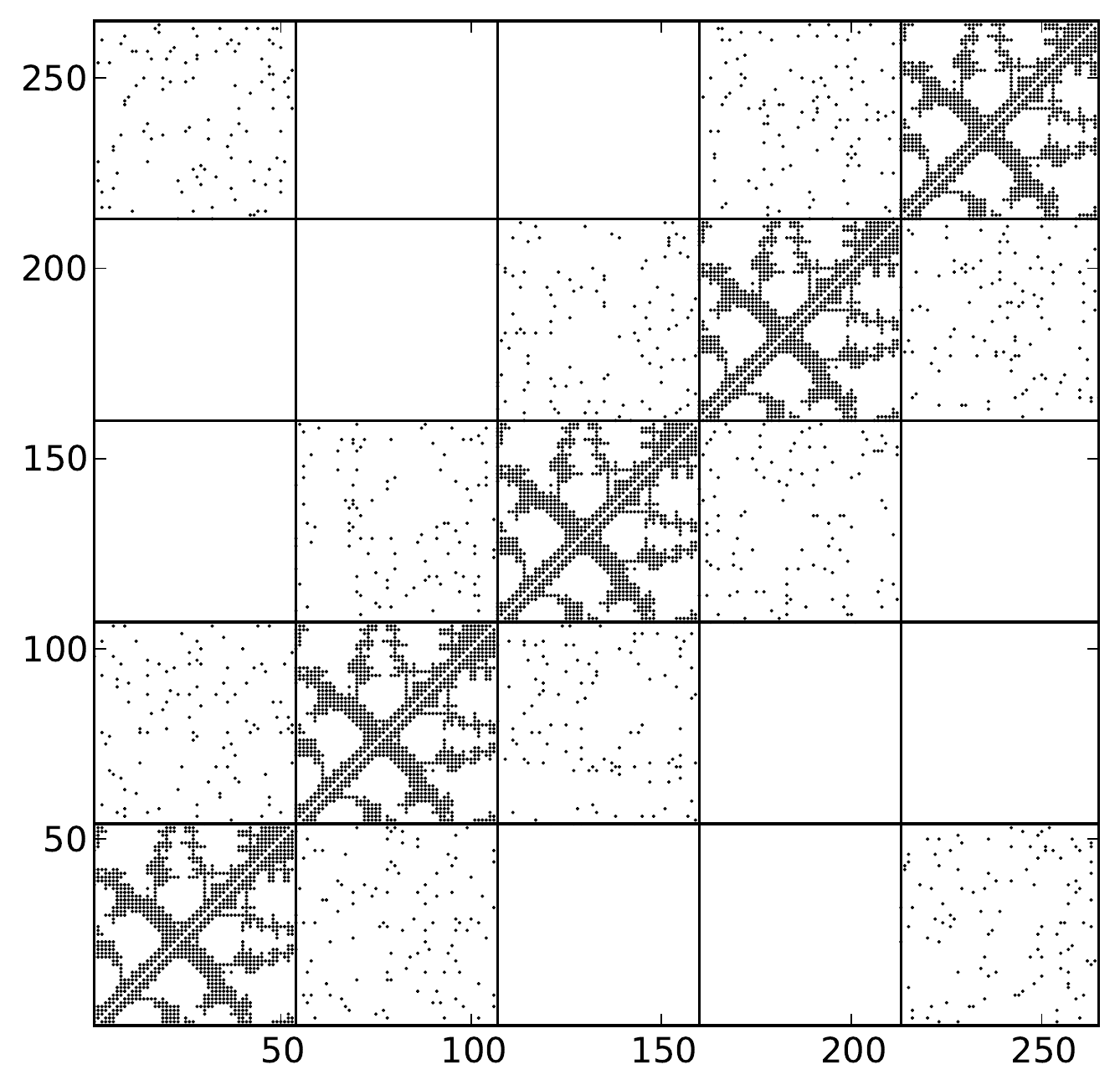}
  \caption{Contact map of the artificial protein complex}
  \label{fig:artificialCM}
\end{figure}

The resulting contact map can be seen in Fig.~\ref{fig:artificialCM}.

2) Couplings for every contact in the contact map were defined.  As a basis for
this, couplings and fields inferred from the PF00014 PFAM alignment
(Kunitz Domain) were used. This inference was done using a masking with the PDB
structure, such that only couplings corresponding to PDB-contacts were allowed
to differ from zero. Given that the same PDB-contacts were used to define the
contacts within one protein in the artificial complex, we could use the
couplings thus inferred without change for the couplings within the artificial
proteins. 

Then we defined the couplings for residue contacts between two proteins. For
every such a resiue contact we chose randomly a coupling of an internal contact
as inferred from the Kunitz domain alignment and assigned it to the residue
contact.

Notice that the 'coupling' between two sites $i$ and $j$ is actually a $21\times21$
matrix $J_{ij}(a,b)$ where $a$ and $b$ can be any of the 21 amino acids. Given
that the internal structure of these matrices might be important we decided to
treat the matrices $J_{ij}$ as single entities and not change their internal
structure.

The fields for every residue, a vector of length 21 for every of the $5 \cdot 53$ residues, were randomly chosen from the inferred fields.

From these couplings and fields, sequences were generated by MC (see
section below) and inferred by plmDCA. Interestingly, a crude
comparison between the histogram of the scores in the artificial model
seem to be very close to that obtained for instance for the LRU case
as shown in Fig.~\ref{fig:artSRUhist}.

In Table \ref{tab:20largest} we compare the ranks of the strongest
inter-protein residue interaction scores in the generating model and the
inferred model. The first column represents the rank of the inter-protein
residue interaction in the generating model, the second column the rank of the
same residue interaction in the inferred model. The model was inferred with the
combined strategy and with 4000 sequences. The numbering is treating the
complex as one large protein.

\begin{table}
\centering
\begin{tabular}{| c | c|}
\hline
Original Rank & Inferred Rank\\
\hline
    1 & 101\\ 
    2 & 13806\\ 
    3 & 10658\\ 
    4 & 64\\ 
    5 & 4\\ 
    6 & 9575\\ 
    7 & 1\\ 
    8 & 15890\\ 
    9 & 6712\\ 
   10 & 1035\\ 
\hline
    7 & 1\\ 
   32 & 2\\ 
   41 & 3\\ 
    5 & 4\\ 
   11 & 5\\ 
11473 & 6\\ 
22464 & 7\\ 
   53 & 8\\ 
 1877 & 9\\ 
   26 & 10\\
\hline
\end{tabular}
\caption{Original vs. inferred rank for the 10 largest original inter-protein
  residue interaction scores and the 10 largest inferred inter-protein residue interaction scores}
\label{tab:20largest}
\end{table}

\begin{figure}
	\begin{subfigure}[b]{0.5\textwidth}
		\centering
		\includegraphics[scale=0.35]{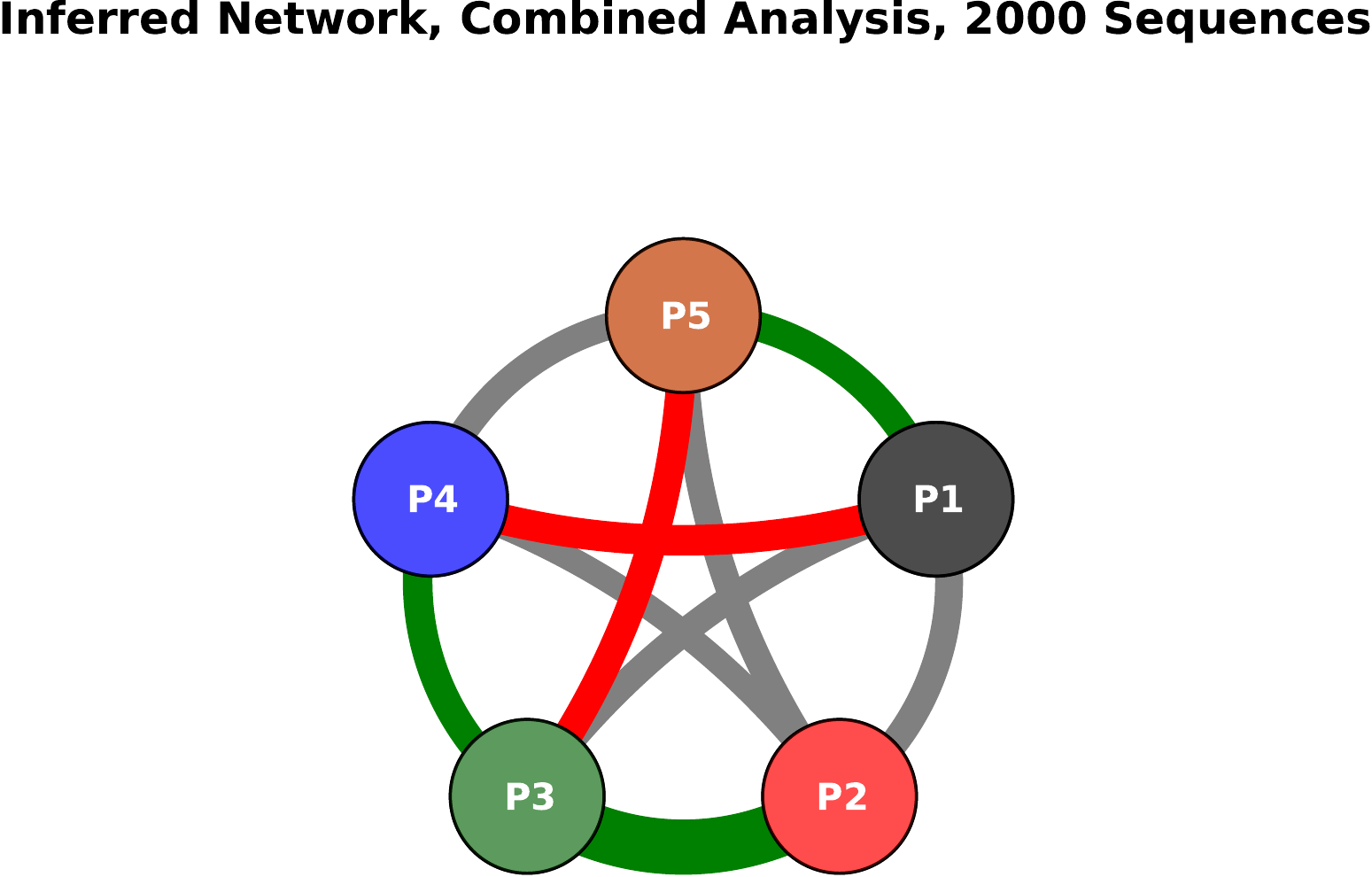}
	\end{subfigure}
	\begin{subfigure}[b]{0.5\textwidth}
		\centering
		\includegraphics[scale=0.35]{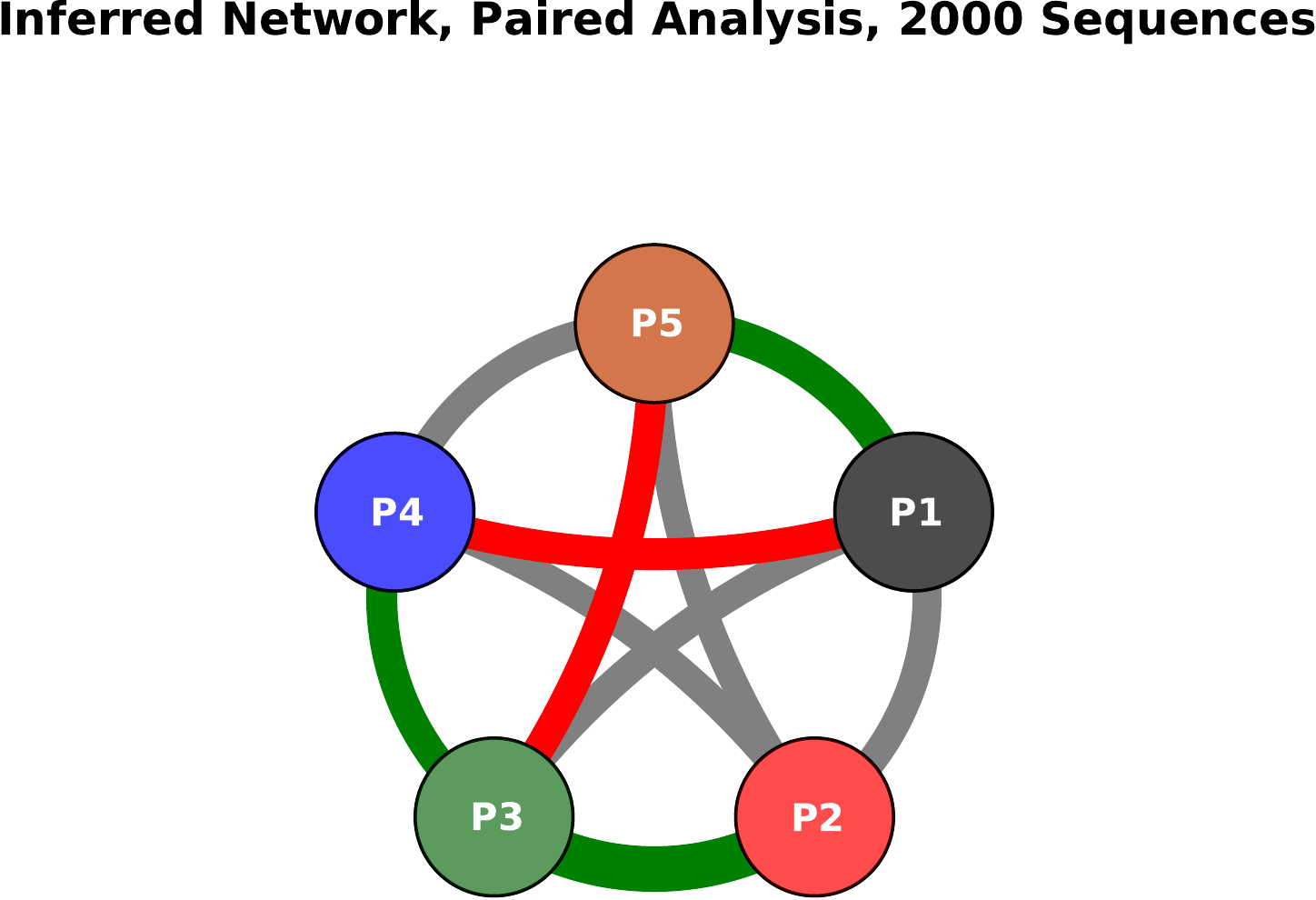}
	\end{subfigure}
	\par\bigskip

	\begin{subfigure}[b]{0.5\textwidth}
		\centering
		\includegraphics[scale=0.35]{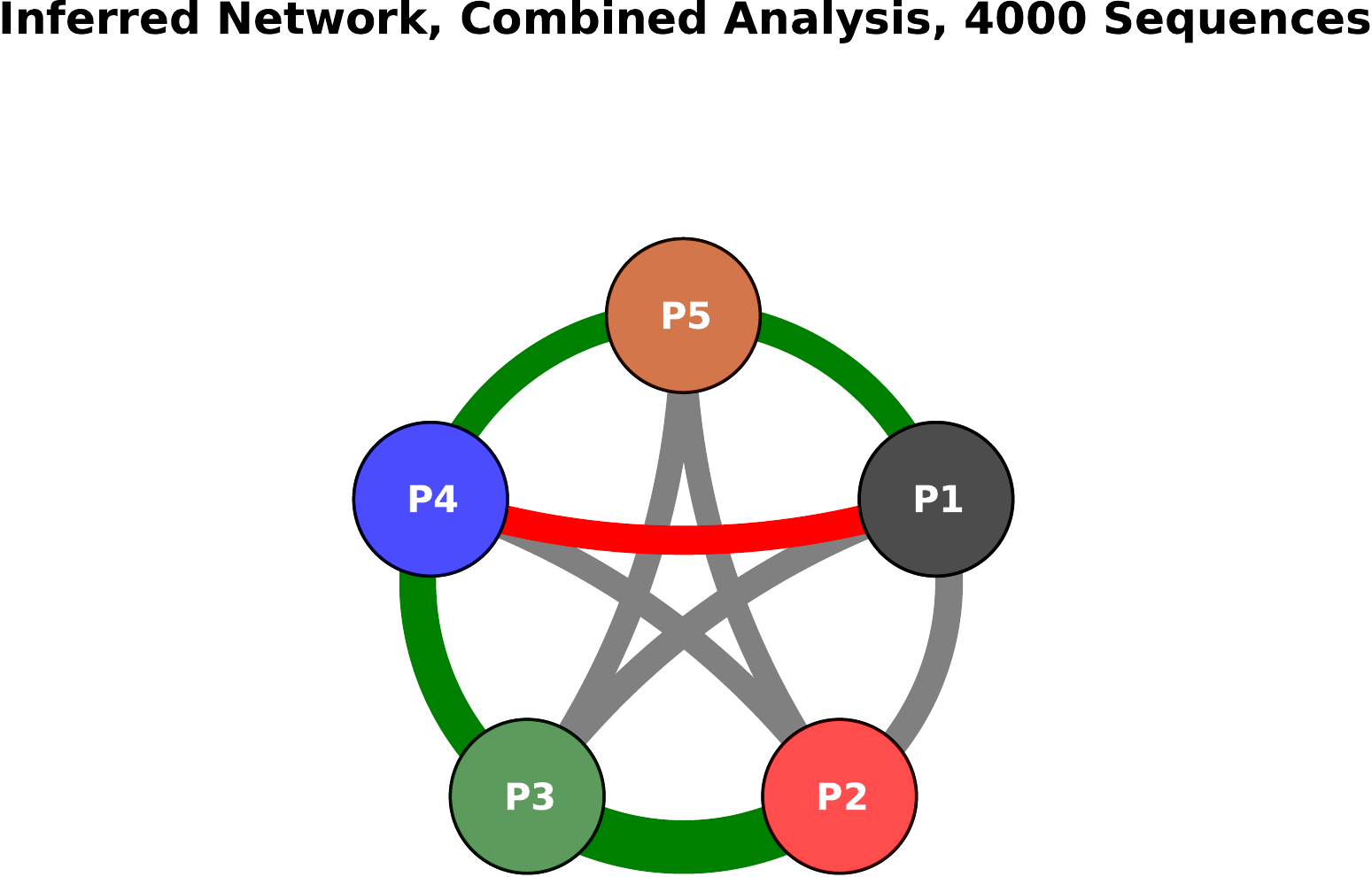}
	\end{subfigure}
	\begin{subfigure}[b]{0.5\textwidth}
		\centering
		\includegraphics[scale=0.35]{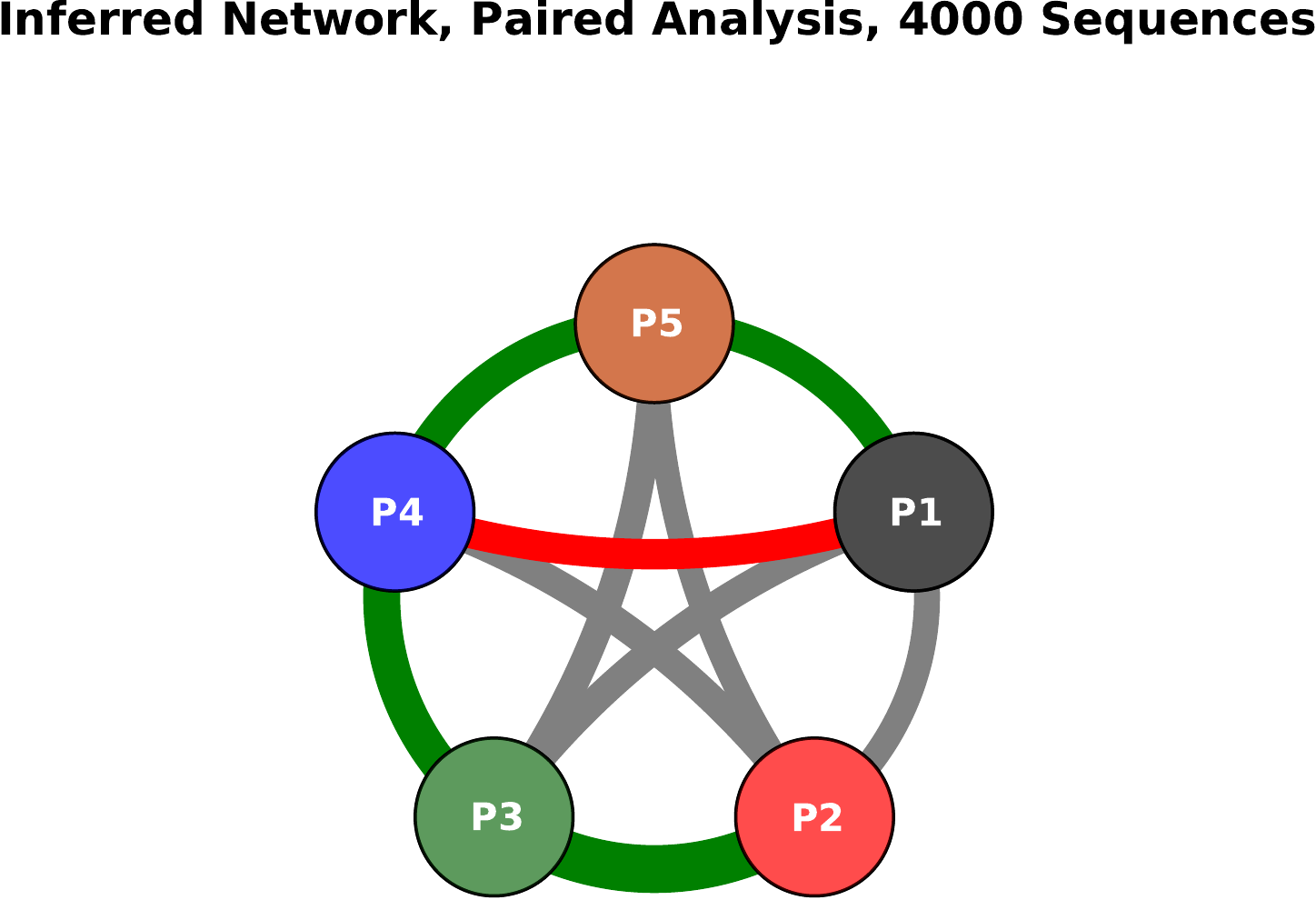}
	\end{subfigure}
	\par\bigskip
	\begin{subfigure}[b]{0.5\textwidth}
		\centering
		\includegraphics[scale=0.35]{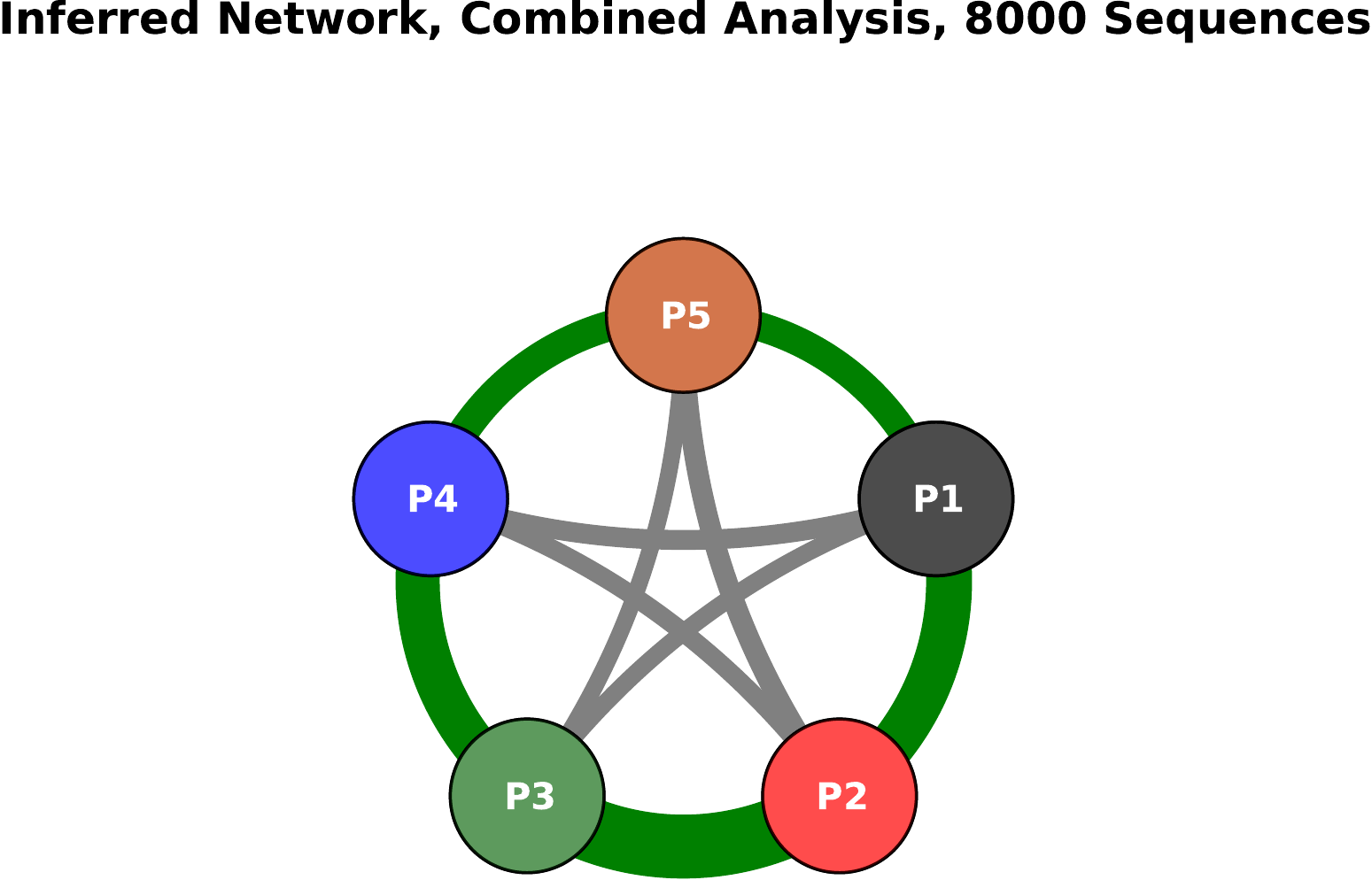}
	\end{subfigure}
	\begin{subfigure}[b]{0.5\textwidth}
		\centering
		\includegraphics[scale=0.35]{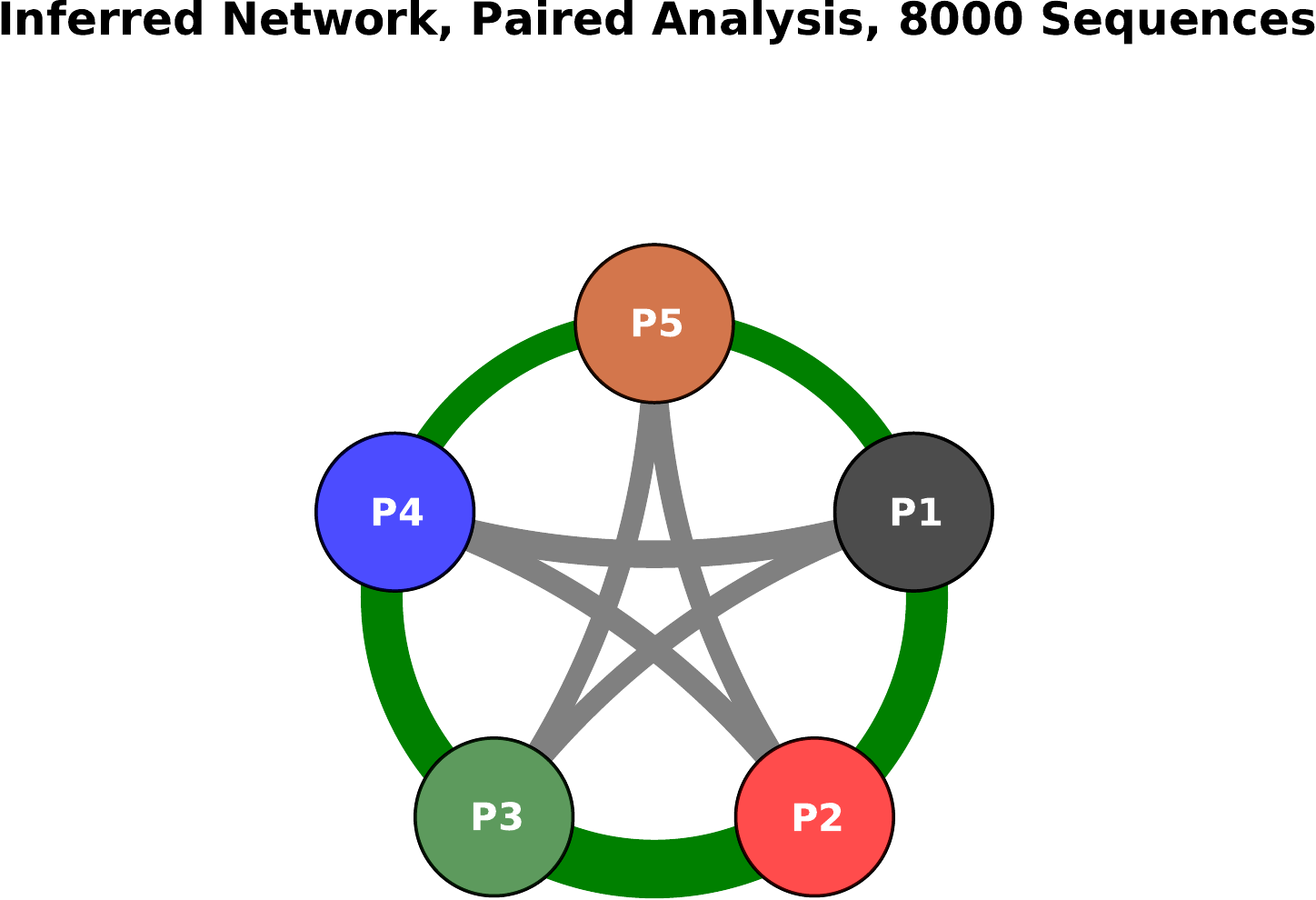}
	\end{subfigure}
	\par\bigskip
	\begin{subfigure}[b]{0.5\textwidth}
		\centering
		\includegraphics[scale=0.35]{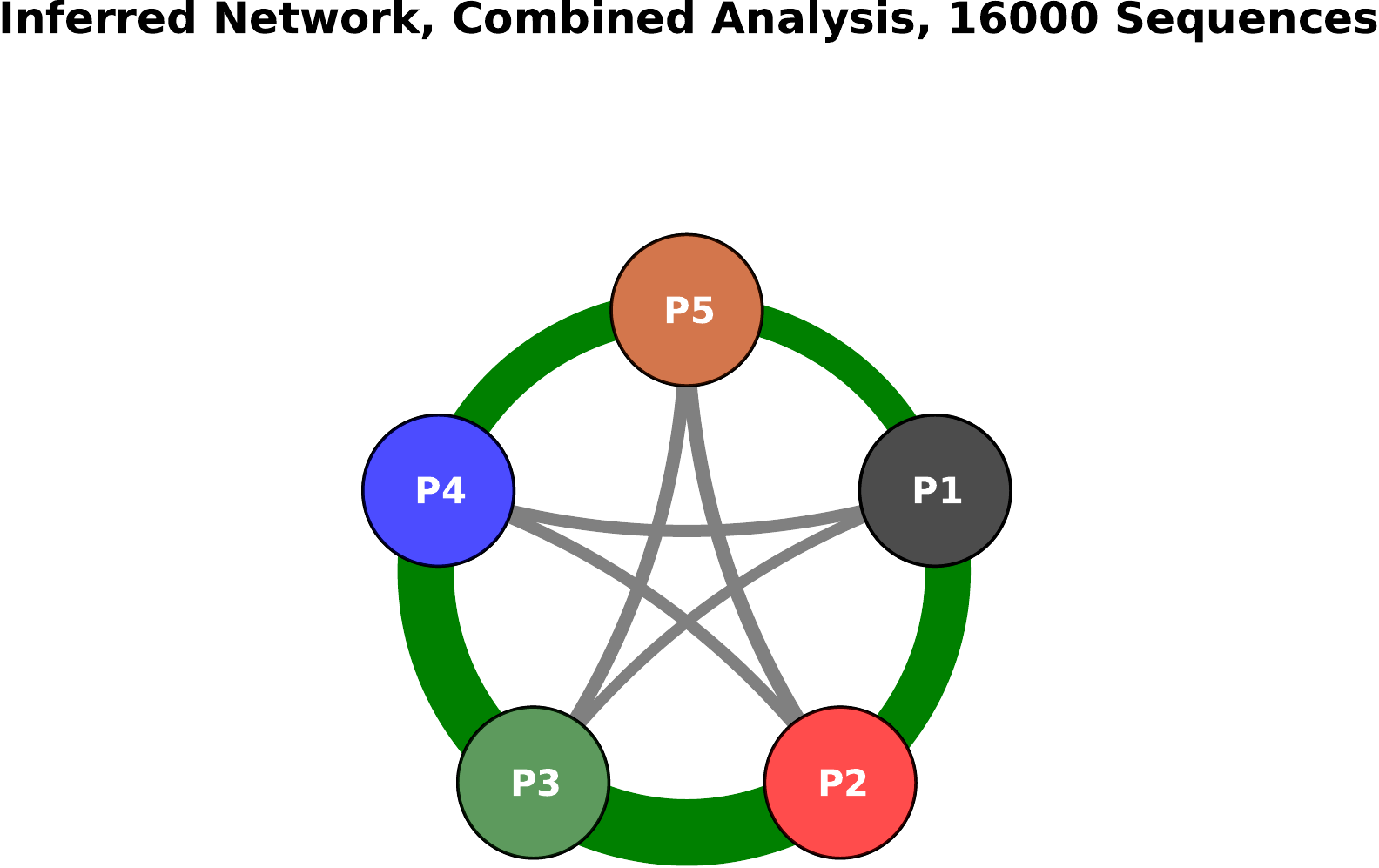}
	\end{subfigure}
	\begin{subfigure}[b]{0.5\textwidth}
		\centering
		\includegraphics[scale=0.35]{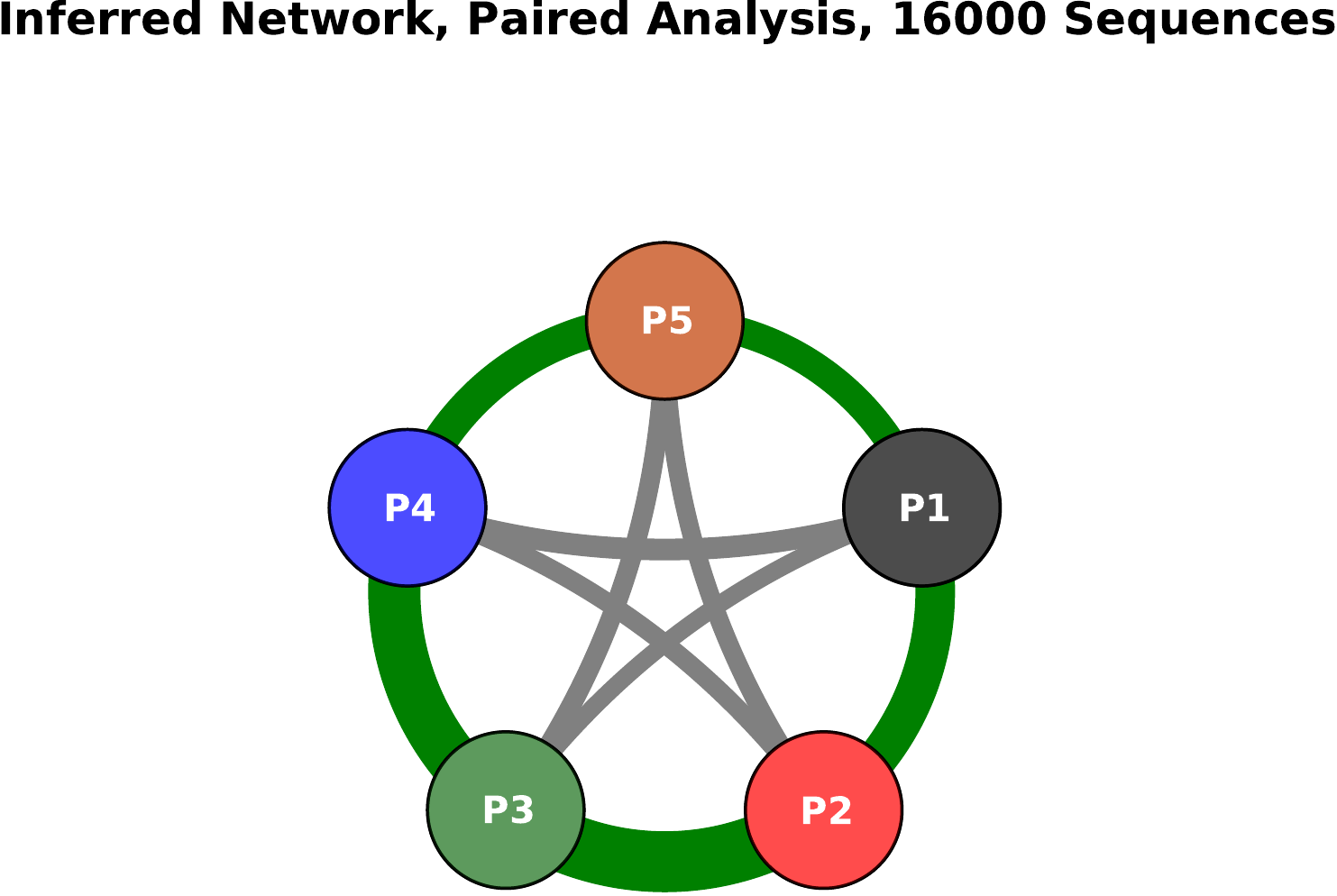}
	\end{subfigure}
	\par\bigskip
	\begin{subfigure}[b]{0.5\textwidth}
		\centering
		\includegraphics[scale=0.35]{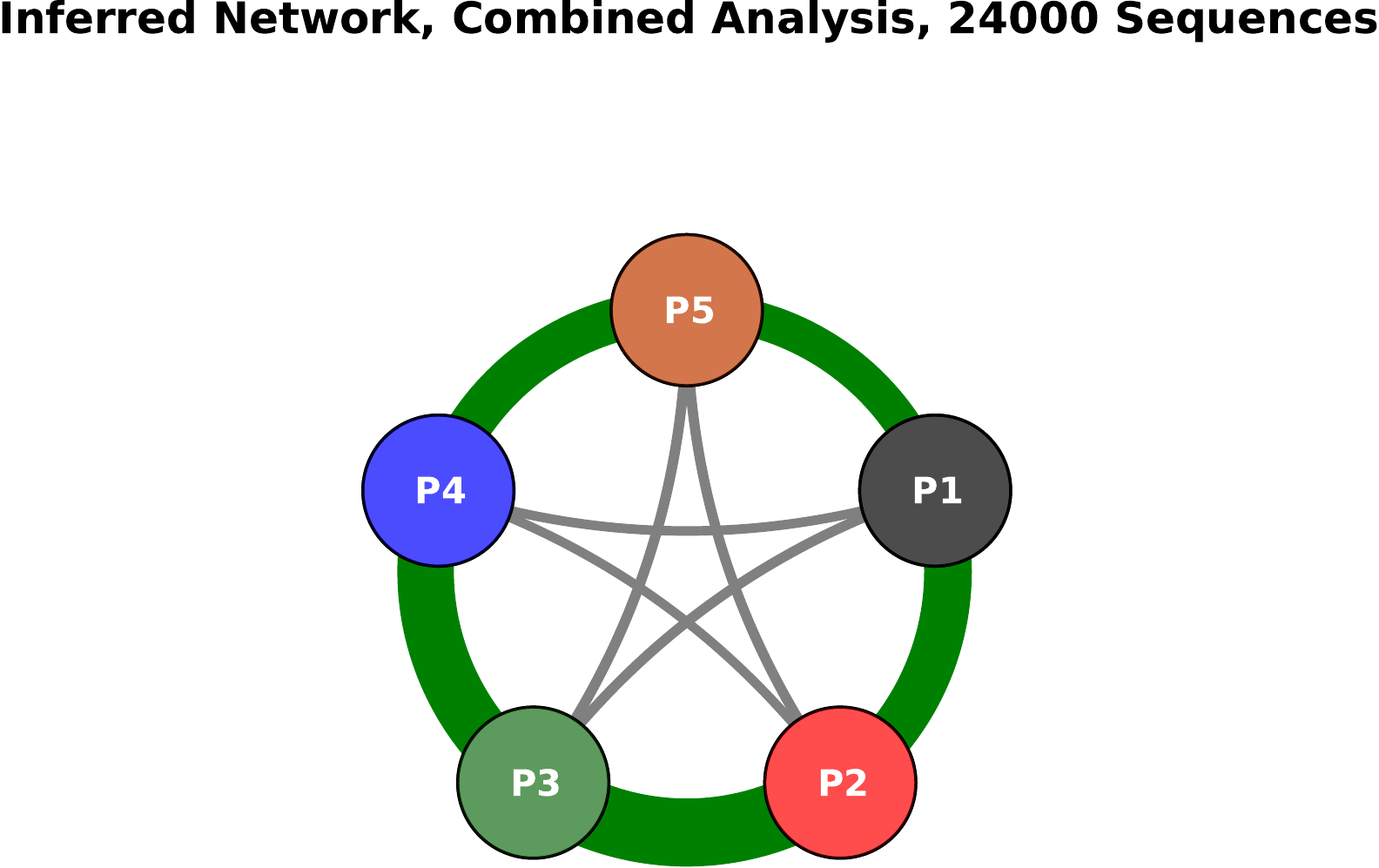}
	\end{subfigure}
	\begin{subfigure}[b]{0.5\textwidth}
		\centering
		\includegraphics[scale=0.35]{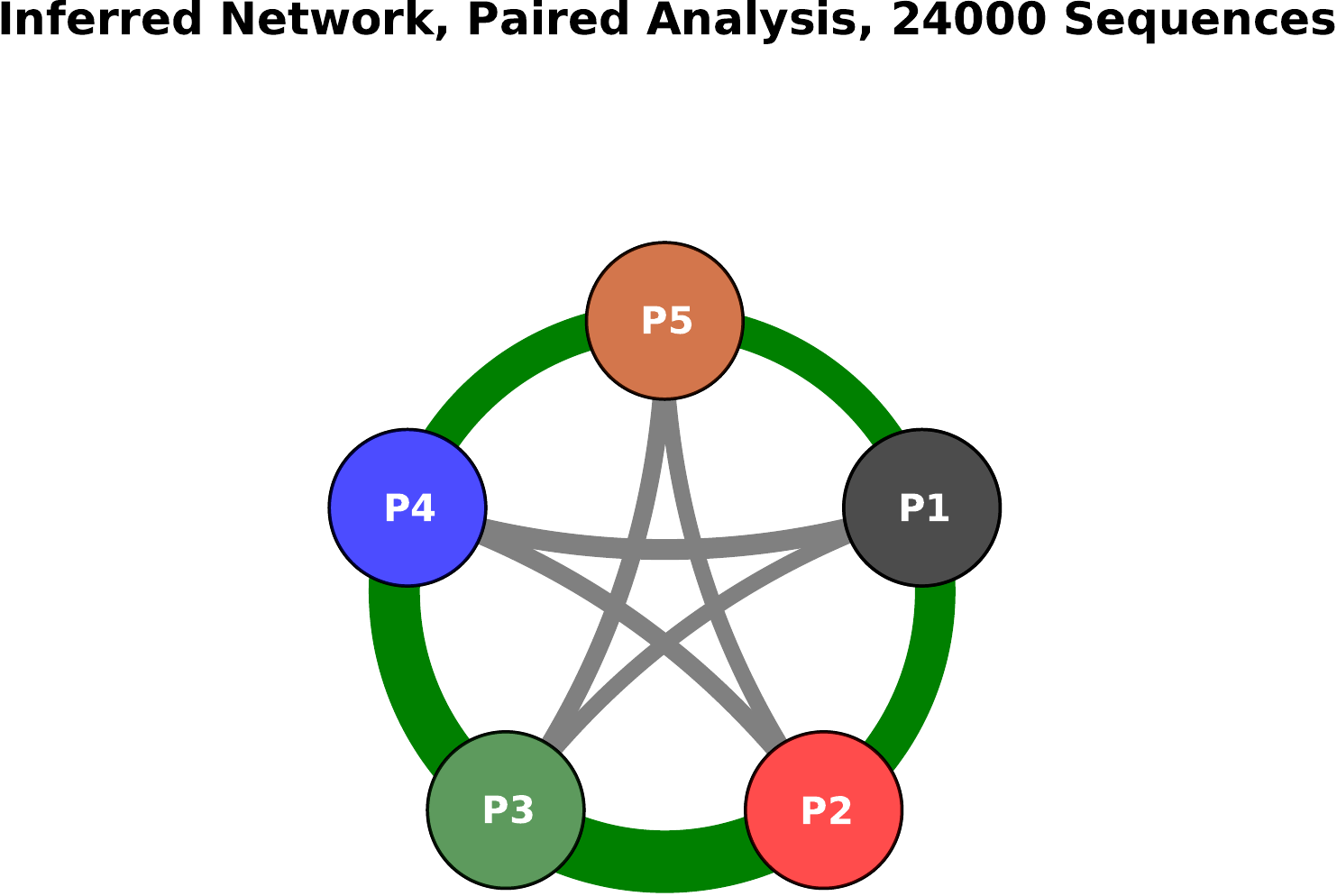}
	\end{subfigure}
\caption{Inferred protein network for different sample sizes; the
  line-thickness is proportional to the inferred interaction scores between
  the proteins (mean of the 4 highest residue interaction scores).
  The thickness has been normalized in the sense
  that the scores have been divided by the mean of the
  scores of the network. The color code is applied for the
  first 5 predictions and shows a green line if the prediction is a
  true positive and a red line if the prediction is a false
  positive. Predictions after the first 5 are grey.\\ \textbf{Combined
    Analysis}: The complete sequences in their whole length were used
  for the inference and calculation of the 
  scores\\ \textbf{Paired Anlysis}: Every protein family was independently
  cut out of the generated sequences and thus a MSA for only this
  protein created. These single MSAs were then paired for all
  protein pairs and used for inference and calculation of the
  scores.}
\end{figure}

\subsection{Monte Carlo Sequence Generation}
Given the parameters of the artificial model, a simple MCMC algorithm
was run to generate samples from the corresponding distribution. We
used one million MC steps to equilibrate the chain and took a sample
every one million steps.

\section{Large scale network inference}

In order to test the approach on a larger scale we created all possible protein pairs from all proteins in the ribosome and the trp operon. The matching procedure was identical to the procedure used in the individual systems. 

\begin{figure}
  \centering
  \includegraphics[width=1.0\columnwidth]{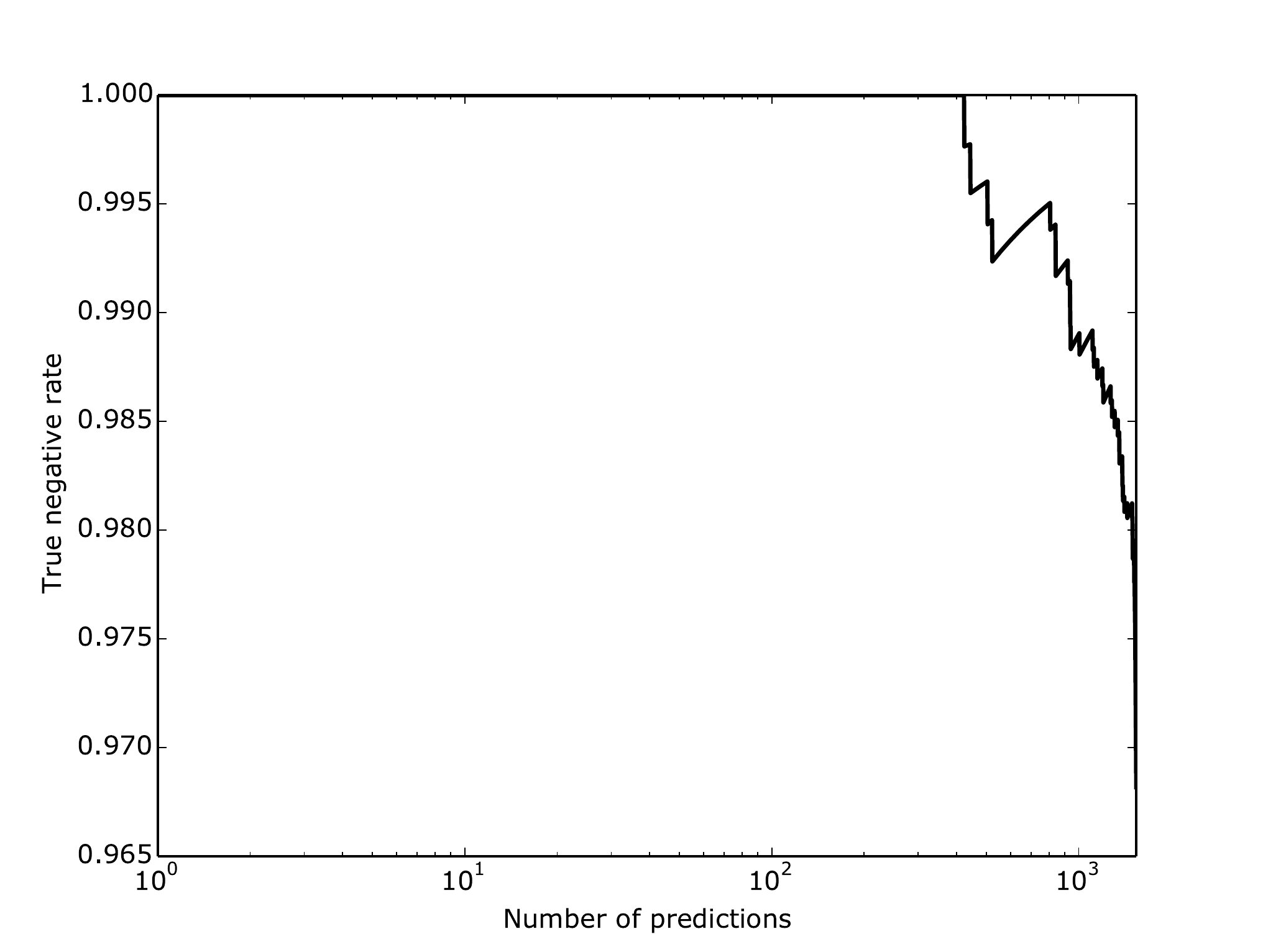}
  \caption{True negative rate; all possible protein pairs between RS,RL and Trp proteins are considered and the protein-protein interaction score is defined as the average of the 4 largest interaction scores on the residue level (as in the main paper). The true negative rate is the fraction of true negatives in the N pairs with the lowest interaction score, where N is the value indicated by the x-axis.}
  \label{fig:true_negative}
\end{figure}

\begin{figure}
  \centering
  \includegraphics[width=1.0\columnwidth]{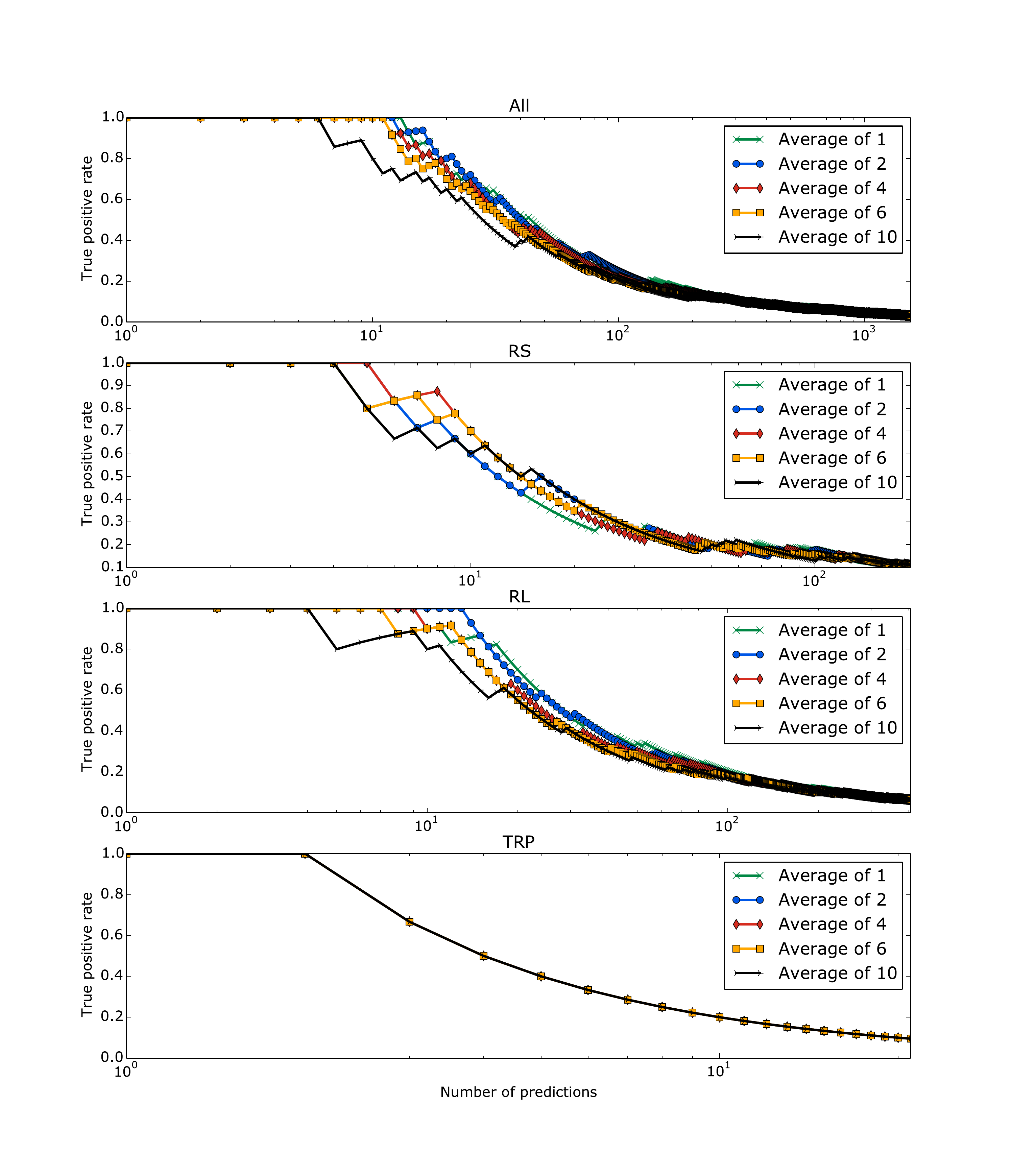}
  \caption{True positive rates at a given number of predictions; All: All possible protein pairs between RS, RL and Trp proteins are considered; RS: Protein pairs within the small ribosomal subunit; RL: Protein pairs within the large ribosomal subunit; Trp: Protein pairs of the Trp operon. Different lines indicate a different number of averaged inter-protein scores on the residue level to get a protein-protein interaction score} 
  \label{fig:rocglobal}
\end{figure}

\begin{figure}
  \centering
  \includegraphics[width=0.95\columnwidth]{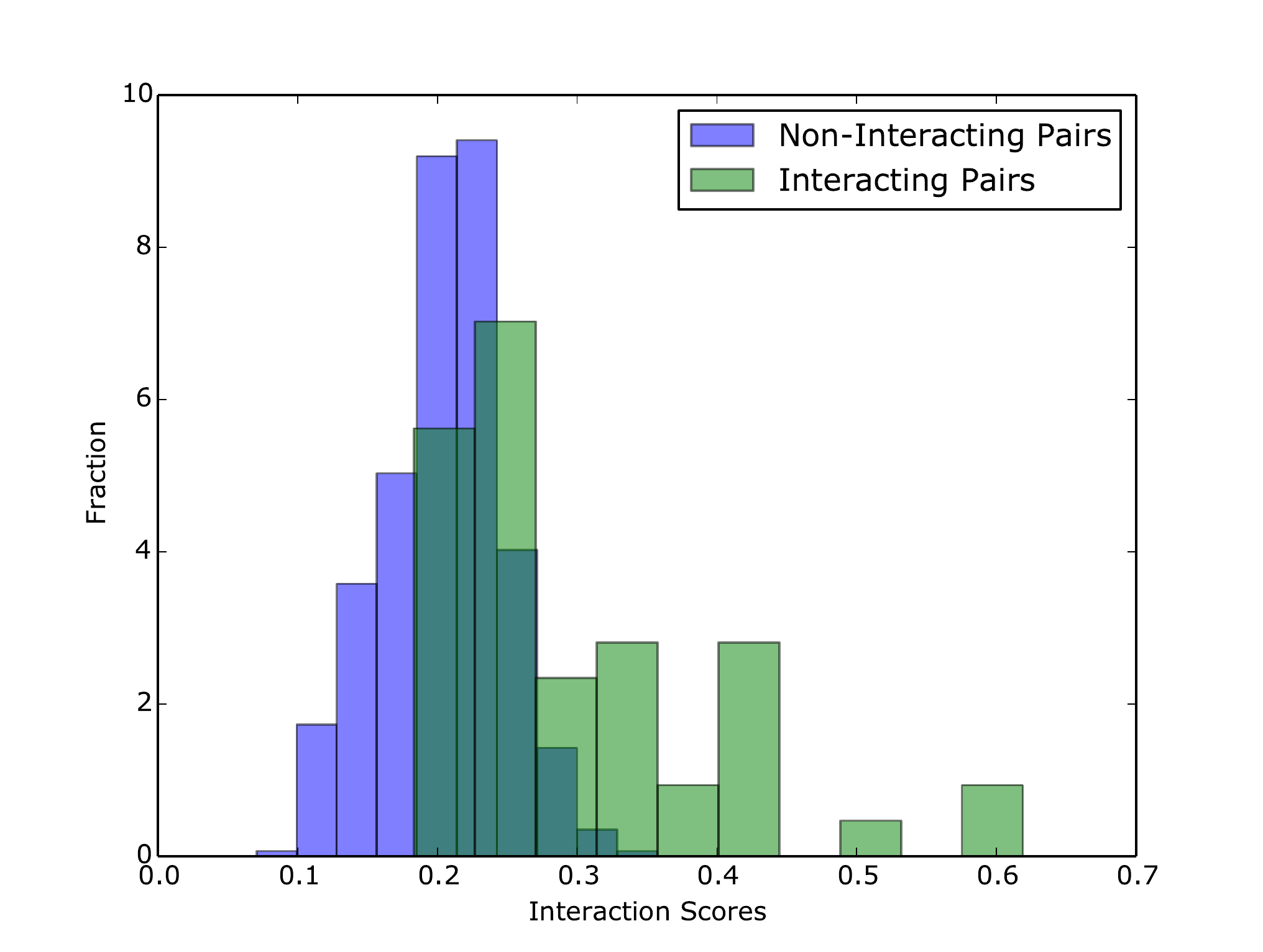}
  \caption{Histograms of interaction scores in the network comprising all possible protein pairs between RS, RL and Trp proteins. The protein-protein interaction scores were calculated averaging the 4 largest inter-protein residue interaction scores (as in the main paper). The histogram shows true positives and true negatives seperately. Both histograms are normalized.}
  \label{fig:2histo}
\end{figure}

\bibliography{main_paper}

\begin{thebibliography}{10}

\bibitem{balakrishnan2011Protein}
S.~Balakrishnan, H.~Kamisetty, J.~G. Carbonell, S.~I. Lee, and C.~J. Langmead.
\newblock {Learning generative models for protein fold families}.
\newblock {\em Proteins: Struct., Funct., Bioinf.}, 79:1061, 2011.

\bibitem{Baldassi2014PLOSONE}
Carlo Baldassi, Marco Zamparo, Christoph Feinauer, Andrea Procaccini, Riccardo
  Zecchina, Martin Weigt, and Andrea Pagnani.
\newblock Fast and accurate multivariate gaussian modeling of protein families:
  Predicting residue contacts and protein-interaction partners.
\newblock {\em PLoS ONE}, 9(3):e92721, 2014.

\bibitem{Julia}
Jeff Bezanzon, Stefan Karpinski, Viral Shah, and Alan Edelman.
\newblock Julia: A fast dynamic language for technical computing.
\newblock In {\em Lang.{NEXT}}, April 2012.

\bibitem{borovinskaya2007structural}
Maria~A Borovinskaya et~al.
\newblock Structural basis for aminoglycoside inhibition of bacterial ribosome
  recycling.
\newblock {\em Nature Struct. Mol. Biol.}, 14(8):727--732, 2007.

\bibitem{braun2008NATUREMETHODS}
Pascal Braun et~al.
\newblock An experimentally derived confidence score for binary protein-protein
  interactions.
\newblock {\em Nature methods}, 6(1):91--97, 2008.

\bibitem{nimwegen2008MSB}
Lukas Burger and Erik Van~Nimwegen.
\newblock Accurate prediction of protein$-$protein interactions from sequence
  alignments using a bayesian method.
\newblock {\em Molecular Systems Biology}, 4(165):165, 2008.

\bibitem{cheng2014toward}
Ryan~R Cheng, Faruck Morcos, Herbert Levine, and Jos{\'e}~N Onuchic.
\newblock Toward rationally redesigning bacterial two-component signaling
  systems using coevolutionary information.
\newblock {\em Poc. Natl. Acad. Sci.}, 111(5):E563--E571, 2014.

\bibitem{UniProt}
The~UniProt Consortium.
\newblock Uniprot: a hub for protein information.
\newblock {\em Nucleic Acids Research}, 43(D1):D204--D212, 2015.

\bibitem{dago2012PNAS}
Angel~E. Dago, Alexander Schug, Andrea Procaccini, James~A. Hoch, Martin Weigt,
  and Hendrik Szurmant.
\newblock Structural basis of histidine kinase autophosphorylation deduced by
  integrating genomics, molecular dynamics, and mutagenesis.
\newblock {\em Poc. Natl. Acad. Sci.}, 109(26):E1733--E1742, 2012.

\bibitem{dandekar1998conservation}
Thomas Dandekar, Berend Snel, Martijn Huynen, and Peer Bork.
\newblock Conservation of gene order: a fingerprint of proteins that physically
  interact.
\newblock {\em Trends in biochemical sciences}, 23(9):324--328, 1998.

\bibitem{valencia2013NATREVGEN}
David de~Juan, Florencio Pazos, and Alfonso Valencia.
\newblock Emerging methods in protein co-evolution.
\newblock {\em Nature Reviews Genetics}, 2013.

\bibitem{aurell2013pl}
M.~Ekeberg, C.~L{\"o}vkvist, Y.~Lan, M.~Weigt, and E.~Aurell.
\newblock Improved contact prediction in proteins: Using pseudolikelihoods to
  infer potts models.
\newblock {\em Physical Review E}, 87(1):012707, 2013.

\bibitem{ekeberg2014fast}
Magnus Ekeberg, Tuomo Hartonen, and Erik Aurell.
\newblock Fast pseudolikelihood maximization for direct-coupling analysis of
  protein structure from many homologous amino-acid sequences.
\newblock {\em arXiv preprint arXiv:1401.4832}, 2014.

\bibitem{gap_paper}
Christoph Feinauer, Marcin~J. Skwark, Andrea Pagnani, and Erik Aurell.
\newblock Improving contact prediction along three dimensions.
\newblock {\em PLoS Comput Biol}, 10:e1003847, 10 2014.

\bibitem{PFAM}
Robert~D. Finn, Alex Bateman, Jody Clements, Penelope Coggill, Ruth~Y.
  Eberhardt, Sean~R. Eddy, Andreas Heger, Kirstie Hetherington, Liisa Holm,
  Jaina Mistry, Erik L.~L. Sonnhammer, John Tate, and Marco Punta.
\newblock Pfam: the protein families database.
\newblock {\em Nucleic Acids Research}, 42(D1):D222--D230, 2014.

\bibitem{finn2011hmmer}
Robert~D Finn, Jody Clements, and Sean~R Eddy.
\newblock Hmmer web server: interactive sequence similarity searching.
\newblock {\em Nucleic acids research}, page gkr367, 2011.

\bibitem{HMMER}
Robert~D. Finn, Jody Clements, and Sean~R. Eddy.
\newblock Hmmer web server: interactive sequence similarity searching.
\newblock {\em Nucleic Acids Research}, 39(suppl 2):W29--W37, 2011.

\bibitem{galperin2000s}
Michael~Y Galperin and Eugene~V Koonin.
\newblock Who's your neighbor? new computational approaches for functional
  genomics.
\newblock {\em Nature biotechnology}, 18(6):609--613, 2000.

\bibitem{harrington2008predicting}
Eoghan~D Harrington, Lars~J Jensen, and Peer Bork.
\newblock Predicting biological networks from genomic data.
\newblock {\em FEBS letters}, 582(8):1251--1258, 2008.

\bibitem{ho2002NATURE}
Yuen Ho et~al.
\newblock Systematic identification of protein complexes in saccharomyces
  cerevisiae by mass spectrometry.
\newblock {\em Nature}, 415(6868):180--183, 2002.

\bibitem{hopf2014elife}
Thomas~A Hopf, Charlotta P~I Sch{\"a}rfe, Jo{\~a}o P G L~M Rodrigues, Anna~G
  Green, Oliver Kohlbacher, Chris Sander, Alexandre M J~J Bonvin, and Debora~S
  Marks.
\newblock Sequence co-evolution gives 3d contacts and structures of protein
  complexes.
\newblock {\em eLife}, 3, 2014.

\bibitem{ito2001PNAS}
Takashi Ito, Tomoko Chiba, Ritsuko Ozawa, Mikio Yoshida, Masahira Hattori, and
  Yoshiyuki Sakaki.
\newblock A comprehensive two-hybrid analysis to explore the yeast protein
  interactome.
\newblock {\em Poc. Natl. Acad. Sci.}, 98(8):4569--4574, 2001.

\bibitem{jones2012PSICOV}
D.~T. Jones, D.~W.~A. Buchan, D.~Cozzetto, and M.~Pontil.
\newblock {PSICOV: precise structural contact prediction using sparse inverse
  covariance estimation on large multiple sequence alignments.}
\newblock {\em Bioinformatics}, 28:184, 2012.

\bibitem{juan2008high}
David Juan, Florencio Pazos, and Alfonso Valencia.
\newblock High-confidence prediction of global interactomes based on
  genome-wide coevolutionary networks.
\newblock {\em Poc. Natl. Acad. Sci.}, 105(3):934--939, 2008.

\bibitem{katoh2002mafft}
Kazutaka Katoh, Kazuharu Misawa, Kei-ichi Kuma, and Takashi Miyata.
\newblock Mafft: a novel method for rapid multiple sequence alignment based on
  fast fourier transform.
\newblock {\em Nucleic acids research}, 30(14):3059--3066, 2002.

\bibitem{MAFFT}
Kazutaka Katoh and Daron~M. Standley.
\newblock Mafft multiple sequence alignment software version 7: Improvements in
  performance and usability.
\newblock {\em Molecular Biology and Evolution}, 30(4):772--780, 2013.

\bibitem{trpGE}
Thorsten Knöchel, Andreas Ivens, Gerko Hester, Ana Gonzalez, Ronald Bauerle,
  Matthias Wilmanns, Kasper Kirschner, and Johan~N. Jansonius.
\newblock The crystal structure of anthranilate synthase from sulfolobus
  solfataricus: Functional implications.
\newblock {\em Proceedings of the National Academy of Sciences},
  96(17):9479--9484, 1999.

\bibitem{marcotte2002predicting}
Cynthia J~Verjovsky Marcotte and Edward~M Marcotte.
\newblock Predicting functional linkages from gene fusions with confidence.
\newblock {\em Applied bioinformatics}, 1(2):93--100, 2002.

\bibitem{marks2011direct}
Debora~S. Marks, Lucy~J. Colwell, Robert Sheridan, Thomas~A. Hopf, Andrea
  Pagnani, Riccardo Zecchina, and Chris Sander.
\newblock Protein 3d structure computed from evolutionary sequence variation.
\newblock {\em PLoS ONE}, 6(12):e28766, 12 2011.

\bibitem{MATLAB2014a}
MATLAB.
\newblock {\em version R2014a}.
\newblock The MathWorks Inc., Natick, Massachusetts, 2014.

\bibitem{morcos2011direct}
Faruck Morcos, Andrea Pagnani, Bryan Lunt, Arianna Bertolino, Debora~S Marks,
  Chris Sander, Riccardo Zecchina, Jos{\'e}~N Onuchic, Terence Hwa, and Martin
  Weigt.
\newblock Direct-coupling analysis of residue coevolution captures native
  contacts across many protein families.
\newblock {\em Poc. Natl. Acad. Sci.}, 108(49):E1293--E1301, 2011.

\bibitem{ovchinnikov2014robust}
Sergey Ovchinnikov, Hetunandan Kamisetty, and David Baker.
\newblock Robust and accurate prediction of residue--residue interactions
  across protein interfaces using evolutionary information.
\newblock {\em eLife}, 3, 2014.

\bibitem{Valencia2002B}
Florencio Pazos and Alfonso Valencia.
\newblock In silico two-hybrid system for the selection of physically
  interacting protein pairs.
\newblock {\em Proteins: Structure, Function, and Bioinformatics},
  47(2):219--227, 2002.

\bibitem{pellegrini1999assigning}
Matteo Pellegrini, Edward~M Marcotte, Michael~J Thompson, David Eisenberg, and
  Todd~O Yeates.
\newblock Assigning protein functions by comparative genome analysis: protein
  phylogenetic profiles.
\newblock {\em Poc. Natl. Acad. Sci.}, 96(8):4285--4288, 1999.

\bibitem{procaccini2011dissecting}
Andrea Procaccini, Bryan Lunt, Hendrik Szurmant, Terence Hwa, and Martin Weigt.
\newblock Dissecting the specificity of protein-protein interaction in
  bacterial two-component signaling: orphans and crosstalks.
\newblock {\em PloS one}, 6(5):e19729, 2011.

\bibitem{Schug2009PNAS}
Alexander Schug, Martin Weigt, Jos\'e~N. Onuchic, Terence Hwa, and Hendrik
  Szurmant.
\newblock High-resolution protein complexes from integrating genomic
  information with molecular simulation.
\newblock {\em Poc. Natl. Acad. Sci.}, 106(52):22124--22129, 2009.

\bibitem{Valencia2002A}
Alfonso Valencia and Florencio Pazos.
\newblock Computational methods for the prediction of protein interactions.
\newblock {\em Current Opinion in Structural Biology}, 12(3):368 -- 373, 2002.

\bibitem{weigt2009identification}
Martin Weigt, Robert~A White, Hendrik Szurmant, James~A Hoch, and Terence Hwa.
\newblock Identification of direct residue contacts in protein--protein
  interaction by message passing.
\newblock {\em Poc. Natl. Acad. Sci.}, 106(1):67--72, 2009.

\bibitem{weigt2009PNAS}
Martin Weigt, Robert~A. White, Hendrik Szurmant, James~A. Hoch, and Terence
  Hwa.
\newblock {Identification of direct residue contacts in protein-protein
  interaction by message passing}.
\newblock {\em Poc. Natl. Acad. Sci.}, 106(1):67--72, 2009.

\bibitem{trpAB}
Michael Weyand, Ilme Schlichting, Anna Marabotti, and Andrea Mozzarelli.
\newblock Crystal structures of a new class of allosteric effectors complexed
  to tryptophan synthase.
\newblock {\em Journal of Biological Chemistry}, 277(12):10647--10652, 2002.

\bibitem{trpCF}
Matthias Wilmanns, John~P. Priestle, Thomas Niermann, and Johan~N. Jansonius.
\newblock Three-dimensional structure of the bifunctional enzyme
  phosphoribosylanthranilate isomerase: Indoleglycerolphosphate synthase from
  escherichia coli refined at 2.0 {\mbox{\normalfont\aa}} resolution.
\newblock {\em Journal of Molecular Biology}, 223(2):477 -- 507, 1992.

\bibitem{wlodawer1984structure}
Alexander Wlodawer, Jochen Walter, Robert Huber, and Lennart Sj{\"o}lin.
\newblock Structure of bovine pancreatic trypsin inhibitor: Results of joint
  neutron and x-ray refinement of crystal form ii.
\newblock {\em Journal of Molecular Biology}, 180(2):301--329, 1984.

\bibitem{Haussler2007}
Chen-Hsiang Yeang and David Haussler.
\newblock Detecting coevolution in and among protein domains.
\newblock {\em PLoS Comput Biol}, 3(11):e211, 11 2007.

\end{thebibliography}

\end{document}